\documentclass[12pt]{article}
\usepackage[affil-it]{authblk}
\usepackage{multirow}
\usepackage{titling}
\usepackage{subcaption}
\usepackage{algorithm}
\usepackage[noend]{algpseudocode}
\usepackage{chngcntr}
\usepackage{apptools}
\usepackage{comment}

\usepackage{epigraph}
\usepackage{graphicx}
\usepackage{color}

\usepackage{setspace}
\usepackage{algorithm}
\usepackage[noend]{algpseudocode}

\usepackage{mdframed}

\usepackage{xr}
\definecolor{clemson-orange}{RGB}{234,106,32}
\definecolor{chicago-maroon}{RGB}{128,0,0}
\definecolor{northwestern-purple}{RGB}{82,0,99}
\definecolor{cornell-red}{RGB}{179,27,27}
\definecolor{sauder-green}{RGB}{171,180,0}
\definecolor{gray}{RGB}{192,192,192}
\definecolor{lawngreen}{RGB}{0,250,154}

\usepackage[round]{natbib}  

\makeatletter
\def\BState{\State\hskip-\ALG@thistlm}
\makeatother

\usepackage{amsmath}
\usepackage{amssymb}
\usepackage{amsthm}
\allowdisplaybreaks

\newcommand{\Z}{{\bb Z}}
\newcommand{\N}{{\bb N}}

\DeclareMathOperator{\argmin}{arg\,min}
\DeclareMathOperator{\argmax}{arg\,max}

\newcommand{\vs}{\mathcal{V}}

\theoremstyle{plain}
\newtheorem{theorem}{Theorem}
\newtheorem{lemma}{Lemma}

\newtheorem{proposition}{Proposition}

\theoremstyle{definition}

\newtheorem{definition}{Definition}

\AtAppendix{\counterwithin{lemma}{section}}
\AtAppendix{\counterwithin{proposition}{section}}
\AtAppendix{\counterwithin{corollary}{section}}

\usepackage[centering,marginparwidth=0.9cm]{geometry}

\newcommand{\mar}[1]{}

\usepackage[colorlinks,citecolor=northwestern-purple,urlcolor=chicago-maroon,linkcolor=chicago-maroon,backref=page]{hyperref}

\usepackage[nameinlink]{cleveref}
\crefname{assumption}{Assumption}{Assumptions}
\crefname{lemma}{Lemma}{Lemmas}
\crefname{theorem}{Theorem}{Theorems}
\crefname{corollary}{Corollary}{Corollaries}
\crefname{proposition}{Proposition}{Propositions}
\crefname{claim}{Claim}{Claims}
\crefname{procedure}{Procedure}{Procedures}
\crefname{algorithm}{Algorithm}{Algorithms}
\crefname{figure}{Figure}{Figures}
\crefname{remark}{Remark}{Remarks}
\crefname{section}{Section}{Sections}
\crefname{procedure}{Procedure}{Procedures}
\crefname{example}{Example}{Examples}
\crefname{definition}{Definition}{Definitions}
\crefname{table}{Table}{Tables}
\crefname{equation}{}{}
\crefname{enumi}{}{}
\crefname{conjecture}{Conjecture}{Conjectures}
\crefname{step}{Step}{Steps}
\crefname{appendix}{Appendix}{Appendices}
\crefname{footnote}{Footnote}{Footnotes}

\begin{document}

\global\long\def\vs{\varphi}%
\global\long\def\pl{\underline{p}}%
\global\long\def\ph{\overline{p}}%
\global\long\def\ql{\underline{q}}%
\global\long\def\qh{\overline{q}}%
\global\long\def\rl{\underline{r}}%
\global\long\def\rh{\overline{r}}%
\global\long\def\ch{\overline{c}}%
\global\long\def\cl{\underline{c}}%
\global\long\def\N{\mathbb{N}}%
\global\long\def\Z{\mathbb{Z}}%
\global\long\def\xih{\overline{\xi}}%
\global\long\def\e{\epsilon}
\global\long\def\s{\sigma}
\global\long\def\p{\pi}
\global\long\def\d{\delta}
\global\long\def\t{\theta}

 \definecolor{cornell-red}{RGB}{179,27,27}
 \newcommand{\yk}[1]{\textcolor{northwestern-purple}{[YK: #1]}}
 \newcommand{\sa}[1]{\textcolor{northwestern-purple}{[SA: #1]}}
\newcommand{\ykk}[1]{\textcolor{blue}{#1}}
\newcommand{\saa}[1]{\textcolor{cornell-red}{#1}}

\title{Prolonged Learning and Hasty Stopping:\\ the Wald Problem with Ambiguity\thanks{We thank David Ahn, Nageeb Ali, Renee Bowen, Martin Cripps, Mark Dean, Francesc Dilm\'e, Daniel Gottlieb, Will Grimme, Yoram Halevy, Nenad Kos, Stephan Lauermann, Bart Lipman, Georg Noeldeke, Jawwad Noor, Pietro Ortoleva, Wolfgang Pesendorfer, Larry Samuelson, Ludvig Sinander, Philipp Strack, Bruno Struvolici, Tomasz Strzalecki, Juuso Toikka, Yangfan Zhou and audiences in various seminars and conferences for helpful comments and discussions. Sarah Auster gratefully acknowledges funding from the German Research Foundation (DFG) through Germany's Excellence Strategy---EXC 2126/1---390838866 and CRC TR 224 (Project B02).}}
\author{Sarah Auster\qquad{}Yeon-Koo Che\qquad{}Konrad Mierendorff\thanks{Auster: Department of Economics, University of Bonn (email: auster.sarah@gmail.com);
Che: Department of Economics, Columbia University (email: yeonkooche@gmail.com);
Mierendorff: Department of Economics, University College London}}
\date{September 2023}

\maketitle
 
\begin{abstract}
 This paper studies sequential information acquisition by an ambiguity-averse decision maker (DM), who decides how long to collect information before taking an irreversible action. The agent optimizes against the worst-case belief and updates prior by prior. We show that the consideration of ambiguity gives rise to rich dynamics:   compared to the Bayesian DM, the DM here tends to experiment excessively when facing modest uncertainty and, to counteract it, may stop experimenting prematurely when facing high uncertainty. In the latter case, the DM's stopping rule is non-monotonic in beliefs and features randomized stopping.\\  
 
\noindent \textbf{Keywords}: Wald problem, ambiguity aversion, prolonged learning, preemptive stopping\\
 \textbf{JEL Classification Numbers:} C61, D81, D83, D91
\end{abstract}

\section{Introduction}
 
The problem of making a decision on an action after deliberating on its merits is ubiquitous in many situations of life. Examples include a firm contemplating a potential new project, a jury deliberating on a verdict, a citizen pondering over his/her political vote, or a scientist evaluating a hypothesis.   The pervasiveness of such a problem makes the framework of central importance, as seen by the extensive research on the subject matter; see \cite{wald1947foundations}, \cite{arrow1949}, \cite{moscarini2001optimal}, \cite{fudenberg2018speed}, \cite{Che2019}, among others. With the exception of a few recent papers, the vast literature studying the problem considers a Bayesian framework. The Bayesian model misses, however, a crucial aspect of the problem: the decision maker (DM) often lacks a clear model or beliefs about the uncertainty she is facing. In fact, Wald, often regarded as the first to formally introduce the framework, was mindful of the ambiguity and suggested a robust approach \cite{wald1947foundations}, which anticipates the maxmin decision rule axiomatized by \citet{gilboa1989maxmin}.

We adopt  the model in \citet{gilboa1989maxmin} by considering a DM who has ambiguous beliefs about the state of the world and evaluates her choices against the worst-case scenario in a set of plausible priors. Specifically, the state is either $L$ or $R$ and the DM must decide between two irreversible actions, $\ell$ and $r$. At each point, the DM can stop and choose an action. Alternatively, she can postpone her action and learn about the state. In the baseline model, the learning takes the form of a Poisson breakthrough that confirms state $R$. Imagine, for instance, a CEO of a biotech company seeking to establish the efficacy of a new drug or a theorist seeking to prove a theorem. The Bayesian optimal strategy in such a situation calls for looking for evidence of state $R$---efficacy of the drug or validity of the proof---for a duration of time until a long stretch of unsuccessful attempts convinces her to declare $R$ to be unlikely and stop for action $\ell$.  

Introducing ambiguity in a dynamic setting raises conceptual issues about updating and dynamic consistency. To avoid the latter, the existing literature often follows a recursive approach, requiring the DM to adopt a prior-by-prior updating rule on a {\it rectangular} set of priors (see \citealt{epstein2003recursive}). 
Rectangularity requires that the set of priors is rich enough to contain a single worst-case belief that guides the DM's conditional choices. We follow a different approach: rather than assuming away dynamic inconsistency, we seek to understand its behavioral implications for a ``sophisticated'' DM who recognizes potential preference reversals and deals with them in a forward-looking and rational manner. To this end, we maintain prior-by-prior updating but allow for changes in the conditional worst-case scenario. 

The paper shows that the consideration of ambiguity and time inconsistency gives rise to rich dynamics in sequential learning. Late in the process, ambiguity increases the DM's incentives to confirm the state, thereby prolonging experimentation. After a period of unsuccessful search for $R$-evidence, the DM contemplates two possibilities: no such evidence exists because the state is $L$ or instead the state is $R$ and she has simply been unlucky. As time progresses, the DM becomes increasingly convinced of the former. Following a sufficiently long period of futile search, the Bayesian DM would thus stop to take action $\ell$. The ambiguity-averse DM, on the other hand, continues to worry about the second possibility and, as a consequence, keeps looking for more proof. Ambiguity aversion thus causes the DM to continue experimenting well past the time at which the Bayesian counterpart would stop. In the context of our example, the biotech CEO or the theorist refuses to see the ``writing on the wall,'' and gets hung up on proving the efficacy of the drug or the validity of the theorem. From the perspective of an outsider, the ambiguity-averse DM would thus be seen to exhibit prolonged indecision.

Early on, however, the DM is concerned about the opposite scenario, namely that $R$ is relatively unlikely and the search for evidence will be in vain. From this vantage point, her later self's refusal to discount $R$ sufficiently enough to stop experimentation in time is seen as misguided. Recognizing the peril of continued indecision, the DM must then intervene and stop before it is too late, namely before the change in the worst-case belief occurs. The DM adopts a mixed strategy in this case, randomizing between experimentation and stopping.  Mixing here hedges against  ambiguity, as is well-known in the literature.  What is novel here is the form it takes; she randomizes in stopping according to a Poisson process at a rate suitably calibrated to hedge against both events.  The stopping occurs around the point of belief switch, which is typically well before the Bayesian DM would find it reasonable to stop. Hence, to an outside observer, the DM would be seen to exhibit premature decisiveness.

 Comparative statics yields further implications of ambiguity.  An increase in ambiguity---as measured by the size of the initial set of priors---increases the tendency for the DM to prolong her experimentation later in the game, which increases her need to preemptively stop earlier in the process and, naturally, the rate at which she does so. These effects, taken together, imply that increased ambiguity ``blunts'' the DM's ability to adapt her learning strategy to the true likelihood of states. In particular, she fails to stop learning when learning becomes unprofitable (i.e., when the true likelihood of $R$ falls low), so that given a sufficiently large ambiguity, the DM's expected experimentation time {\it increases} rather than decreases (as would occur under a Bayesian DM) when $R$ becomes unlikely. The combined effects also imply that the stopping time is more dispersed with larger ambiguity. 

Finally, we extend the baseline model to the case of incremental learning with diffusion signals. Although some details of the solution change, we show that the main features of the model---excessive learning and preemption via randomized stopping---persist in this case.

In sum, our analysis captures two behavioral traits---protracted indecision and hasty, often unwarranted, decisiveness. The other main contribution of the paper is methodological. Analyzing dynamic decision-making under ambiguity is a priori difficult because the associated time inconsistency problem makes the dynamic programming machinery inapplicable. We characterize the solution to our problem via a set of novel Hamilton-Jacobi-Bellman (HJB) conditions  that account for potential differences in preferences of current and future selves. Our HJB characterization  thereby restores the dynamic programming logic in the time-inconsistent environment and hence serves as a powerful toolkit that can make analysis tractable without compromising sharp prediction. We expect the approach, and the micro-foundation justifying it, to be valuable beyond our setting.\\

\emph{Related Literature.}  We incorporate ambiguity into an optimal stopping framework \`a la \citet{wald1947foundations}. Optimal stopping problems in the Bayesian framework with Poisson signals have been analyzed in \citeauthor{peskir2006optimal} (\citeyear{peskir2006optimal}, ch.VI), with economic applications including \cite{Che2019} and \cite{nikandrova2018dynamic}. Other applications adopt a drift-diffusion model, e.g. \cite{moscarini2001optimal}, \cite{fudenberg2018speed}, while \cite{zhong2017optimal} studies the problem under flexible choice of information.

The current paper is part of a growing literature on optimal stopping problems under ambiguity. Most of the existing literature adopts a recursive approach, thereby precluding the possibility of changes in the worst-case scenario and time inconsistency; examples include \cite{epstein2007learning}, \cite{riedel2009optimal}, \cite{miao2011risk},  and \cite{cheng2013optimal}.  Closest to our paper in this literature is \cite{epstein2022optimal}, which studies a Wald problem in a drift-diffusion framework with recursive preferences. Under the recursive approach, there is a single prior in the initial set that minimizes the DM's expected payoff at every point in time. Due to this feature, standard dynamic programming methods are applicable and the optimal stopping rule continues to be described by two stopping boundaries, as in the Bayesian benchmark. By contrast, dynamic inconsistency figures prominently in our model, which we analyze by developing a saddle-point version of the dynamic programming technique.

There is also a growing literature looking at the implication of dynamic inconsistency due to ambiguity in different applications. For example, \cite{bose2009dynamic}, \cite{bose2014mechanism}, \cite{ghosh2021sequential} and \cite{auster2022robust} consider auctions/mechanisms with ambiguity-averse agents, while \cite{kellner2018endogenous} and \cite{beauchene2019ambiguous} study ambiguity and updating in communication and persuasion settings.

Finally, there is a literature studying dynamic inconsistent stopping in other behavioral frameworks. For instance, \cite{barberis2012model}, \cite{xu2013optimal}, \citeauthor{ebert2015until} (\citeyear{ebert2015until}, \citeyear{ebert2018never}), and \cite{henderson2017randomized} analyze stopping problems under prospect theory. \citeauthor{christensen2018finding} (\citeyear{christensen2018finding,christensen2020time}) study stopping problems for a class of time-inconsistent models, which can capture endogenous habit formation, non-exponential discounting, and mean-variance utility. 

\section{Model}\label{sec:model}

A DM must choose between two actions, $r$ and $\ell$, whose payoffs depend on an unknown state $\omega\in\{R,L\}$. Specifically, the DM's  payoffs satisfy $u_r^R>u_{\ell}^R$ and $u_{\ell}^L>u_{r}^L$, where  $u_a^{\omega}\in\mathbb R$ denotes her payoff from choosing action $a\in\{r,\ell\}$ in state $\omega\in\{R,L\}$. The DM thus wants to ``match'' the action with the state. We further assume  $
u_r^R>u_r^L$ and $u_{\ell}^R<u_{\ell}^L.$  This means that whether a state is desirable or not depends on the action that the DM takes, a feature that makes ambiguity nontrivial.\footnote{If this assumption does not hold, the ambiguity-averse DM minimizing over a set of probabilities is behaviorally indistinguishable from a Bayesian DM, whose prior is equal to the element of the set that maximizes the likelihood of the ``bad'' state.}
Let $U_a(p)=pu_a^R+(1-p)u_a^L$ denote the DM's payoff associated with action $a$ when the probability of state $R$ is $p$.

The DM may delay her action and acquire information. We model the DM's information acquisition problem as a stopping problem in continuous time. At each point in time $t\geq0$, the DM decides whether to wait for more information or whether to stop and take an immediate action, $\ell$ or $r$.  If the DM waits (or ``experiments''), she incurs a flow cost $c>0$ per unit of time. Payoffs are realized when the DM stops and takes an action. Information takes the form of $R$-evidence which is received only in state $R$ with arrival rate $\lambda>0$. Observing news thus conclusively reveals state $R$, while the absence of news is a signal favoring state $L$.\\

\emph{Bayesian benchmark.} Considering the benchmark case of a Bayesian DM, let $p_t$ denote the probability that the DM assigns to state $R$ at time $t$. In the absence of news, the DM's belief drifts ``leftward'' according to the law of motion:
\begin{align} \label{eq:law}
\dot{p}_t=\eta(p_t):=-\lambda p_t(1-p_t).
 \end{align}
Intuitively, the DM becomes pessimistic toward $R$ when she looks for $R$-evidence but is unable to find any.

As is well known, the Bayesian optimal information acquisition rule can be described by two thresholds, $p^B_{\ell}$ and $p^B_{r}$. If $p_t\leq p^B_{\ell}$ or $p_t\geq p^B_{r}$, the DM is sufficiently confident in the state and takes the appropriate action without gathering any additional information. If instead $p_t$ belongs to $(p^B_{\ell},p^B_{r})$, the DM experiments until either she receives $R$-evidence or the Bayesian update of $p_t$ reaches the threshold $p^B_{\ell}$.
\[
0\underbrace{\vphantom{p^B_{\ell}}\text{------------}}_{\text{action }\ell}p^B_{\ell}\underbrace{\vphantom{\ph^{1}}\hspace{-5pt}\longleftarrow\longleftarrow\longleftarrow\longleftarrow\longleftarrow\longleftarrow\longleftarrow\longleftarrow}_{\text{experimentation}}p^B_{r}\underbrace{\vphantom{\ph^{1}}\text{------------}}_{\text{action }r}1
\]
Let $\phi(p; p', U_{\ell}(p'))$ denote the value to the Bayesian DM of experimenting until the belief $p$ reaches $p'<p$, assuming that if she receives $R$-evidence during the experimentation, she realizes $u^R_r$, while if she does not receive such evidence until her belief reaches $p'$, she stops and takes  action $\ell$. The standard HJB conditions characterize the DM's value function as a solution to an ODE:
\begin{equation} \label{eq:Bayesian-ODE}
    c=\lambda p(u_r^R-\phi(p))+\phi'(p)\eta(p), \forall p>p'.
\end{equation}
Intuitively, the flow cost of experimentation (LHS) equals the rate of value increase due to possible breakthrough news and belief updating in its absence (RHS). Optimal stopping calls for $p'$ to be chosen optimally, so the value of the stopping problem is 
\begin{equation} \label{eq:Bayesian-Stopping Value}
\Phi(p)=\max_{p'\in[0,p]}\phi(p;p',U_{\ell}(p')),
\end{equation} 
and the left stopping boundary $p_{\ell}^B$ is the associated maximizer.\footnote{ One can characterize the optimal stopping boundary $p_{\ell}^B$ by invoking value matching and smooth pasting.  Note also that if $c$ is sufficiently large, then the optimal solution is to stop immediately; \ that is, $p'=p$.} Finally, since the DM has the option to choose $r$ at any point in time, the DM with belief $p$ enjoys a payoff of 
\begin{equation} \label{eq:Bayesian-Value}
\Phi^*(p)=\max\{\Phi(p),U_r(p)\},
\end{equation} 
with the DM indifferent between experimentation and action $r$ at the right stopping boundary $p_r^B$. We will call $\Phi^*(p)$ the \emph{Bayesian value function} throughout. Its characterization is relegated to \Cref{app:Bayesian}.

\begin{figure}
\begin{centering}
\includegraphics[width=0.6\textwidth]{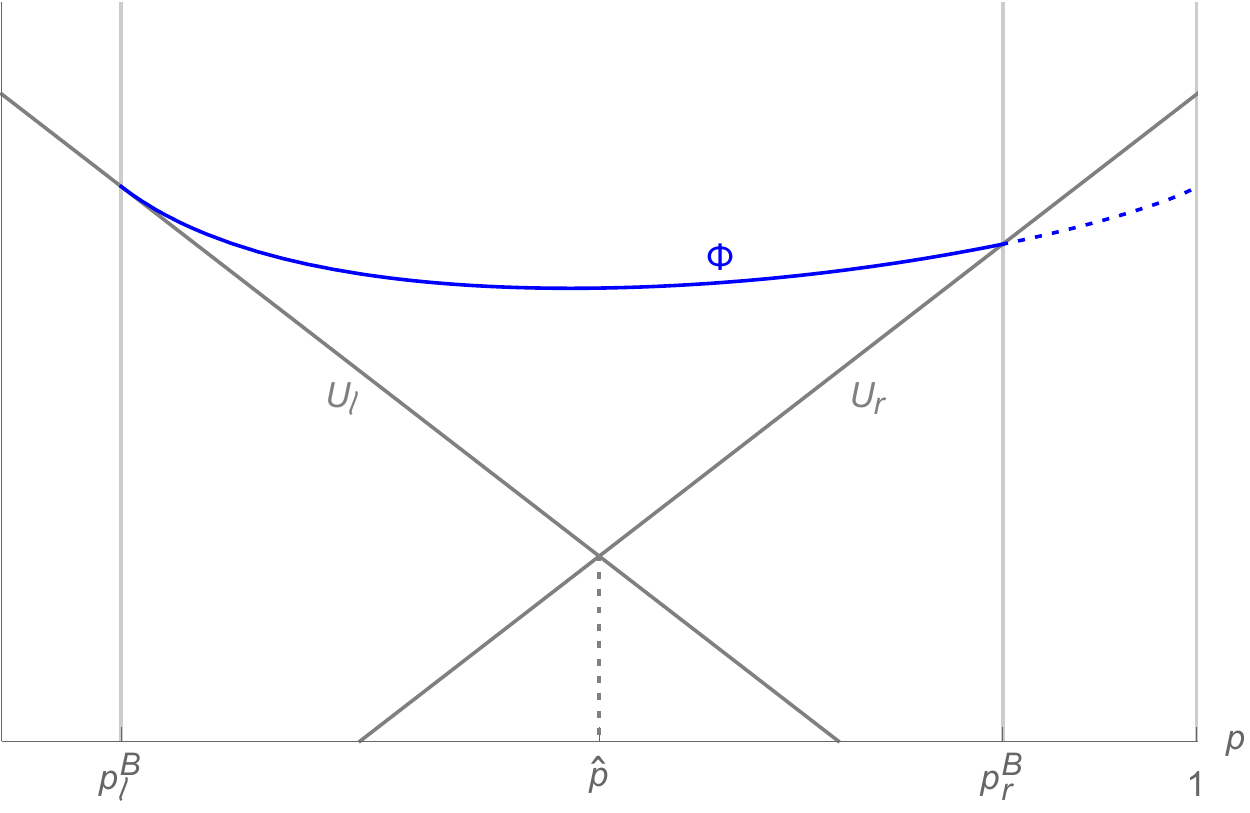}
\par\end{centering}
\caption{Bayesian optimal stopping rule.\label{fig:Bayesian}}
\end{figure}

\cref{fig:Bayesian} illustrates graphically the value of the Bayesian DM in a typical situation. The downward sloping line shows the DM's expected payoff from taking action $\ell$ as a function of the probability $p$ she assigns to state $R$. Similarly, the upward-sloping line represents the DM's expected payoff from taking action $r$.  The crossing point $\hat p$ is the belief at which the DM is indifferent between actions $\ell$ and $r$. The blue curve depicts $\Phi$, i.e. the value from experimenting with stopping belief $p_{\ell}^B$. The optimal value $\Phi^*$ can be seen as the upper envelope of $\Phi$ and $U:=U_{\ell}\vee U_r$. Naturally, the experimentation range $(p_{\ell}^B,p_r^B)$ depends on the flow cost c; the range shrinks as $c$ increases and disappears when 
\begin{eqnarray}\label{eq:c-bar}
c\ge \overline{c}:=\lambda\frac{(u_{r}^{R}-u_{\ell}^{R})(u_{\ell}^{L}-u_{r}^{L})}{(u_{r}^{R}-u_{\ell}^{R})+(u_{\ell}^{L}-u_{r}^{L})}.
\end{eqnarray}
  The threshold $\overline c$ is the value of $c$ at which $\Phi(\hat p)=U(\hat p)$. \\

\noindent\emph{Ambiguity and Updating.} We assume that the DM faces ambiguity over the likelihood of state $R$ and that her preferences are represented by the maxmin expected utility model \citep{gilboa1989maxmin}. According to this model, the DM seeks to maximize her payoff guarantee across a set of priors, described by an interval of probabilities on state $R$: $\mathcal{P}_0=[\pl_0,\ph_0]$. We assume $0<\pl_0\leq\ph_0<1$. 

A key question is how the DM updates her ambiguous beliefs as information arrives. We assume that the DM uses \emph{Full Bayesian Updating}: she updates each element of $\mathcal{P}_0$ according to Bayes rule and considers the worst case over the set of posterior beliefs obtained in this way. The set of posteriors at time $t>0$ is thus given by an interval $\mathcal{P}_t:=[\pl_t,\ph_t]$, where $\pl_t$ is the Bayesian update of the initial lower bound $\pl_0$ and $\ph_t$ is the Bayesian update of the initial upper bound $\ph_0$. 

As time progresses and no breakthrough occurs, the interval $\mathcal{P}_t$ drifts to the left. In the process, the interval expands in size as it moves away from the right boundary and shrinks as it approaches the left boundary  (ultimately, both $\pl_t$ and $\ph_t$ converge to zero). 
 Rescaling the belief in terms of a log-likelihood ratio keeps the interval ``size'' constant.  Hence, from now on, we shall refer to
\[
\Delta:=\ln\left(\frac{\ph_0}{1-\ph_0}\right)-\ln\left(\frac{\pl_0}{1-\pl_0}\right)
\]   as {\it the degree of ambiguity}. It is easy to verify that $\Delta$ does not vary as one updates beliefs over time.\\

\noindent\emph{Preference reversals.} As is well known, the Full Bayesian Updating rule can lead to violations of dynamic consistency.\footnote{Beyond the issue of dynamic inconsistency, Full Bayesian Updating, also referred to as  {\it prior-by-prior updating}, has been criticized in the literature for implying a possible situation in which ``all news is bad news'' (see \cite{gul2021evaluating} and \cite{shishkin2023ambiguous}). Since in our setting $R$-evidence is always good news, this type of situation does not arise here.}  This feature will play a crucial role in our model. To explain the source of dynamic inconsistency in our setting, it will be useful to first consider the DM's optimal information acquisition rule from the perspective of time $0$; we will then show why, after updating the belief set, the DM may have the incentive to deviate from it. 

\begin{figure}
     \centering
     \begin{subfigure}[b]{0.45\textwidth}
         \centering
         \includegraphics[width=\textwidth]{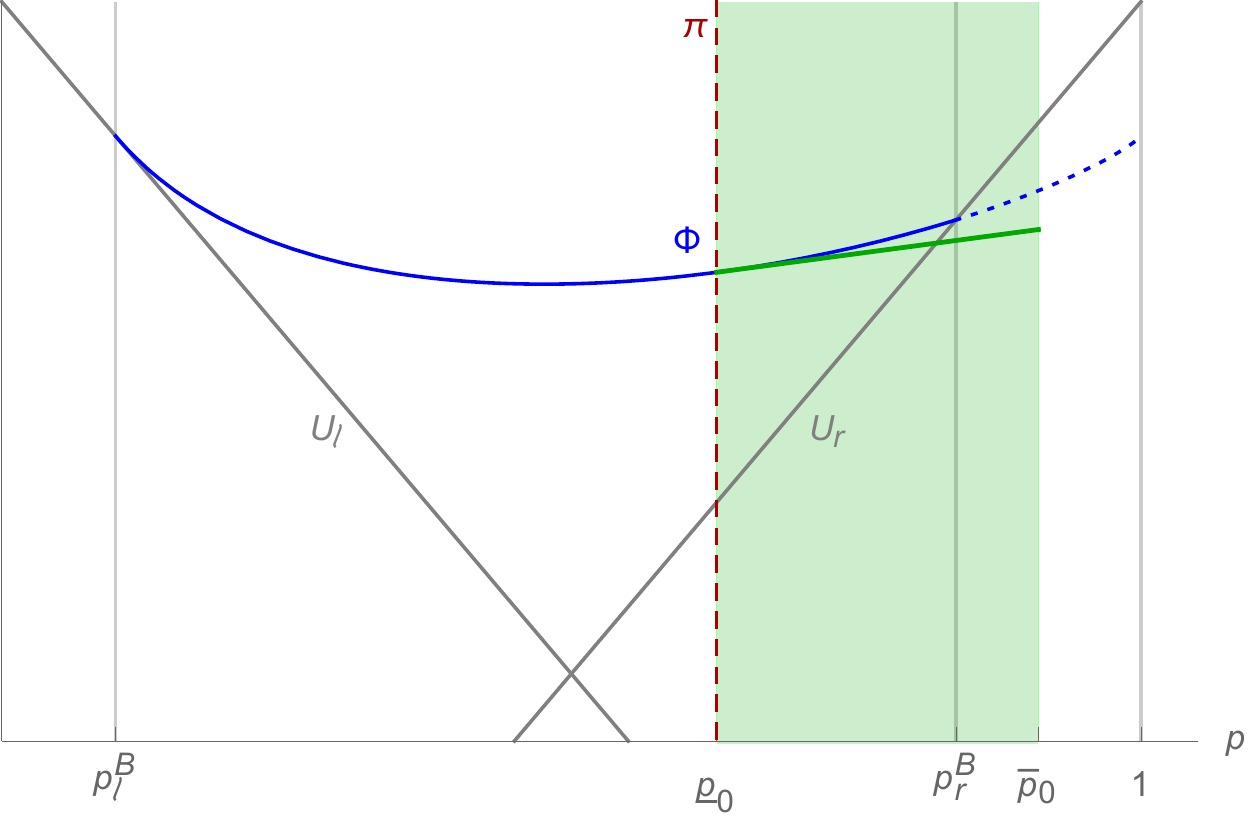}
     \end{subfigure}
     \begin{subfigure}[b]{0.45\textwidth}
         \centering
         \includegraphics[width=\textwidth]{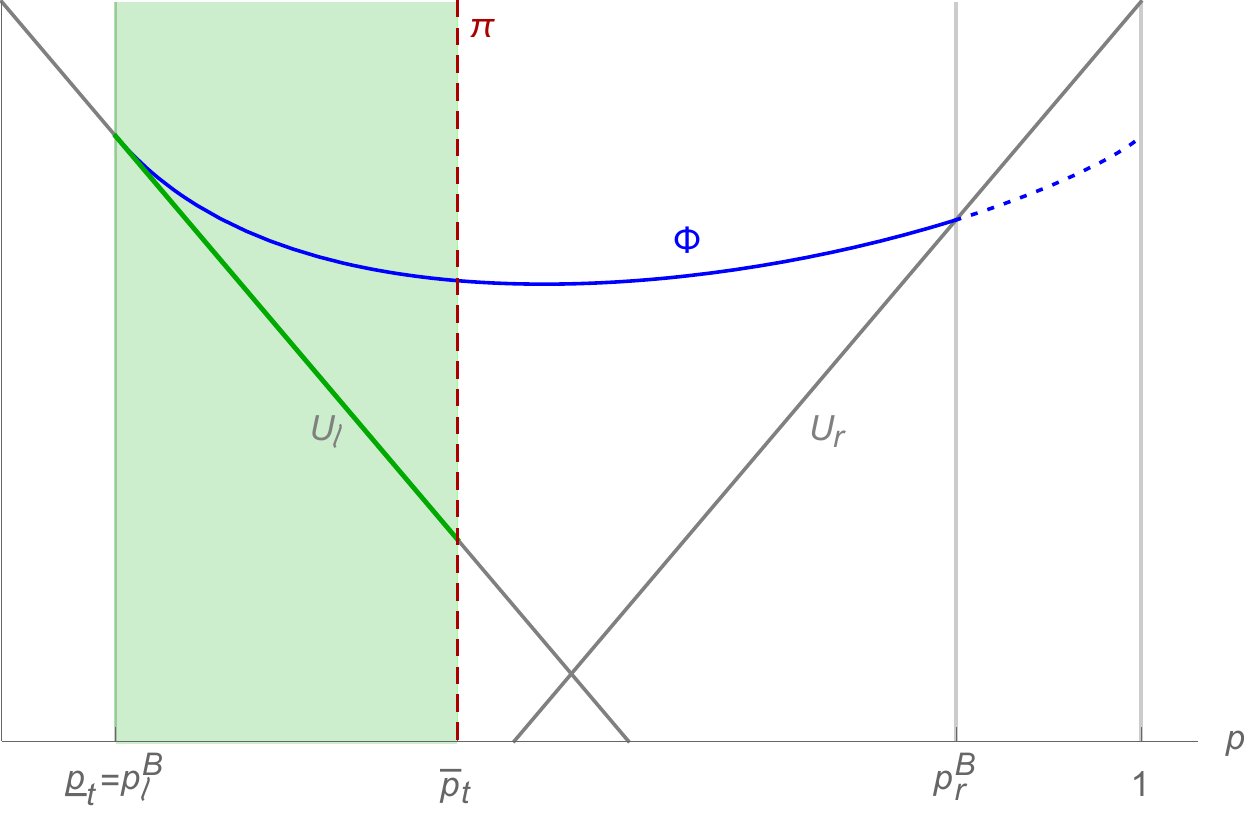}
     \end{subfigure}
     \caption{Worst-case scenario (indicated by the red dashed line) at the start of experimentation (left panel) and at the ex-ante optimal stopping time (right panel).}
     \label{fig:naive}
\end{figure}

The ex-ante optimal rule solves the DM's optimization problem in a hypothetical setting where she can commit to a dynamic strategy that maximizes her expected payoff against the worst prior in $\mathcal{P}_0$. This optimization problem is clearly independent of the assumed updating rule and thus constitutes an important benchmark. Finding the solution amounts to analyzing a zero-sum game between the DM and adversarial nature choosing $p\in \mathcal{P}_0$ to minimize the DM's expected payoff. Its equilibrium---a saddle point---describes the DM's maxmin optimal value obtained from any information acquisition rule. \Cref{app:commitment} establishes the existence of a saddle point. Given a saddle point, its value also equals the value of the minmax problem in which nature first chooses a prior in $\mathcal{P}_0$ and the DM then selects the best strategy for the chosen prior (see \cite{Osborne1994} Proposition 22.2, for example). It follows that the minmax value is simply the lowest Bayesian value $\Phi^*$ within $\mathcal{P}_0$, or 
$$
 \min_{p\in \mathcal{P}_0}\Phi^*(p).$$
 This value coincides with the Bayesian value at the left-most belief $\pl_0$ if $\mathcal{P}_0$ is sufficiently far to the right, at the right-most belief $\ph_0$ if $\mathcal{P}_0$ is sufficiently far to the left, and at an interior belief if $\mathcal{P}_0$ contains the global minimizer of $\Phi^*$ (see \cref{fig:naive} for the first two cases). For later purposes, we denote the minimizer of $\Phi^*$ by 
$$p_*:=\arg\min_{p\in [0,1]} \Phi^*(p).$$

While the maxmin commitment value is easily obtained, the saddle point strategy on the part of the DM requires a little care.   As we show in \Cref{app:commitment}, if $p_*< p^B_r$ or $p^B_r\not\in \mathcal{P}_0$, then the maxmin strategy is indeed the Bayesian optimal strategy for the worst belief in $\mathcal{P}_0$.  If $p_*= p^B_r\in \mathcal{P}_0$, however, nature's indifference, which is needed for the saddle point requirement, calls for the DM to randomize.  If $c<\overline c$, then the DM must randomize between $r$ and experimentation. If instead $c\geq\overline c$, experimentation is too costly and the DM mixes between actions $r$ and $\ell$ with probabilities $\hat \rho$ and $1-\hat \rho$, respectively, where $\hat \rho$ satisfies 
\begin{eqnarray}\label{eq:rho-hat}
\hat\rho U_r'(p)+(1-\hat\rho)U_{\ell}'(p)=0.
\end{eqnarray}

To see now why the described solution may be incompatible with the preferences of the DM's later selves, suppose the interval $\mathcal{P}_0$ is sufficiently far to the right so that $\pl_0$ minimizes $\Phi^*$ over $\mathcal{P}_0$ (see the left panel of \cref{fig:naive}). If $\pl_0<p^B_{r}$, the DM's optimal (commitment) plan at time $0$ is to experiment until either she receives $R$-evidence or  the Bayesian update $\pl_t$ of $\pl_0$ reaches $p^B_{\ell}$. Once the intended stopping point $t$ where $\pl_t=p^B_{\ell}$ is reached, however, the worst-case belief is no longer given by $\pl_t$ but instead by $\ph_t$, as illustrated in the right panel of  \cref{fig:naive}. According to $\ph_t$, the optimal plan (as prescribed by the commitment solution) is to acquire more information, until the Bayesian update of $\ph_t$ reaches the stopping boundary $p^B_{\ell}$. Hence, later selves of the DM will renege on the original plan  prescribed by the commitment solution  and {\it experiment excessively} from the perspective of the original self.

A dynamic inconsistency arises as a result of a switch in the worst-case scenario. When the DM experiments, she is concerned about two events: 1) not receiving $R$-evidence in state $L$ and 2) taking action $\ell$ in state $R$. At the start of the experimentation phase, pessimism about the former event dominates and the worst-case belief is given by $\pl_t$, which minimizes the likelihood of $R$. When the DM gets closer to the intended stopping point, however, the concern about taking the wrong action looms large and the worst-case belief switches to $\ph_t$, i.e. the belief maximizing the likelihood of $R$. The new worst-case belief calls for more experimentation, thereby creating a conflict with the preferences of earlier selves. A sophisticated DM understands that later selves deviate from the commitment plan by experimenting for too long and optimizes accordingly. We next analyze the behavior of such a DM.

\section{Sophisticated Stopping Rule}\label{sec:main}

We now analyze the dynamic behavior of a sophisticated DM.  The dynamic inconsistency problem requires us to analyze the DM's stopping problem as an intrapersonal game in which a current self plays against future selves.

\subsection{Intrapersonal game.}

At each point in time $t\geq0$, the DM decides between continuing to gather information and stopping, taking as given the choices of her future selves. If a breakthrough occurs, all posteriors in the updated set are equal to one, so it is optimal for the DM to stop and take action $r$. Taking this part of the strategy as given, it is without loss of generality to focus on the Markov strategy that depends  on  the set of posteriors  rather than time.  Indeed, in the absence of $R$-evidence, there is a one-to-one mapping between calendar time and the set of posteriors. Having fixed $\Delta$ as a primitive, we can index the possible sets of posteriors by their upper bound $\ph\in(0,1)$. The right-most belief in the set, $\ph$, is thus our state variable. We denote the set of posteriors at state $\ph$ by
\[
\mathcal{P}(\ph):=\left\{p\leq\ph:\ln\left(\frac{\ph}{1-\ph}\right)-\ln\left(\frac{p}{1-p}\right)\leq\Delta\right\},
\]
and define $\pl(\ph):=\min \mathcal{P}(\ph)$ as the lower bound of this set.  Clearly, it follows that  $\pl_t=\pl(\ph_t)$.

A Markov strategy is a measurable function $\sigma:(0,1)\rightarrow[0,\infty)\times[0,1]\times[0,1]$, with $\sigma(\ph)=(\nu(\ph),m(\ph),\rho(\ph))$. The strategy specifies a \emph{stopping rate} $\nu(\ph)$, an {\it instantaneous stopping probability}
$m(\ph)$, and the {\it probability of taking action $r$ conditional on stopping} $\rho(\ph)$. Intuitively, densities in the distribution of stopping times are accommodated by interior stopping rates $\nu(\ph)$, while atoms in the stopping time distribution are captured by positive values of $m(\ph)$.\footnote{One can show that our Markov strategy is equivalent to an alternative formulation that specifies the DM's strategy by a family of stopping times, or more precisely by a family  $\{F_t\}_{t\ge 0}$ of distributions satisfying some restrictions such as right-differentiability, where $F_t(\tau)$ is the probability that the DM stops by time $t+\tau$ conditional on not having received any breakthrough news by time $t$. See \Cref{sec:stopping time} for details.}  To ensure that a strategy $\sigma$ induces a well-defined stopping time, we impose the following {\it admissibility} restrictions on strategies: (1) every connected set $\mathcal P\subseteq[0,1]$ with $m(\ph)=1$ for all $\ph\in \mathcal P$ contains its supremum; (2) there is a countable set $M^{\sigma}\subset(0,1)$ such that $m(\ph)\in(0,1)$ only if $\ph\in M^{\sigma}$.\footnote{These conditions correspond to the requirement that the stopping-time distribution is right-continuous and differentiable at the points of continuities. Specifically, condition (1) ensures that the stopping time is well-defined when the belief drifts toward a (left) stopping boundary.  Meanwhile, Condition (2) corresponds to the fact that a distribution function can have at most countably many points of discontinuities.}

Fixing a strategy $\sigma$, we denote by $V^{\sigma}(p,\ph)$ the DM's expected payoff at state $\ph$ when the strategy is evaluated at belief $p$. The detailed expression of $V^{\sigma}(p,\ph)$ can be found in \Cref{sec:strategies}; see \cref{eq:value}.

Nature's goal is to minimize $V^{\sigma}(p,\ph)$ at each state $\ph$. Since $V^{\sigma}(p,\ph)$ is a convex combination of the payoffs that strategy $\sigma$ yields in state $R$ and in state $L$ with coefficient $p$, the DM's value $V^{\sigma}(p,\ph)$ is linear in the first argument. Nature's problem thus has a  corner solution, except when $V^{\sigma}_p(\cdot,\ph)=0$. Letting $\pi(\ph)\in \mathcal{P}(\ph)$ denote nature's choice at state $\ph$, we define a solution concept as follows.

\begin{definition}\label{def:pseudo}
A strategy $\sigma=(\nu,m,\rho)$ constitutes a  \emph{pseudo equilibrium} if for each state $\ph\in(0,1)$, there exists some $\pi=\pi(\ph)$ with
\begin{equation}
\pi(\ph)\in\arg\min_{p\in \mathcal{P}(\ph)}V^{\sigma}(p,\ph) \label{eq:worst-case-belief_proof1}
\end{equation}
such that\footnote{Recall that $U=U_{\ell}\vee U_r$ is the maximum stopping payoff from actions $\ell$ and $r$.}
\begin{align}
& U(\pi)\leq V^{\sigma}(\pi,\ph);\label{eq:UvsValue}\\
&U(\pi)=V^{\sigma}(\pi,\ph) \mbox{ if } \nu(\ph) \mbox{ or } m(\ph)>0;\label{eq:opt-stop}\\
&\rho(\ph)\in\arg\max_{\rho\in[0,1]}\rho U_r(\pi)+(1-\rho)U_{\ell}(\pi).\label{eq:opt-action}
\end{align}
\end{definition}

A strategy $\sigma$ is a pseudo equilibrium if, given nature's choice $\pi$, no self of the DM has the incentive to deviate.\footnote{Note that the conditions in  \cref{def:pseudo} and later in \cref{def:intra} make no reference to the initial state $\ph_0$.  This means that the same strategy should be employed, \emph{regardless of  the initial state.} While this is, in principle, a restriction, this restriction entails no loss except for one knife-edge case.  See \cref{ft:eq} for details.} Due to the continuous-time nature of our game, many  pseudo-equilibria exist. For instance,  it is a pseudo equilibrium for the DM to stop at every state and choose the optimal action against the worst-case belief. If some self of the DM were to deviate and refuse to stop, her infinitesimally close future selves would stop immediately. Since experimentation for an instant yields a zero probability of a breakthrough, the deviation is never profitable. However, the prescribed strategy is implausible and simply an artifact of the continuous time nature of our game.

To rule out such implausible strategies, we focus on the pseudo equilibria in which the DM can control her action for a vanishingly small amount of  time. Formally, for any pseudo equilibrium $\sigma$ we imagine a DM  who can commit at each time $t$ her action for the period $[t,t+\epsilon)$ against adversarial nature, anticipating that $\sigma$ will follow after $t+\epsilon$. We then require $\sigma$ to be obtained as a limit of such $\epsilon$-commitment solutions as $\epsilon\to 0$.\footnote{This refinement is similar in effect to taking a limit of the behavior in discrete times models as the time interval shrinks to zero; in such models,  the DM's experimentation during a unit period  has a non-negligible probability of producing a breakthrough. The current refinement accomplishes the same effect within a continuous time framework.} Such a vanishing control over future behavior avoids total coordination failure among the different selves.  As justified in \Cref{sec:microfound-HJB} more precisely, the refinement introduces a  version of dynamic programming conditions.

To begin, for a value function $V: [0,1]^2 \to \mathbb{R}$, define a functional  
\begin{align*}
G(m,\nu,\rho,p,\ph,V,dV)=&m\left[U_{\rho}(p)-V(p,\ph_{-})\right]+(1-m)\left[-c+\nu(U_{\rho}(p)-V(p,\ph_{-}))\right.\\
&\left.+p\lambda(u_r^R-V(p,\ph_{-}))+V_p(p,p_{-})\eta(p)+V_{\ph}(p,\ph_{-})\eta(\ph)\right],
\end{align*}
where $dV$ is the gradient of the value function,  $V(p,\ph_{-}):=\lim_{\ph'\uparrow\,\ph}V(p,\ph')$,
and $V_x(p,p_{-}):=\lim_{\ph'\uparrow\,\ph}\partial V(p,\ph')/\partial x$.\footnote{We use the left limits of the value function and its partial derivatives
	(a) to make sure the conditions are well defined even at points where
	the derivatives do not exist and (b) since the belief drifts downwards
	and the value function in the HJB equation reflects the continuation
	value.}
Recall $\eta(p)=-\lambda p(1-p)$ denotes the law of motion. The functional $G$ has a familiar interpretation from the dynamic programming literature. Its first term captures a value increase when the DM stops with a point mass $m$; she collects $U_{\rho}(p)-V(p,\ph_{-})$. The second term (inside the square brackets) captures a value increase when the DM ``continues.''  In this case, at the flow cost $c$, the DM engages in a flow stopping at rate $\nu$ and experimentation yielding a breakthrough at rate $\lambda p$: her value increases by $U_{\rho}(p)-V(p,\ph_{-})$ in the former case and by $u_r^R-V(p,\ph_{-})$ in the latter case. The last two terms track the change of value arising from the updating of the state $\ph$ and the updating of the current worst-case belief.  This part of the HJB is non-standard and accounts for dynamically-inconsistent belief changes. To compare, in the standard case without ambiguity, the DM's belief $p$ is the only payoff relevant variable. Hence, the first term $V_p^{\sigma}(p,\ph)$ would be present, while the second-term $V_{\ph}^{\sigma}(p,\ph)$ would not. In the current case, the worst-case belief tomorrow may differ from tomorrow's update of today's worst-case belief,\footnote{Specifically, a dynamic inconsistency arises when $\pi(\ph +\eta(\ph) dt)\ne \pi(\ph) + \eta(\pi(\ph))dt$. } making it necessary to account for changes both in the state $\ph$, which determines the continuation strategy, and in the belief $p$ with which the strategy is evaluated.

For each $\ph \in (0,1)$, consider a saddle-point version of Hamilton-Jacobi-Bellman equations:\footnote{Similar to the commitment solution, a saddle point here characterizes the maxmin optimality of the DM's strategy (for a vanishingly near future). Equations \eqref{eq:HJB0}-\eqref{eq:HJBpi} resemble Isaac's condition, which characterizes the equilibrium of dynamic zero-sum games played by two long-run players (see \cite{bardi/capuzzo-dolceta:97}, page 446, and \cite{laraki-solan}).  The fact that our DM is not a time-consistent long-run player differentiates our condition in its function and purpose from Isaac's condition.}
\begin{align}
0 &=G(m(\ph),\nu(\ph),\rho(\ph),\pi(\ph),\ph,V,dV),\label{eq:HJB0} \\  
\sigma(\ph) =(m(\ph),\nu(\ph),\rho(\ph))& \in\arg\max_{m,\nu,\rho}G(m,\nu,\rho,\pi(\ph),\ph,V,dV), \label{eq:HJBm} \\
\pi(\ph) &\in\arg\min_{p\in \mathcal{P}(\ph)}G(m(\ph),\nu(\ph),\rho(\ph),p,\ph,V,dV). \label{eq:HJBpi}
\end{align}
Condition \cref{eq:HJB0} captures the optimality requirement embodied in the value function: the variation from the optimal value for the DM is precisely zero at the equilibrium.  Condition \cref{eq:HJBm} requires the DM's flow decision to be optimal against nature's choice $\pi$, taking her future strategy as given.  This is not required in \cref{def:pseudo}, as is seen in the implausible ``always stopping'' behavior.  Finally, \cref{eq:HJBpi} chooses the belief to be worst for the DM against the flow action, taking the continuation strategy into consideration.  Effectively, equations \cref{eq:HJB0}, \cref{eq:HJBm}, and \cref{eq:HJBpi} represent the saddle-point version of HJB equations. We define our equilibrium as follows:  
\begin{definition} \label{def:intra}
A pseudo equilibrium $\sigma=(\nu,m,\rho)$, together with $\p$, is an  \emph{intrapersonal equilibrium}, or simply an \emph{equilibrium}, if  it satisfies \cref{eq:HJB0}, \cref{eq:HJBm}, and \cref{eq:HJBpi} for each $\ph\in (0,1)$.
\end{definition}

The precise microfoundation for this solution concept based on vanishing commitment power is provided in \Cref{sec:microfound-HJB}. It is important to note that the purpose of our HJB equations differs from the  standard one, which is to characterize the long-term optimal strategy. Recall that our DM cannot control her future selves, so her strategy is typically not long-term optimal  (even given her adversarial belief). Instead, our HJB equations characterize the (vanishingly) near-term maxmin-optimality of her strategy, as explained above, thereby refining the set of potential equilibria. This use of HJB conditions for dynamic decisions under ambiguity, and the micro-foundation justifying them, are of independent interest that we believe are applicable beyond the particular setting we consider here. While time inconsistency makes the dynamic programming machinery inapplicable, the HJB characterization restores the dynamic programming logic in a saddle-point and near-term optimality sense.   This characterization serves as a powerful toolkit that can make analysis tractable without compromising sharp prediction, as we argue is the case here.

\subsection{Main Result}\label{sec:main-result}

\begin{figure}
\begin{centering}
\includegraphics[width=0.6\textwidth]{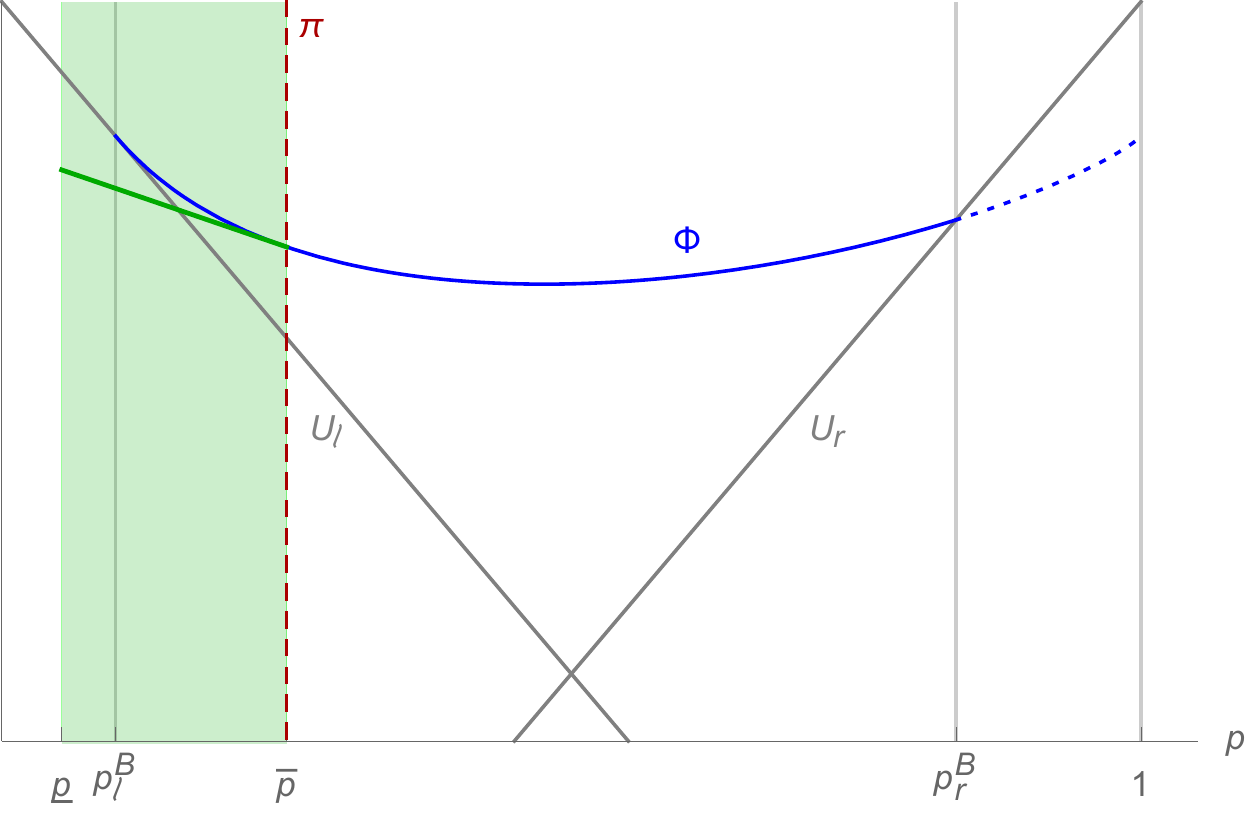}
\par\end{centering}
\caption{The value segment (in green) and worst-case belief for $\ph\in(\ph_1,\ph_2)$.\label{fig:region2}}
\end{figure}

\begin{figure}
\begin{centering}
\includegraphics[width=0.6\textwidth]{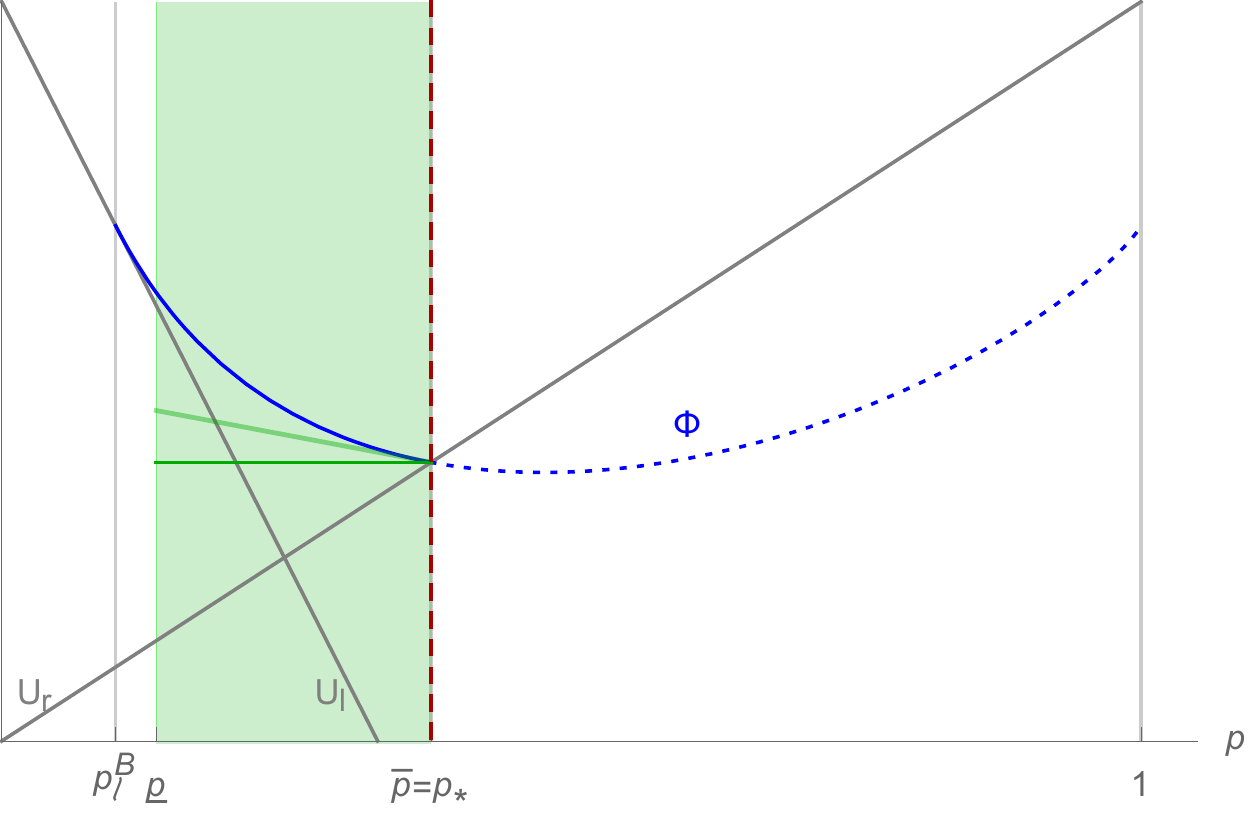}
\par\end{centering}
\caption{Case $\Phi'(p_*)<0$. The value segment for $\ph\nearrow p_*$ and $\ph=p_*$.\label{fig:region3'}}
\end{figure}

To state our main result, we introduce an additional threshold for the flow cost $c$:
\[
\underline{c}:=\frac{\delta_r}{\delta_r+\delta_{\ell}}\ch,
\]
where $\delta_a=|u_a^R-u_a^L|$ denotes the payoff difference across the two states given the DM's action $a\in\{\ell,r\}$. As can be easily seen, $\underline c$ is strictly smaller than $\overline c$, i.e. the Bayesian threshold  defined in \cref{eq:c-bar}. We will focus on the case in which the Bayesian value function $\Phi^*$ attains its minimum at $p_*<p_r^B$ such that ${\Phi}'(p_*)=0$, as graphed in \cref{fig:region2}. This case, labeled \emph{Case 1}, occurs when the value of outright action $r$ is not too high.  The opposite case, labeled \emph{Case 2}, in which $p_*=p_r^B$, or ${\Phi}'(p_*)<0$ (as graphed in \cref{fig:region3'}) will be treated at the end of this section.  This is also where we discuss the conditions under which the equilibrium we describe is unique (and under which it is not). 

\begin{theorem}\label{thm:main_result} The following intrapersonal equilibrium exists.
\begin{description}
\item[(i)] Suppose $c\geq\overline c$. The DM does not acquire information. If $\ph\leq\hat p$, she takes action $\ell$; if $\pl(\ph)\ge \hat p$, she takes action $r$; if $\hat p\in(\pl(\ph),\ph)$, she randomizes between $\ell$ and $r$ with probability $\hat\rho$.
\item[(ii)] Suppose $c<\overline c$ and $\Phi'(p_*)=0$. For each $c\in(0,\overline c)$ there is a threshold $\Delta_c\in(0,+\infty]$ with $\Delta_c=+\infty$ if and only if $c\leq\underline c$ such that the intrapersonal equilibrium is described as follows:
\medskip
\begin{enumerate}
\item [(a)] If $\Delta\leq\Delta_c$, there exist cutoffs $0<\ph_1<\ph_2\le\ph_3<\ph_4<1$ with $\ph_1=p_{\ell}^B$ and $\ph_2=p_*$  such that\footnote{We leave $\rho$ unspecified if $m=\nu=0$.}
\[
(m(\ph),\nu(\ph),\rho(\ph))=
\begin{cases}
(1,0,0)\\
(0,0,\cdot)\\
(0,\nu^*(\ph),0)\\
(0,0,\cdot)\\
(1,0,1)
\end{cases}\;\pi(\ph)=\begin{cases}
 \ph & \text{if }\ph\in[0,\ph_{1}]\\
 \ph & \text{if }\ph\in(\ph_{1},\ph_{2}]\\
\pi^*(\ph) & \text{if }\ph\in(\ph_{2},\ph_{3})\\
 \pl(\ph) & \text{if }\ph\in[\ph_{3},\ph_{4})\\
 \pl(\ph) & \text{if }\ph\in[\ph_{4},1],
\end{cases}
\]
where $\nu^*(\ph)>0$ and $\pi^*(\ph)\in(\pl(\ph),\ph)$,  for $\ph\in(\ph_2,\ph_3)$, 
and $\ph_{2}<\ph_{3}$ if and only if $\Phi(p^{*})<U_{\ell}(\pl(p^{*}))$.
\medskip
\item [(b)]  If $\Delta>\Delta_c$, the equilibrium has the same structure as in (a) except that 
\[
(m(\ph),\nu(\ph),\rho(\ph))=\left(1,0,\hat\rho\right)\quad\text{and}\quad \pi(\ph)=\hat p,
\]
for $\ph\in[\ph_3,\ph_4)$.
\end{enumerate}
\end{description}   
\end{theorem}

We structure the discussion of \cref{thm:main_result} according to the level of experimentation costs. If $c\geq\overline{c}$, the DM does not experiment, and her equilibrium strategy coincides with the commitment solution described in \cref{prop:commitment}.  This is a consequence of the facts that 1) given $c\geq\overline{c}$,  the DM's commitment solution involves no experimentation, as we have seen in \cref{sec:model}, and 2) implementing this solution requires no commitment. The DM's value in this case is $V^{\sigma}(\pi(\ph),\ph)=\min_{p\in \mathcal{P}(\ph)}U(p)$. 

Our main focus will lie on the case $c\leq\underline c$.  Under this restriction, the equilibrium is always described by \cref{thm:main_result}-(ii)(a) and can be illustrated as follows:
\[
0\underbrace{\vphantom{\ph^{1}}\text{------------}}_{\text{action }\ell}\ph_{1}\underbrace{\vphantom{\ph^{1}}\hspace{-5pt}\longleftarrow\longleftarrow\longleftarrow\longleftarrow}_{\text{pure exp.}}\ph_{2}\underbrace{\vphantom{\ph^{1}}\hspace{-5pt}\longleftarrow\longleftarrow\longleftarrow\longleftarrow\longleftarrow}_{\text{mixed stopping \ensuremath{(\ell)}}}\ph_{3}\underbrace{\vphantom{\ph^{1}}\hspace{-5pt}\longleftarrow\longleftarrow\longleftarrow\longleftarrow}_{\text{pure exp.}}\ph_{4}\underbrace{\vphantom{\ph^{1}}\text{------------}}_{\text{action }r}1
\]
 When $\ph$ is sufficiently extreme ($\ph\leq\ph_1$ or $\ph\geq\ph_4$), the DM takes immediate action without gathering additional information, just like the Bayesian DM. For the remaining states, however, the characterization differs qualitatively from the benchmark case. In particular, if $\Delta$ is sufficiently large, there is a middle region of states where the DM randomizes between experimentation and action $\ell$ at a positive rate. As a result, the region of pure experimentation is no longer convex. 

To derive the solution we leverage the fact that in the absence of breakthrough news, the set of beliefs drifts to the left.  Hence, we explain the logic of the equilibrium strategy via backward induction, starting from a far left set of beliefs.\\

\noindent\emph{Region 1: $\ph\le \ph_1$.}
The first threshold $\ph_1$ coincides with the Bayesian stopping boundary $p^B_{\ell}$. If $\ph\leq p^B_{\ell}$, then stopping followed by action $\ell$ is optimal for all beliefs in $[\pl(\ph),\ph]$ and thus constitutes the equilibrium strategy for our ambiguity-averse DM. The worst-case belief under this strategy is $\ph$.\\

\noindent\emph{Region 2: $\ph\in (\ph_1, \ph_2]$.} In Region 2, the right-most belief $\ph$ continues to be the worst-case belief for all $\ph\in(\ph_1,\ph_2]$, and the DM follows the Bayesian optimal stopping rule with respect to $\ph$. Since $\ph>\ph_1=p_{\ell}^B$, this strategy entails experimentation until either a breakthrough occurs or the Bayesian update of $\ph$ reaches $p^B_{\ell}$. The right boundary of this region is $\ph_2:=p_*$, the belief at which the Bayesian value function attains its minimum.

Fix any state $\ph\in (\ph_1, \ph_2]$. We can evaluate the expected payoff $V^{\sigma}(p,\ph)$ of strategy $\sigma$  as a function of the belief  $p$. The set of values corresponding to possible alternative  beliefs, $\left\{\left(p,V^{\sigma}(p,\ph)\right):p\in \mathcal{P}(\ph)\right\}$, forms a linear segment, which we will call \emph{value segment} throughout (see for instance \cref{fig:region2}). It is useful to track the movement of the value segment, as time moves backward or forward. Note that the value segment is tangent to the value function $\Phi$ at $\ph$, due to the fact that the stopping rule is Bayesian optimal with respect to $\ph$.\footnote{Bayesian optimality of the strategy with respect to $\ph$ clearly implies $V^{\sigma}(\ph,\ph)=\Phi(\ph)$. This, together with linearity of $V^{\sigma}(p,\ph)$ in the first argument and the fact that $\Phi(p)\geq V^{\sigma}(p,\ph)$ holds for all $p\in(p^B_{\ell},p^B_{r})$, implies that the value segment is tangent to $\Phi$ at $\ph$.}  Tangency at this point implies that the value segment is downward sloping and, hence, that $V^{\sigma}(p,\ph)$ is indeed minimized at $p=\ph$. Hence, $\pi(\ph)=\ph$ is nature's optimal choice against $\sigma(\ph)$, and the prescribed $\sigma(\ph)$, being  Bayesian optimal for $p=\ph$, is a best response  against $\pi(\ph)$.  The profile $(\sigma(\ph), \pi(\ph))$ thus forms a saddle point for state $\ph$.  
 
\begin{figure}
     \centering
     \begin{subfigure}[b]{0.45\textwidth}
         \centering
         \includegraphics[width=\textwidth]{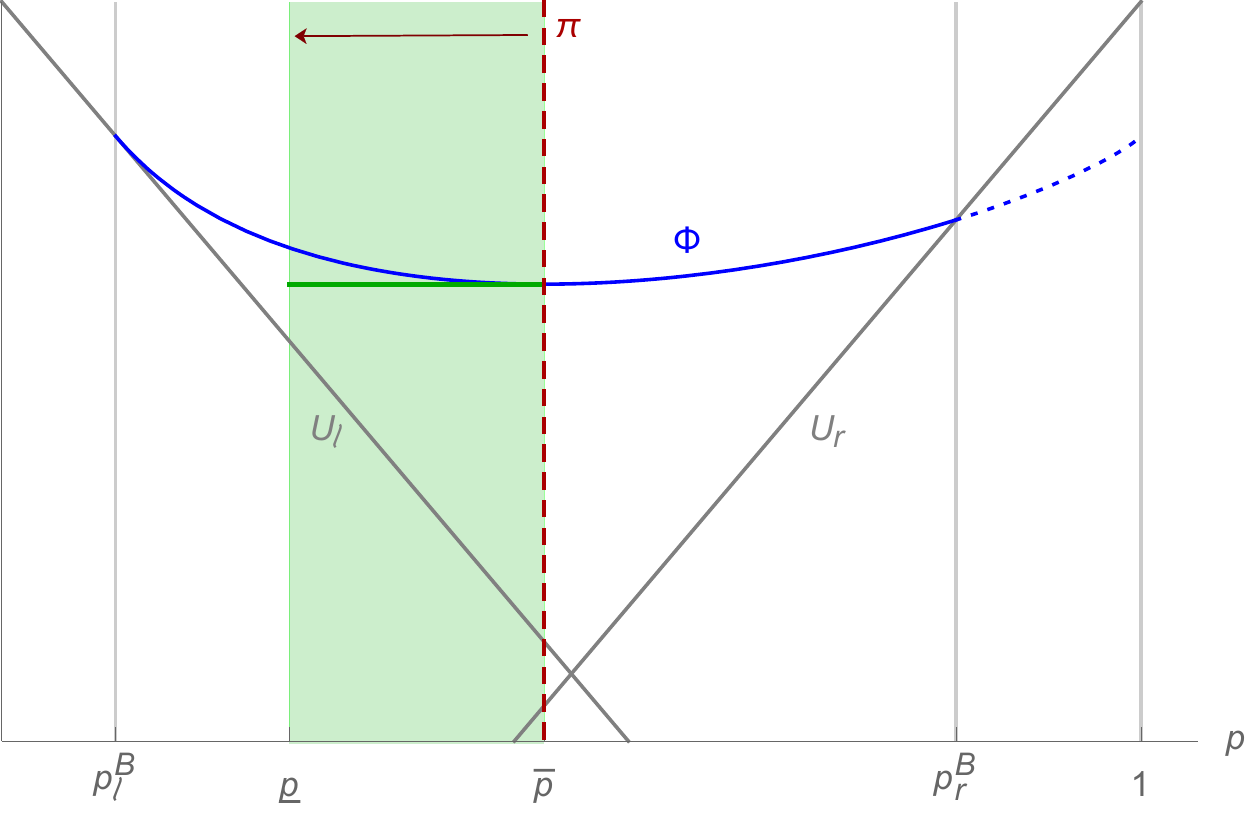}
     \end{subfigure}
     \begin{subfigure}[b]{0.45\textwidth}
         \centering
         \includegraphics[width=\textwidth]{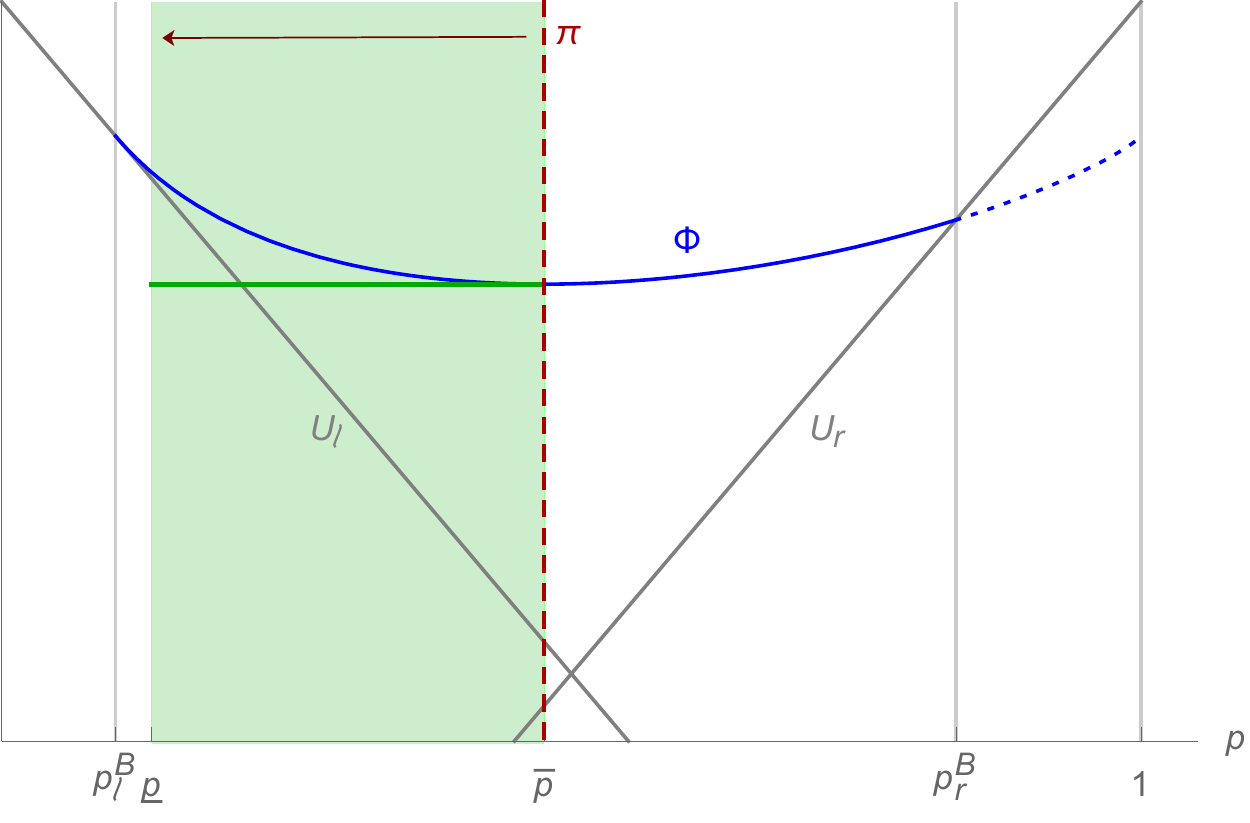}
     \end{subfigure}
     \caption{The value segment and worst-case belief for $\ph=\ph_2$ with $V^{\sigma}(\pl(p_{*}),p_{*})\geq U_{\ell}(\pl(p_{*}))$ (left panel) and $V^{\sigma}(\pl(p_{*}),p_{*})< U_{\ell}(\pl(p_{*}))$ (right panel).}
     \label{fig:p2}
\end{figure}

At $\ph=p_{*}=:\ph_{2}$, the value segment becomes flat (see \Cref{fig:p2}), as we assumed $\Phi'(p_*)=0$,  so all posteriors  in $[\pl(\ph),\ph]$ minimize the DM's expected payoff. \\

\noindent\emph{Regions 3: $\ph\in (\ph_2, \ph_3]$.} Moving further back in time, suppose the state $\ph$ is slightly higher than $p_*$ and the DM contemplates experimenting according to the strategy prescribed in Region 2. The value segment becomes now upward sloping, implying that the worst-case scenario is described by $\pl(\ph)$ rather than $\ph$. We then distinguish two cases. If the degree of ambiguity $\Delta$ is sufficiently small and, hence, the value segment is sufficiently ``short,'' experimentation is optimal at the worst-case belief $\pl(\ph)$ (see \cref{fig:p2}, left panel).\footnote{Formally, we have $V^{\sigma}(\pl(p_{*}),p_{*})\geq U_{\ell}(\pl(p_{*}))$.} The randomization region, Region 3, is empty here, and we directly transition to Region 4 at $\ph=p_*$. 

\begin{figure}
\begin{centering}
\includegraphics[width=0.6\textwidth]{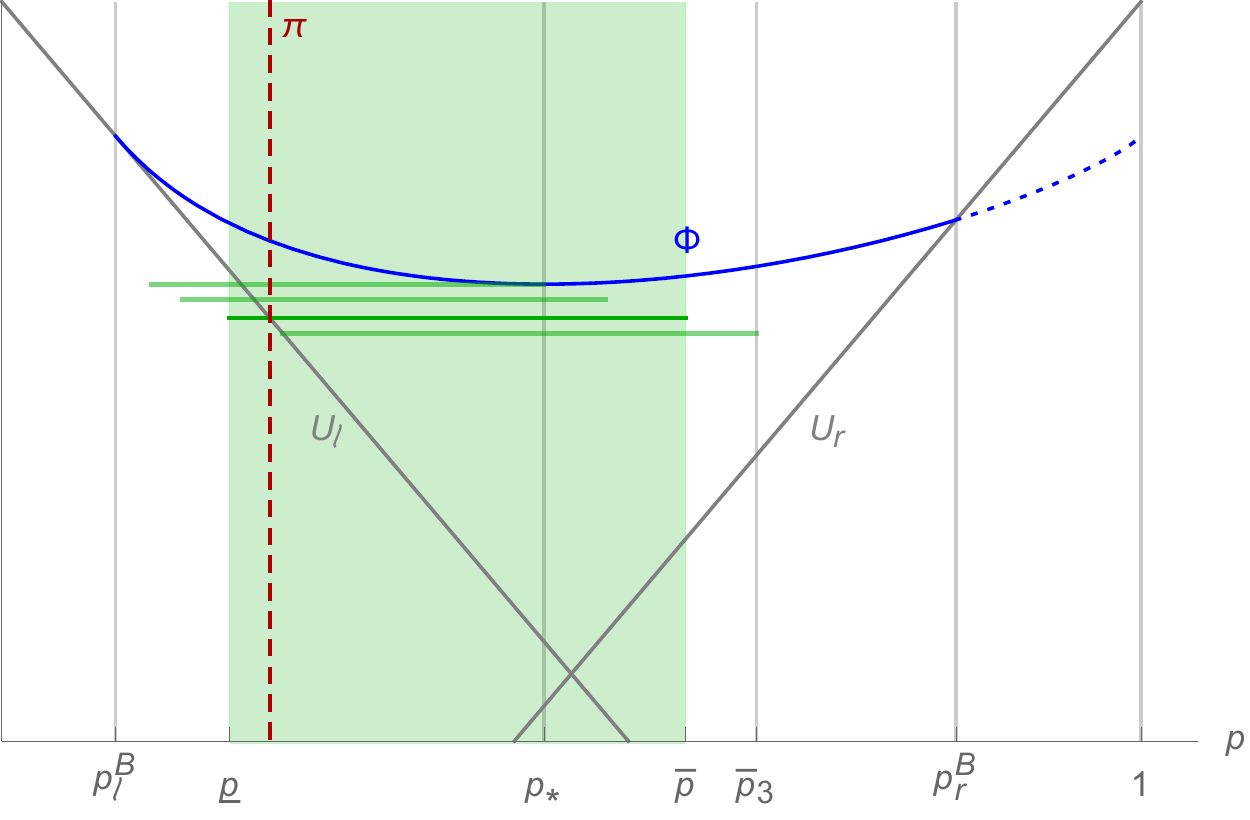}
\par\end{centering}
\caption{The value segment and worst-case belief for $\ph\in(\ph_2,\ph_3)$.}\label{fig:region3}
\end{figure}

More interestingly, suppose $\Delta$ is sufficiently large such that the value segment crosses the stopping payoff line $U_{\ell}$ from below.  Then, at state $\ph=p_*$ experimentation is dominated by stopping followed by $\ell$ according to the worst-case belief $\pl(p_*)$ (see \cref{fig:region3}, right panel). Choosing $\ell$, however, cannot be an equilibrium either, as nature would now choose $\ph$, which makes it optimal for the DM to experiment. The intrapersonal equilibrium is thus in mixed strategies, where the DM randomizes between experimentation and action $\ell$.

The DM's value and the equilibrium stopping rate in the mixed stopping region can be derived via our HJB conditions. Randomization between experimentation and action $\ell$ requires that the coefficient of $\nu$ in \cref{eq:HJBm} must vanish and thus: 
 \begin{equation}\label{eq:ind-DM}
 V(\pi (\ph), \ph)= U_{\ell}(\pi (\ph)).
 \end{equation}
Graphically, the stopping payoff $U_{\ell}$ and the value segment must cross at the worst-case belief $\pi(\ph)$, making the DM indifferent between experimentation and action $\ell$. Since the required belief $\pi(\ph)$ will be in the interior of $[\pl(\ph),\ph]$, for nature to optimally choose it, the value segment must remain flat so that all posteriors in $[\pl(\ph),\ph]$ minimize the DM's expected payoff. We can thus write $V (p,\ph)=\hat V(\ph), \forall p.$ Substituting this into the DM's indifference condition \eqref{eq:ind-DM}, we obtain nature's choice  
\begin{align}\label{eq:pi-r3}
\pi(\ph) & =\frac{u_{\ell}^{L}-\hat{V}(\ph)}{u_{\ell}^{L}-u_{\ell}^{R}}.
\end{align} 
The randomization by the DM in turn implies that the derivative of the objective in  \cref{eq:HJBm} with respect to $p$ must vanish.  This fact, together with $V (p,\ph)=\hat V(\ph), \forall p$, yields the equilibrium stopping rate  
\begin{align}\label{eq:nu-r3}
\nu(\ph)=\frac{\lambda(u_{r}^{R}-\hat{V} (\ph))}{u_{\ell}^{L}-u_{\ell}^{R}}.
\end{align}
Using \eqref{eq:pi-r3} and \eqref{eq:nu-r3}, the value function $\hat{V}(\ph)$ is derived by substituting for $\pi$ and $\nu$ in the HJB condition \cref{eq:HJB0}, which gives rise to the following differential equation:
 \begin{equation}
	\tag{\ensuremath{\widehat{\text{ODE}}}}\ph(1-\ph)\hat{V}^{\prime}(\ph)=\frac{\left(u_{r}^{R}-
		\hat{V}(\ph)\right)\left(u_{\ell}^{L}-\hat{V}(\ph)\right)}{u_{\ell}^{L}-u_{\ell}^{R}}-\frac{c}{\lambda}.
	\label{eq:ODE_Vhat}
\end{equation}
Together with the boundary condition $\hat V(\ph_2)= \Phi(\ph_2)$, \eqref{eq:ODE_Vhat} admits a unique solution, describing the DM's value in Region 3.

The reason that the DM stops according to a Poisson process, and thus with vanishing probability (rather than positive probability), is because when $\ph\approx p_*$, the value segment is already arbitrarily close to flat, so it takes a vanishingly small stopping probability  to turn the value segment completely flat. The same situation and intuition are repeated as we move backward in time: for a positive range of states, labeled Region 3, the DM repeatedly hedges between experimentation and Poisson stopping at a positive rate $\nu(\ph)$. 

 Intuitively, the repeated hedging by the DM can be seen as a resolution of conflicts between current and future selves. On the one hand, the DM is highly uncertain about the state and thus finds it optimal to acquire information. In particular, the DM worries about action $\ell$ being the wrong choice (state $R$) and would thus like to gather more information before taking it. On the other hand, the DM is aware that the journey of experimentation will take on its own life, with the future selves doing too much of it from the current self's perspective. Prolonged experimentation is particularly detrimental  in state $L$ where the costly search for evidence is futile. The randomized stopping hedges the DM against both events: at each state in the mixing region, the DM's continuation value is independent of the underlying state. This resolves the conflict and restores the delicate balance between the multiple selves. 

While at first glance it may seem odd that the randomized stopping arises in the middle region around $p_*$, this is no coincidence. As noted earlier, it is precisely where preference reversal and  dynamic inconsistency  arise.  As will be seen, this seemingly peculiar feature of the equilibrium proves quite robust across different specifications of the learning environment and technologies.\\
 
\begin{figure}
     \centering
     \begin{subfigure}[b]{0.45\textwidth}
         \centering
         \includegraphics[width=\textwidth]{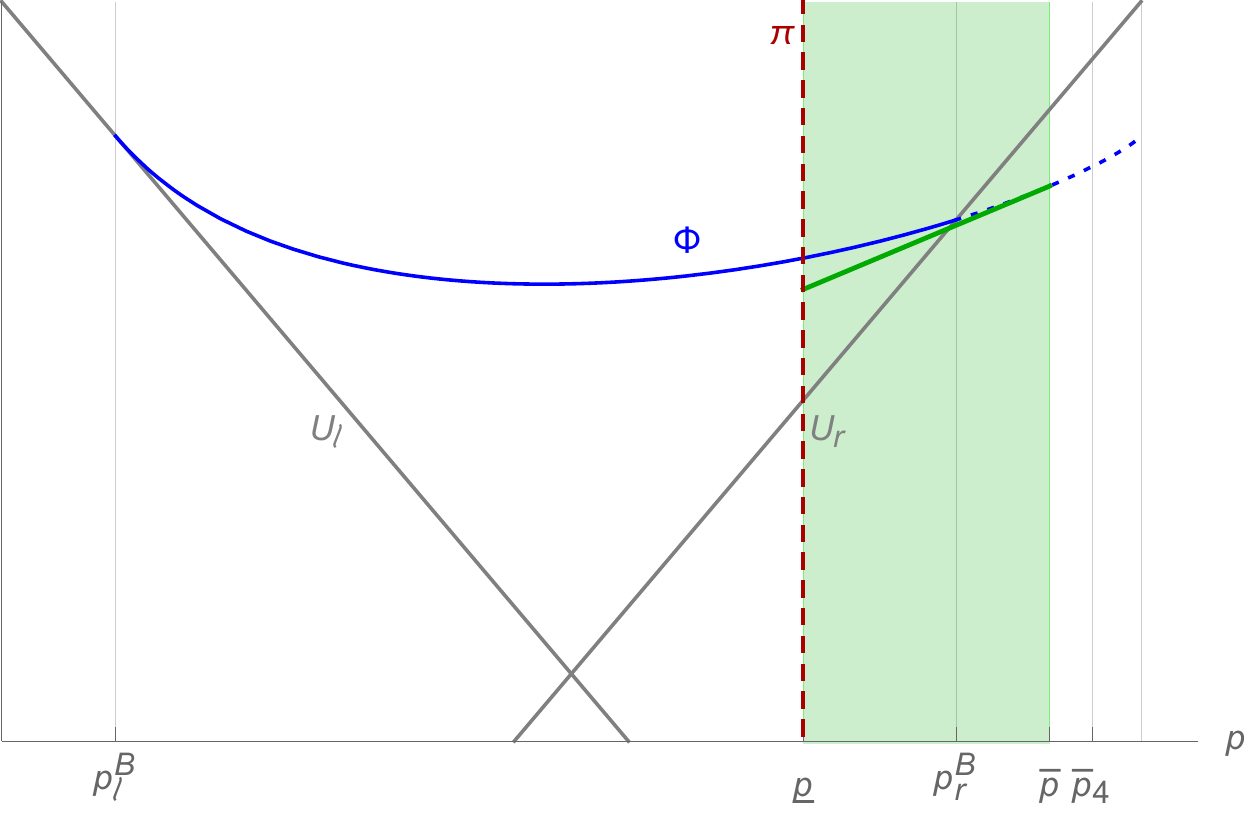}
              \end{subfigure}
     \begin{subfigure}[b]{0.45\textwidth}
         \centering
         \includegraphics[width=\textwidth]{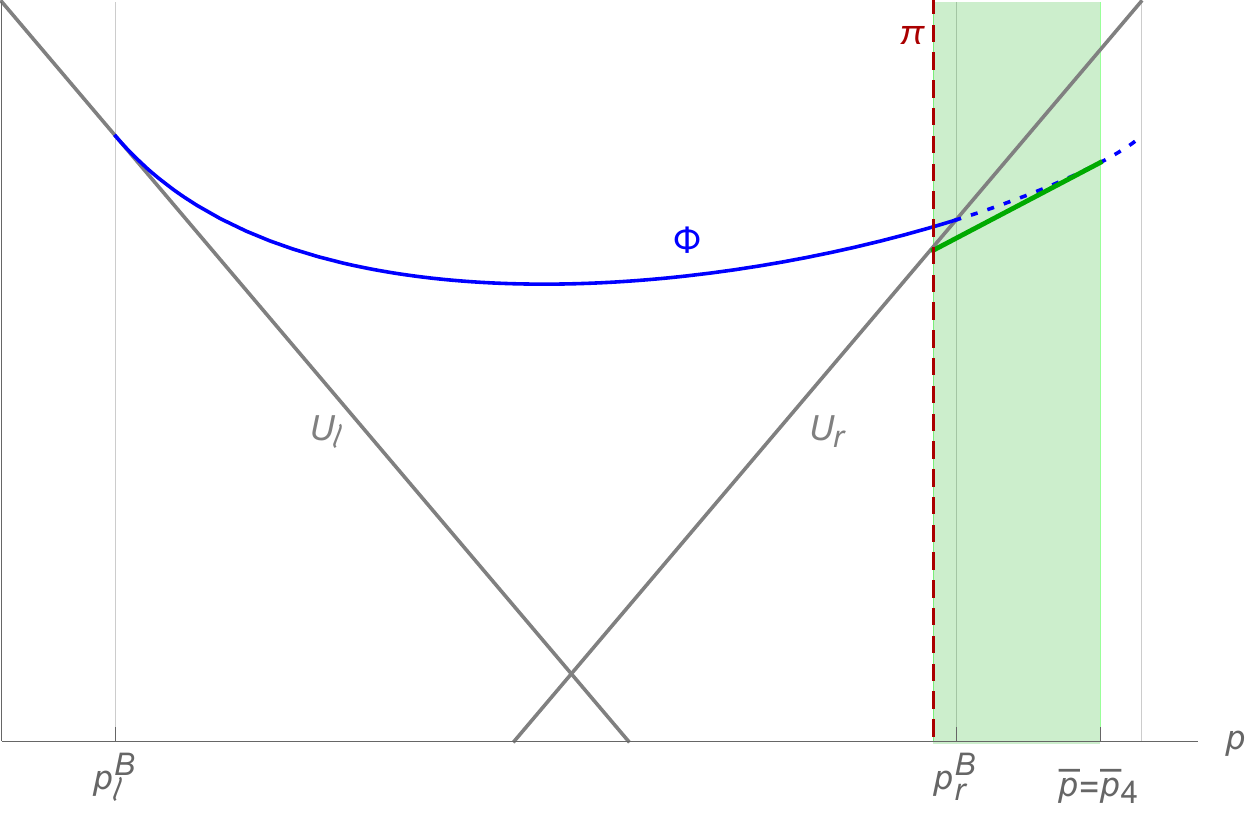}
     \end{subfigure}
     \caption{The value segment and worst-case belief for $\ph\in(\ph_3,\ph_4)$ in the case of $\Delta$ small.}
     \label{fig:region4}
\end{figure}

 \noindent\emph{Regions 4: $\ph\in (\ph_3, \ph_4]$.} As already mentioned, when ambiguity $\Delta$ is small, Region 3 is empty, and Regions 2 and 4 collapse to a single experimentation region.  Its right boundary, $\ph_4$, occurs at the point where, according to the worst-case belief $\pl(\ph)$, action $r$ yields the same expected payoff as experimentation, i.e. $U_r(\pl(\ph_4))=V^{\sigma}(\pl(\ph_4),\ph_4)$.  Graphically, $\ph_4$ can be seen as the point where the left endpoint of the value segment crosses the stopping payoff $U_{r}$ (see right panel of \cref{fig:region4}). It is important to notice that at $\ph=\ph_4$, the value of $\pl(\ph)$ lies strictly to the left of the Bayesian stopping boundary $p^B_{r}$.  This follows from the fact that the DM's future selves follow a rule that is optimal with respect to $\ph$ but not with respect to $\pl(\ph)$. From the perspective of $\pl(\ph)$, the value of experimentation is thus strictly lower than the Bayesian value $\Phi(\pl(\ph))$.  This means that the DM with worst-case belief $\pl(\ph)$ is more inclined to stop than the Bayesian DM with the corresponding belief.  

Next, consider the case in which $\Delta$ is sufficiently large so that Region 3 with mixed stopping is nonempty. The right boundary of the mixed stopping region, $\ph_3$, is given by the state $\ph$ at which the value segment separates from $U_{\ell}$ and experimentation becomes again optimal for all posteriors in $[\pl(\ph),\ph]$ (see \cref{fig:region3}). To the right of $\ph_3$, we then have a second region of pure experimentation, Region 4. The value segment becomes upward-sloping in this region, so $\pl(\ph)$ becomes the worst-case belief. As with the case of small $\Delta$, the right boundary of Region 4 is determined as the point where the left end of the value segment reaches the stopping payoff $U_r$. Notice, however, that the value segment in the experimentation region remains below the Bayesian value function. Compared to the  case of small $\Delta$, the possibility of randomization in future states reduces the value of experimentation further, which pushes the stopping boundary $\ph_4$ towards the middle.  Intuitively, the DM here is discouraged from experimentation, and is thus more inclined to stop, not only because it leads to excessive experimentation in the distant future, but also because it leads to inefficient stopping in the intermediate future.\\

We now complete the characterization by addressing several remaining cases and issues.  The reader interested in the behavioral implications of the baseline characterization may skip to the next section.\\

\noindent\emph{Intermediate experimentation costs.} Consider next the case in which $c\in(\underline c,\overline c)$. \cref{thm:main_result} shows that for each $c\in(\underline c,\overline c)$ there exists a threshold $\Delta_c\in\mathbb R$ such that for $\Delta\leq\Delta_c$ the equilibrium continues to be described by case (a) of \cref{thm:main_result}. If instead $\Delta>\Delta_c$, the fourth region features randomization between actions $\ell$ and $r$ rather than experimentation.
\[
0\underbrace{\vphantom{\ph^{1}}\textbf{------------}}_{\text{action }\ell}\ph_{1}\underbrace{\vphantom{\ph^{1}}\hspace{-5pt}\longleftarrow\longleftarrow\longleftarrow\longleftarrow}_{\text{pure exp.}}\ph_{2}\underbrace{\vphantom{\ph^{1}}\hspace{-5pt}\longleftarrow\longleftarrow\longleftarrow\longleftarrow\longleftarrow}_{\text{mixed stopping \ensuremath{(\ell)}}}\ph_{3}\underbrace{\vphantom{\ph^{1}}\hspace{-5pt}\textbf{---------------}}_{\text{mixing between $\ell$ and $r$}}\ph_{4}\underbrace{\vphantom{\ph^{1}}\textbf{------------}}_{\text{action }r}1
\]

The equilibrium behavior for the first three regions is the same as in the case with low experimentation costs. In the third region, however, where the DM randomizes between experimentation and action $\ell$, the value segment may drop as low as the point at which the two stopping payoff lines, $U_{\ell}$ and $U_r$, cross. The state at which this occurs constitutes our new boundary $\ph_3$. The DM's value at this point is equal to $\min_pU(p)=U(\hat p)$, the minimal expected payoff that the DM can guarantee to herself by randomizing between actions $r$ and $\ell$ with probabilities $\hat\rho$ and $1-\hat\rho$, respectively.  At states to the right of this point, randomizing between immediate actions with $\hat\rho$ strictly dominates experimentation with randomized stopping, giving rise to the fourth region $[\ph_3,\ph_4)$. Its right boundary is the state $\ph$ that satisfies  $\pl(\ph)=\hat p$. Further to the right, where $\pl(\ph)>\hat p$, the DM optimally takes action $r$. The solution for intermediate costs $c\in(\underline c,\overline c)$ and large ambiguity $\Delta>\Delta_c$ can thus be viewed as a combination of the previous two solutions.\\

\noindent\emph{Case 2.} We now turn to Case 2, which is characterized by $\Phi'(p_*)<0$. The formal characterization of the intrapersonal equilibrium for this case is relegated to \cref{ssec:case2}. Compared with the case $\Phi'(p_*)=0$, there is only one significant change. At the right boundary of the first experimentation region $(\ph_1,\ph_2)$, the stopping distribution has an atom: the DM takes action $r$ with positive probability $m(\ph_2)$ and experiments with the complementary probability. 
\[
0\underbrace{\vphantom{\ph^{1}}\text{------------}}_{\text{action }\ell}\ph_{1}\underbrace{\vphantom{\ph^{1}}\hspace{-5pt}\longleftarrow\longleftarrow\longleftarrow\longleftarrow}_{\text{pure exp.}}\underbrace{\ph_{2}}_{m>0}\underbrace{\vphantom{\ph^{1}}\hspace{-5pt}\longleftarrow\longleftarrow\longleftarrow\longleftarrow}_{\text{mixed stopping \ensuremath{(\ell)}}}\ph_{3}\underbrace{\vphantom{\ph^{1}}\hspace{-5pt}\longleftarrow\longleftarrow\longleftarrow\longleftarrow\longleftarrow}_{\text{pure exp. or mixing $\ell/r$}}\ph_{4}\underbrace{\vphantom{\ph^{1}}\text{------------}}_{\text{action }r}1
\]
Moving backwards in time through the experimentation region $(\ph_1,\ph_2)$, the value segment remains downward sloping as the state $\ph$ approaches $\ph_2=p_{*}$. The unique worst-case belief in the limit as $\ph\nearrow p_*$ is thus $\ph$. For states to the right of $p_*$, the Bayesian optimal strategy under $\ph$ is to take action $r$ rather than experimenting. Taking action $r$ in those states does, however, not constitute an intrapersonal equilibrium, as the worst-case belief under this strategy would be $\pl(\ph)$ rather than $\ph$. The DM must therefore experiment with positive probability in states to the right of $p_*$. To make experimentation optimal, the value segment must be weakly upward sloping for states belonging to an interval $(p_*,p_*+\varepsilon)$. This property requires the DM to take action $r$ with strictly positive probability at state $\ph=p_*$. Graphically, the DM's randomization flips the left end of the value segment downwards, so that it is flat at $\ph=p_*$, as illustrated in \cref{fig:region3'}.\footnote{\label{ft:eq}The knife-edge case $\ph_0=p_*$ (in Case 2) is the only instance where the requirement that the DM's strategy constitutes an equilibrium regardless of the initial state actually restricts the set of equilibria.  In this instance, requiring equilibrium conditions \emph{only} for  subsequent states $\ph\leq\ph_0$ would leave some flexibility in picking the   randomization between experimentation and action $r$ at state $\ph_0=p_*$. Indeed, as long as the value segment remains (weakly) downward sloping, all equilibrium conditions would be satisfied for the states to the left of $\ph_0=p_*$. It is the equilibrium conditions for states  $\ph> p_*$ that require the value segment to become flat at state $p_*$.}\\

\noindent\emph{Uniqueness.}  In \Cref{sec:uniqueness}, we prove that the intrapersonal equilibrium is unique when either $c<\cl$ or  $c\ge\ch$. When costs are intermediate, however, there is a class of intrapersonal equilibria. These differ from the one described in \Cref{thm:main_result} only in the position of their stopping boundary on the right. More specifically, recalling that $\hat p$ is the belief at which the DM is indifferent between actions $\ell$ and $r$, fix any $q$ such that $\pl(q)<\hat p\leq q$. We then construct a strategy profile $(\tilde \s,\tilde \p)$ indexed by $q\in[\hat p,\underline p^{-1}(\hat p))$ as follows:
\[
\tilde\s(\ph)=(\tilde m(\ph),\tilde\nu(\ph),\tilde\rho(\ph))=
\begin{cases}
\s(\ph)\\
(1,0,\hat\rho)\\
(1,0,1) 
\end{cases}\;\tilde\pi(\ph)=\begin{cases}
 \pi(\ph) & \text{if }\ph\leq q\\
 \hat p & \text{if }q<\ph<\pl^{-1}(\hat p)\\
\pl & \text{if }\pl^{-1}(\hat p)\leq\ph,
\end{cases}
\]
where  $(\s,\p)$ is the intrapersonal equilibrium defined in \Cref{thm:main_result}. In words, the DM follows the same strategy as in \Cref{thm:main_result} for states $\ph\le q$ but stops immediately in states $\ph>q$, followed by the mixed action $\hat \rho$ if $\hat p\in (\pl(\ph), \ph)$ and action $r$ if $\hat p\le \pl(\ph)$.  Since $c<\ch$, this new strategy  yields a strictly lower payoff  for the DM in the states just to the right of $q$. In fact, the value function jumps down at the ``switching'' state $\ph=q$ to the point where the two stopping payoffs cross, so the DM is worse off from this strategy relative to the one in \Cref{thm:main_result}. Nevertheless, the strategy profile---there is a continuum of such indexed by $q\in [\hat p, \pl^{-1}(\hat p))$---satisfies all requirements of our solution concept, including the HJB conditions. The intuition is that, even though the HJB conditions capture control over vanishingly proximate future actions, these conditions are not strong enough. Given $c\ge\underline c$, there always exists some $\epsilon<\ph-q$ such that deviating only until the state falls to $\ph-\epsilon$ will not generate a strictly higher payoff.

 Nevertheless, we view these equilibria as implausible. First, we see that $\tilde \s$, regardless of $q$, is Pareto worse than $\s$ for all selves of the DM evaluated at the worst-case beliefs. Hence, $\tilde\s$ will be eliminated if the multiple selves of the DM can coordinate on the selection of equilibria, as is often invoked in the name of  ``consistent planning;'' see \cite{siniscalchi2011dynamic}.\footnote{More precisely, \cite{siniscalchi2011dynamic} requires the current self to break a tie in favor of her earlier self. See \cite{ebert2018never} for a related approach.}  Second, we could strengthen the notion of vanishing commitment to eliminate the equilibria. Namely, instead of requiring the equilibrium strategy profile to be a limit of the $\e$-commitment solution {\it pointwise} for each $\ph$, one could require the limit convergence to be {\it uniform} across all $\ph$'s. The profile $(\tilde\s,\tilde \p)$ cannot survive this stronger test since for any $\epsilon>0$, there will always exist a state $\ph\in (q, q+\e)$, such that the $\e$-commitment solution for the DM at that state will entail a deviation to $\s$ to enjoy a discontinuous payoff jump.\footnote{One could adopt this stronger requirement as a solution concept. We do not take this approach for two reasons. First, the uniform convergence requirement is not easy to check since it requires one to verify the equi-continuity property of the value function. Second, the requirement jeopardizes the existence in the more general Poisson model we discuss in \cref{sec:general-poisson}. The unique intrapersonal equilibrium presented in \Cref{thm:2news-L} involves a discontinuous value function when the cost is intermediate, which does not survive the stronger refinement.}\\

\noindent\emph{Knightian Uncertainty.}  The extreme case in which $\pl=0$ and $\ph=1$, known as Knightian uncertainty,  constitutes an important theoretical benchmark. Naturally, the equilibrium characterization for this case is obtained as the limit of the solution we characterized above for $(\pl_0,\ph_0)\to(0,1)$.

\begin{proposition}\label{prop:Knightian}
Assume $\mathcal{P}_0=[0,1]$. If $c\ge \cl$, the DM mixes between immediate actions $\ell$ and $r$ with probability $\hat \rho$.  Otherwise the DM mixes between experimentation and stopping followed by $\ell$ at a stationary rate
\begin{align*}
\tilde\nu=\frac{\lambda}{\delta_{\ell}}(u_{r}^{R}-\tilde u),\text{ where } \tilde u=\frac{u_{r}^{R}+u_{\ell}^{L}}{2}-\sqrt{\left(\frac{u_{r}^{R}-u_{\ell}^{L}}{2}\right)^{2}+\frac{c}{\lambda}\delta_{\ell}}.
\end{align*}
This solution is the limit of the equilibrium strategy characterized in \cref{thm:main_result} (\cref{thm:case2} for Case 2) for the case $0<\pl_0<\ph_0<1$ as $(\pl_0,\ph_0)\to(0,1)$.
\end{proposition}

As  $(\pl_0,\ph_0)\to(0,1)$, Region 3 featuring randomized stopping takes over the entire region to the right of $p_*$. Starting from $\ph_0$ close to one, the DM will then spend an arbitrarily long time in Region 3. 
In the extreme case, when $[\pl_0, \ph_0]=[0,1]$, the DM's set of priors remains unchanged in the absence of breakthrough news. Hence, under any Markovian strategy, the DM either stops immediately or stops at a constant rate. The latter is optimal if and only if  $c\le \cl$.  

We can interpret the Knightian solution as a form of hedging by the DM to ensure her payoff equalizes between the states. Indeed, it is easy to verify that the stationary stopping rate $\tilde\nu$ satisfies the following equality:
\begin{eqnarray}\label{eq:stationary}
u_{\ell}^L-\frac{c}{\tilde\nu}=\frac{\lambda}{\tilde\nu+\lambda}u_r^R+\frac{\tilde\nu}{\tilde\nu+\lambda}u_{\ell}^R-\frac{c}{\tilde\nu+\lambda}.
\end{eqnarray}
The LHS shows the DM's payoff from stopping at a constant rate $\tilde\nu$ when the state is $L$. In state $L$ the DM never observes a breakthrough and thus always ends up taking the correct action. The downside is a relatively high expected cost of experimentation equal to $c/\tilde\nu$. The RHS shows the DM's payoff  when the state is $R$. Now, the DM receives $R$-evidence before she stops experimenting, with positive probability  $\lambda/(\tilde\nu+\lambda)$. With the complimentary probability, however, the DM takes the wrong action and receives the lower payoff $u_{\ell}^R$. The payoff gross of experimentation costs is then lower in state $R$ than in state $L$, which is compensated by the DM experimenting for shorter periods (in expectation) and therefore incurring a lower expected cost of experimentation equal to $c/(\tilde\nu+\lambda)$.\\

\noindent\emph{The role of randomization.} Whether randomization helps to hedge against ambiguity depends on the DM's internal view on how the uncertainty unfolds (see \citet{saito2015preferences} and \citet{ke2020randomization}). We adopt the assumption that the DM views the game against nature as one with simultaneous moves. Hence, in the DM's mind, nature cannot condition her choice on the outcome of the DM's randomization. If the DM takes on an even more pessimistic view, believing that nature can adjust her choice \emph{after} the DM's randomization outcome, any realization of a mixed strategy will be evaluated according to the worst-case belief for this realization, so the DM has no incentives to randomize. 

Modifying the setting in this way clearly changes the characterization of our solution for the case when ambiguity is large. It does, however, not erase the DM's desire to stop preemptively in anticipation of a long experimentation phase when ambiguity is large.  As we will show, stopping  occurs again at intermediate sets of beliefs where the worst-case belief is about to switch. But instead of an extended period of randomized stopping for an interval of states,  now  stopping occurs at an isolated state $\ph^*$. Anticipating this, earlier selves of the DM are happy to experiment until that point. Even though the game ends once the state reaches $\ph^*$, if the DM would find herself at state $\ph<\ph^*$, she would again be experimenting.  In this sense, the non-monotonic pattern of stopping persists even without randomization. To keep the proof short, we restrict attention to the symmetric case where $u^R_r=u^L_{\ell}=:\delta$ and $u^R_{\ell}=u^L_{r}=0$.

\begin{proposition}\label{prop:random}
Suppose the DM is restricted to pure strategies. Assume $\Phi'(p_*)=0$, $c<\ch$ and payoffs are symmetric. If $\Delta$ is sufficiently large, then there is a state $\ph^*\in(0,1)$ such that the DM takes action $\ell$ at $\ph^*$ and experiments on a left and right neighborhood of $\ph^*$.
\end{proposition}

To understand this result, let us meet the DM on the journey backward in time, at the exact moment when the worst-case belief is about to switch, which is captured by the right panel of \Cref{fig:p2}.\footnote{Note the DM's equilibrium behavior in Regions 1 and 2 is unchanged.}  Moving further back in time from that moment, the value segment associated with experimentation becomes upward-sloping, so the DM, as before, finds herself in a dilemma where further experimentation is met with the adversarial belief $\pl(\ph)$ whereas stopping with $\ell$ is met with $\ph$. With nature now wielding the second-mover advantage, the DM can no longer protect herself from these adversarial beliefs via randomization. Instead, the DM must choose the lesser of two bad choices. As long as $\ph>p_*$ is relatively close to $p_*$, the value segment is sufficiently flat so that the value of experimentation at $\pl(\ph)$ is higher than the stopping payoff $U_{\ell}$ at belief $\ph$. Experimentation thus remains the better option for an interval of time before $p_*$ is reached. However, going further back in time, and assuming that ambiguity is sufficiently large, there comes a state $\ph^*$ at which the value of experimentation at its worst belief $\pl(\ph^*)$ is as low as $U_{\ell}(\ph^*)$. At that state, stopping becomes optimal for the DM. As noted above, given that the DM stops at $\ph^*$, earlier selves are willing to experiment until $\ph^*$ is reached. The equilibrium strategy thus prescribes experimentation to the left and right of $\ph^*$.

\section{Behavioral Implications of Ambiguity}

In this section, we perform comparative statics to further explore the behavioral implications of ambiguity.\\

\noindent\emph{Prolonged Learning and Indecision.} We first investigate the DM's incentive to prolong her experimentation late in the game and how it varies with ambiguity.  
For the latter purpose, we denote by $\theta$ the {\it true} probability of $R$ and consider the case where the DM's set of priors $\mathcal{P}$ is centered around $\theta$. Specifically, for each $\theta\in(0,1)$ and each $\Delta\geq 0$, we define
\[
\mathcal{P}_{\Delta}(\theta):=\left\{p\in[0,1]:\textstyle{\left|\ln\left(\frac{p}{1-p}\right)-\ln\left(\frac{\theta}{1-\theta}\right)\right|\leq\frac{\Delta}{2}}\right\}
\]
as the set of priors with midpoint $\theta$ in terms of the log-likelihood ratio and ambiguity level $\Delta$. We then denote by $T_{\Delta}(\theta)$ the expected length of experimentation when the DM's set of priors is $\mathcal{P}_{\Delta}(\theta)$ and the true probability of state $R$ is $\theta$. We have the following result. 

\begin{proposition} [Length of experimentation] \label{prop:prolong}  Suppose $c<\ch$ so that experimentation occurs under some $\theta$. Let $0\leq\Delta<\Delta'$ and define $\overline\theta:=\sup\{\theta: T_{\Delta}(\theta)>0\}$.  There exists some $\hat\theta\in(0,\overline\theta]$ such that $T_{\Delta'}(\theta)\geq T_{\Delta}(\theta)$ if  $\theta<\hat \theta$ and $T_{\Delta'}(\theta)\le T_{\Delta}(\theta)$ if $\theta\in [\hat\theta,\overline\theta)$.  Further, $T'_{\Delta'}(\theta)\leq T'_{\Delta}(\theta)$ for all $\theta<\hat\theta$. 
\end{proposition}

\begin{figure}
\begin{centering}
\includegraphics[width=0.6\textwidth]{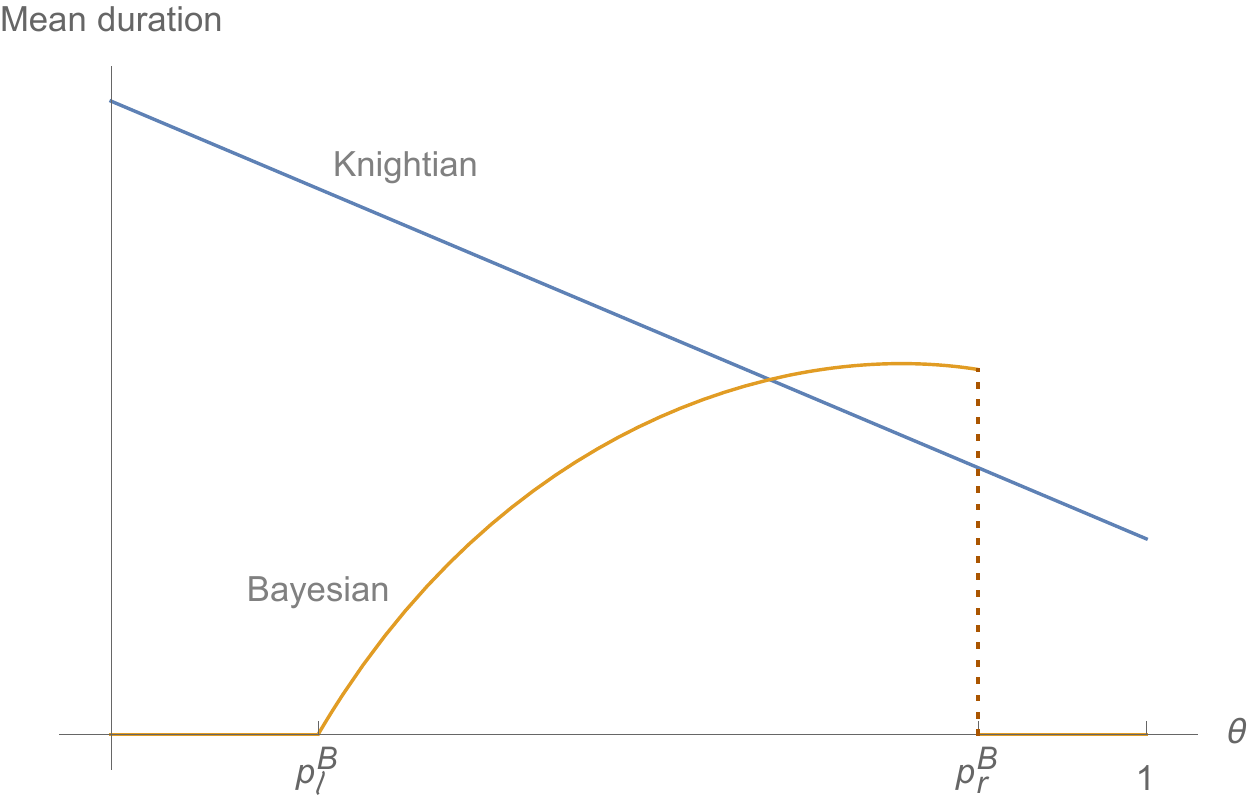}
\par\end{centering}
\caption{Mean sampling time for the Bayesian case and the case of Knightian uncertainty.
\label{fig:CS1}}
\end{figure}

In words,  an increased ambiguity leads the DM to prolong her experimentation when the likelihood $\theta$ of state $R$ is sufficiently small. For a benchmark, consider the Bayesian DM with 
accurate prior $p_0=\theta$ and no ambiguity ($\Delta=0$). The length of experimentation for the Bayesian DM is concave and strictly increasing in $\theta$ for $\theta$ not too large,\footnote{See Proposition 1 of \cite{Che2019}.} as illustrated in \Cref{fig:CS1}.  There are two competing forces.  On the one hand, as $\theta$ falls, experimentation is less likely to produce conclusive $R$-evidence; this tends to lengthen the experimentation.  On the other hand, the DM adjusts her learning strategy to shorten its duration as her (accurate) prior $\theta$ falls.  For $\theta$ sufficiently small, the latter effect dominates the former effect, so the overall learning time decreases as $R$ becomes less likely.

As stated in \Cref{prop:prolong}, ambiguity makes the DM's strategy less responsive to the change in $\theta$. The reduced responsiveness to $\theta$ corresponds to the behavioral trait of ``indecisiveness.''  In the  extreme case of $\Delta\to\infty$ (i.e., Knightian ambiguity), the learning strategy becomes completely unresponsive to the true likelihood $\theta$, so that, as  $\theta$ falls, her learning time actually increases (see \Cref{fig:CS1}). The reason is that even after an arbitrarily long duration of unsuccessful experimentation, the DM's extreme ambiguity  prevents her from dismissing  $R$ as  sufficiently unlikely and stopping experimentation.\\

\noindent\emph{Premature Decisiveness.}   While the Bayesian optimal stopping time is deterministic conditional on not receiving $R$-evidence, a sufficiently large ambiguity leads to randomized stopping at earlier points in time. Therefore, the ambiguity-averse DM exhibits early ``decisiveness'' not shown by her Bayesian counterpart. 
At the same time, an increased ambiguity also leads to prolonged experimentation, as seen in \Cref{prop:prolong}, which increases the likelihood of delayed stopping. The combined effect is the increased dispersion of the stopping times.  To formalize this result, consider two prior sets $\mathcal P$ and $\mathcal Q$ with $\mathcal P\subset \mathcal Q$.   Let $F_{\mathcal P}:[0,+\infty)\to[0,1]$ denote the cumulative distribution function over stopping times when the initial set of priors is $\mathcal{P}$. Hence, given prior set $\mathcal{P}$, $F_{\mathcal{P}}(t)$ is the probability that the DM stops before time $t$. 

\begin{proposition}[Dispersion of stopping times.]\label{prop:CS}
Assume $c\leq\underline c$ and $\Phi'(p_*)\geq0$. Let $\mathcal{P}$ and $\mathcal{Q}$ be such that $\mathcal{P}\subseteq \mathcal{Q}$. There exists a time $\hat t$ such that $F_{\mathcal{Q}}(t)\geq F_{\mathcal{P}}(t)$ for all $t<\hat t$ and $F_{\mathcal{Q}}(t)\leq F_{\mathcal{P}}(t)$ for all $t>\hat t$.
\end{proposition}

\begin{figure}
\begin{centering}
\includegraphics[width=0.6\textwidth]{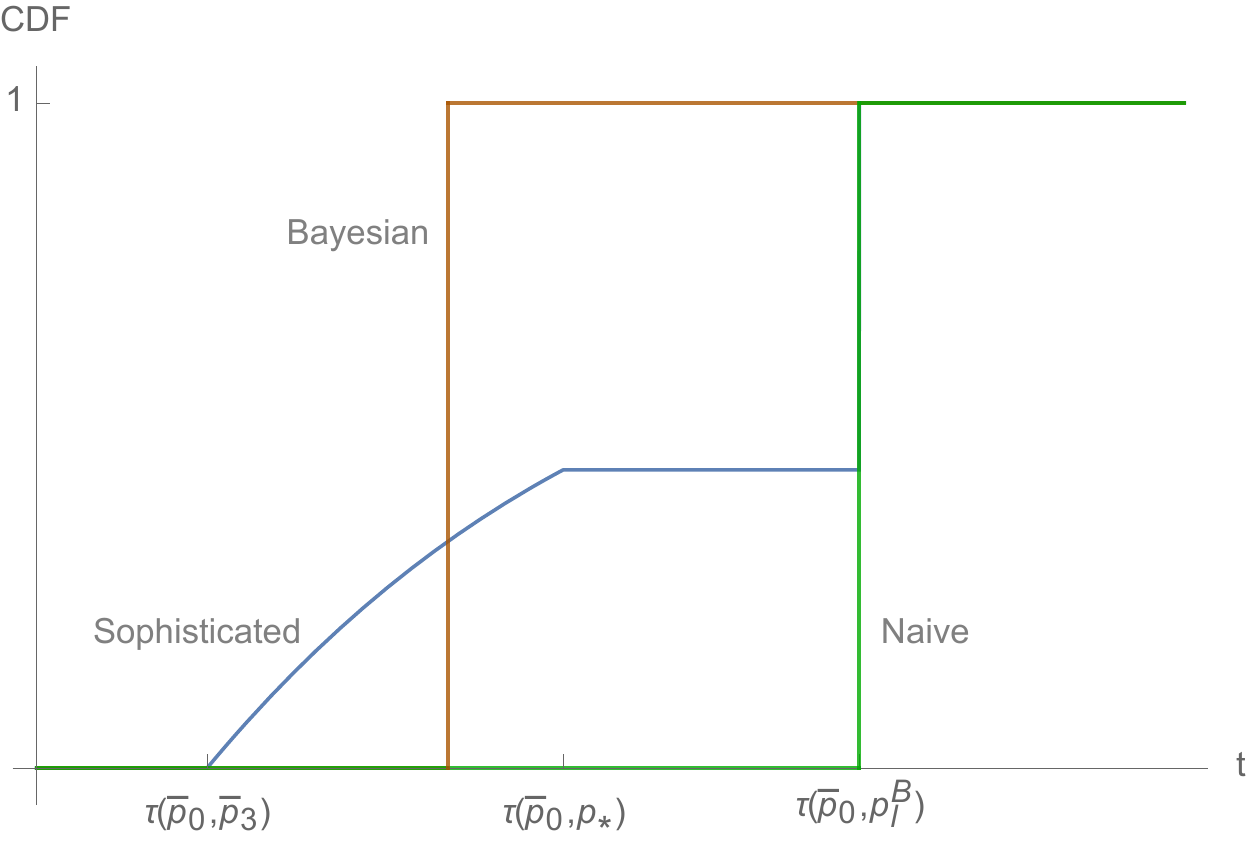}
\par\end{centering}
\caption{Cumulative distribution of stopping times (conditional on $\omega=L$) for the Bayesian DM and the ambiguity-averse DM when sophisticated (in blue) and naive (in green). In the graph, $\tau(p,q)$ denotes the time it takes for belief $p$ to be updated to $q$ in the absence of $R$-evidence.
\label{fig:CS2}}
\end{figure}

\cref{prop:CS} shows that the stopping time distribution under the two nested sets $\mathcal{P}$ and $\mathcal{Q}$ exhibits a single crossing property. This property indicates that the distribution over stopping times under the larger set $\mathcal{Q}$ is more dispersed than the one under $\mathcal{P}$. Indeed, whenever the mean stopping time under $\mathcal{P}$ is weakly greater than under $\mathcal{Q}$, then $F_{\mathcal{P}}$ second-order stochastically dominates $F_{\mathcal{Q}}$ (when the means are equal, $F_{\mathcal{Q}}$ constitutes a mean-preserving spread of $F_{\mathcal{P}}$).  The single crossing property implies that the range of potential stopping points spreads as ambiguity becomes larger: an expansion of the initial set of priors leads to an earlier first potential stopping point and a later last potential stopping point. The property is illustrated in \cref{fig:CS2}, where we plot the cumulative probability of stopping conditional on state L for the Bayesian DM (orange graph) and the ambiguity-averse DM (blue graph).\\

\noindent\emph{Can the two traits coexist?}  According to our theory, ambiguity-aversion leaves  potentially testable behavioral markers---prolonged indecision and impulsive decisiveness.  While seemingly contradictory, there is a coherent logic that ties these two traits together. For an illustration, consider again the theorist seeking to prove a new theorem with unknown validity.  According to our theory, ambiguity---with regard to the unknown validity of the theorem---makes it difficult for the theorist to dismiss the validity of the theorem even after many unsuccessful attempts to prove it.  Hence, ambiguity might lead the DM to a costly process of continued trial that could be seen by the outsider as 
a ``wild goose chase."  Recognizing that such an obsessive pursuit is in the offing---perhaps also reminded by how such endeavors ended in the past---the DM may simply refrain from getting into the situation by stopping early; the theorist may prematurely give up after making a few unsuccessful attempts at proving the theorem.\footnote{
\cite{barkley-levenson2016} find indecision and impulsivity to be positively correlated among the respondents of their behavioral survey, and reach an interpretation uncannily close to our theory:  ``Our findings suggest ... indecisiveness and impulsivity can be viewed as two sides of the same coin. Both indecisiveness and impulsivity are maladaptive behavioral responses to difficulty engaging with a decision.
We surmise that these responses arise from a common desire
to avoid negative affect that some individuals experience when making choices. ... they may behave in an impulsive manner, hastily choosing in order to avoid unpleasant deliberation over opportunity costs.''}\\

\noindent\emph{The role of sophistication.} While the feature of prolonged learning as a response to ambiguity does not hinge on the DM's sophistication, premature decisiveness is driven by the DM's ability to correctly forecast her future behavior and respond accordingly. Indeed, a fully naive DM---one who acts according to the best plan given by the current worst-case scenario but fails to carry out the plans of earlier selves---only exhibits the first behavioral trait and will thus always experiment longer than the corresponding Bayesian DM. The sophisticated DM, on the other hand, will engage in preemptive stopping when ambiguity is sufficiently large, so she will tend to stop earlier than her naive counterpart.  Note, however, that both DMs share the same final stopping points, so with positive probability both stop at the same time. We illustrate this feature in \cref{fig:CS2}:  the CDF for the naive DM first-order stochastically dominates that of the sophisticated DM.

It is also interesting to contrast the sophisticated and naive DM with a DM who has full commitment power. As we saw in \cref{sec:model}, the commitment solution is given by the Bayesian optimal stopping rule for the prior in $[\pl_0,\ph_0]$ that minimizes the Bayesian value function $\Phi^*$. Assuming $\ph_0>p_*$, the graph illustrating the commitment solution is thus qualitatively similar to that of the Bayesian DM in \cref{fig:CS2}.  Specifically, a DM with full commitment will stop before the naive DM does. The comparison with a sophisticated DM is ambiguous, except that the latter's stopping time is more dispersed in the sense established in \Cref{prop:random}.

\section{Incremental Learning}\label{ext:incremental}

So far, we have focused on Poisson learning, where news is one-sided and conclusive. Another canonical case is the incremental learning model in which the signal follows a diffusion process. In this section, we modify the baseline model and consider this alternative information technology. For simplicity, we restrict attention to the case where payoffs are symmetric across states: $u_r^R=u_{\ell}^L$, $u_r^L=u_{\ell}^R$.

Suppose the DM receives a signal $X\in\mathbb R$ about the state $\omega\in\{L,R\}$ at a flow cost of $c$ per unit time, which follows the process:
$$d X_t=\mu_{\omega} dt +\sigma dB_t,$$
where $\mu_R>\mu_L$.  As usual, it is more convenient to convert $X$ into a belief, particularly into a log-likelihood ratio $Z$,
where
$$
dZ_t= \frac{\psi}{\sigma} dX_t -\frac{\psi}{\sigma} \left( \frac{\mu_R+\mu_L}{2}\right)dt= \pm \frac{\psi^2}{2}dt +\psi dB_t,
$$
and $\psi:=(\mu_R-\mu_L)/\sigma$ is the signal-to-noise ratio (see \cite{daley2012waiting,daley2020bargaining} for example).  

\begin{figure}
\centering
\begin{subfigure}{.49\textwidth}
  \centering
  \includegraphics[width=.9\textwidth]{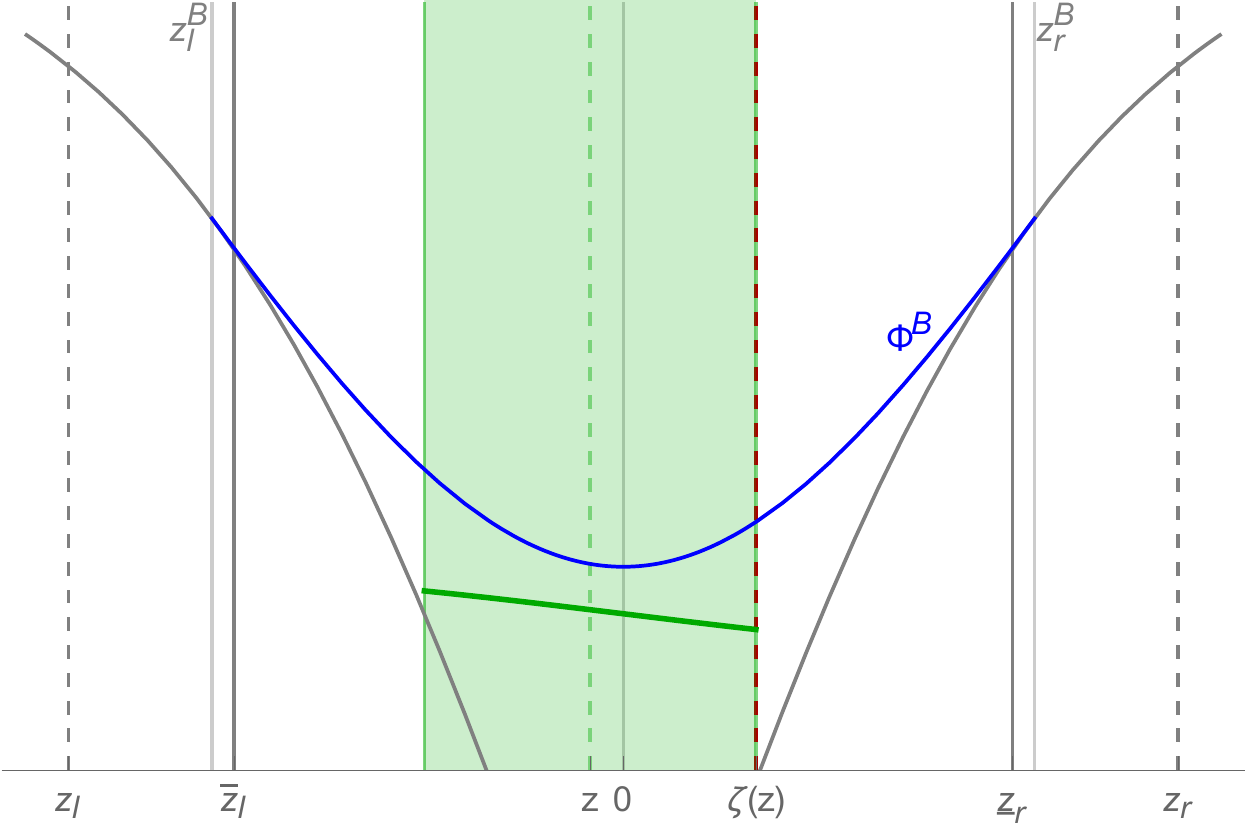}
\end{subfigure}
\begin{subfigure}{.49\textwidth}
  \centering
  \includegraphics[width=.9\textwidth]{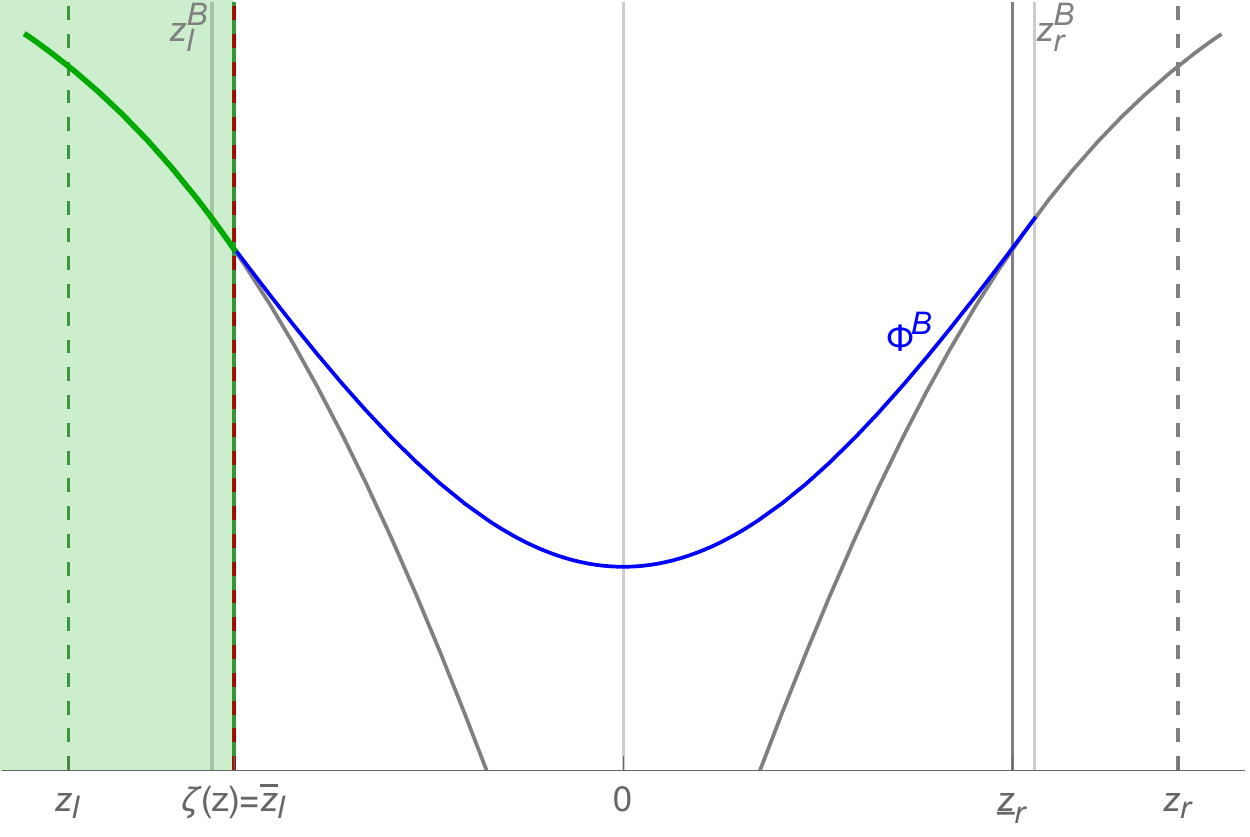}
\end{subfigure}
\caption{Agent's value when experimenting (left) and when stopping (right), with worst-case belief $\zeta(z)$ and boundaries $\overline z_{\ell}:=z_{\ell}+\Delta/2$ and $\underline z_r:=z_r-\Delta/2$. The Bayesian value function is $\Phi^B(z):=\phi(z|z_{\ell}^B,z_r^B)$.}
\label{fig:Brownian-small}
\end{figure}

As before, the Bayesian optimal behavior is characterized by two stopping boundaries (in terms of log-likelihood ratio), $z_{\ell}$ and $z_{r}$, such that the DM stops and chooses  $\ell$ if $z\le z_{\ell}$, stops and chooses $r$ if $z\ge z_r$, and experiments if $z\in (z_{\ell}, z_r)$. Let $\phi(z|z_{\ell},z_{r})$ be the value of this strategy when  the DM's prior is $z$, where $z=0$ corresponds to the belief of $p=1/2$. The Bayesian optimal stopping boundaries are denoted by $(z_{\ell}^B,z_{r}^B)$, where $z_{\ell}^B=-z_r^B<0$ due to the symmetry of payoffs. As before, the stopping boundaries are graphically depicted by tangent points of the Bayesian value function at the stopping payoffs (which are nonlinear due to the change of variable); see \Cref{fig:Brownian-small}.\footnote{This value function and the Bayesian optimal stopping boundaries, $z_{\ell}^B$ and $z_r^B$, can be derived using standard methods.}

Consider now an ambiguity-averse DM whose prior set is represented by an interval  $\mathcal{Y}(z):=[z-\Delta/2, z+\Delta/2]$. In contrast to before, let us index the ``state'' by the midpoint $z$ and the deviation from the midpoint in terms of the log-likelihood ratio by $\gamma\in[-\Delta/2,\Delta/2]$.  Upon suitable modification, the intra-personal equilibrium is characterized much like in the baseline model.\\

\noindent\emph{Small $\Delta$.} If ambiguity is sufficiently small, there are again two stopping boundaries $z_{\ell}<0<z_r$ (where $z_r=-z_{\ell}$) such that the DM experiments if and only if the state $z$ belongs to the interval $(z_{\ell}, z_{r})$. Suppose the DM follows this strategy.  Then, at state $z$, the DM with belief $(z+\gamma)\in \mathcal{Y}(z)$ perceives an expected payoff of
\[
V(z+\gamma,z)=\phi(z+\gamma|z_{\ell}+\gamma,z_r+\gamma).
\]
 
 Close to the stopping boundary, the DM's worst-case belief is given by the ``inner-most" belief of the set, just as in the Poisson model. The two stopping boundaries $z_{\ell}$ and $z_r$ are thus determined by the smooth-pasting conditions with respect to these beliefs:
\begin{eqnarray*}
 \phi'\left(z_{\ell}+\textstyle{\frac{\Delta}{2}}\,\left|\,z_{\ell}+\textstyle{\frac{\Delta}{2}},z_r+\textstyle{\frac{\Delta}{2}}\right.\right)&=&U_{\ell}'\left(z_{\ell}+\textstyle{\frac{\Delta}{2}}\right)\\
 \phi'\left(z_r-\textstyle{\frac{\Delta}{2}}\,\left|\,z_{\ell}-\textstyle{\frac{\Delta}{2}},z_r-\textstyle{\frac{\Delta}{2}}\right.\right)&=&U_{r}'\left(z_r-\textstyle{\frac{\Delta}{2}}\right),
 \end{eqnarray*}
 where, with slight abuse of notation, $U_a,a=\ell,r$ denotes the DM's stopping payoff as a function of the log-likelihood ratio. Stopping for action $\ell$ occurs at state $z_{\ell}$ where the DM is indifferent between stopping and experimenting {\it at her right-most belief} $z_{\ell}+\Delta/2$, while stopping for action $r$ occurs at state $z_{r}$ where the DM is indifferent between stopping and experimenting {\it at her left-most belief} $z_{r}-\Delta/2$. 
 
 A notable difference relative to the Poisson model is that the stopping beliefs, $z_{\ell}+\Delta/2$ and $z_r-\Delta/2$, no longer coincide with the corresponding Bayesian stopping boundaries, $z_{\ell}^B$ and $z_r^B$. Instead, we have $z_{\ell}^B<z_{\ell}+\Delta/2<0<z_r-\Delta/2<z_r^B$, so the DM stops experimenting before stopping becomes Bayesian optimal for all beliefs in the set.\footnote{
\cite{epstein2022optimal} obtain a qualitatively similar result for the case of rectangular sets of priors.} 
To see this, consider the case where the DM's right-most belief is close to $z_{\ell}^B$. If the DM continues experimentation, there is a small chance that the state drifts all the way to the other side of the right stopping boundary. The DM understands that if this happens, she will stop only when the left-most belief reaches the stopping boundary. From the viewpoint of the right-most belief, there is hence excessive experimentation close to the right stopping boundary, which reduces the value of experimenting. The same applies to the other side. The argument further implies that in the experimentation region, there is no belief in $[z-\Delta/2,z+\Delta/2]$ for which the equilibrium strategy is Bayesian optimal. This means that the value segment for the ambiguity-averse DM is strictly below, and thus never touches,  the Bayesian value function, as illustrated in \cref{fig:Brownian-small}. \\

\begin{figure}
  \centering
\begin{subfigure}{.49\textwidth}
  \centering
  \includegraphics[width=.9\textwidth]{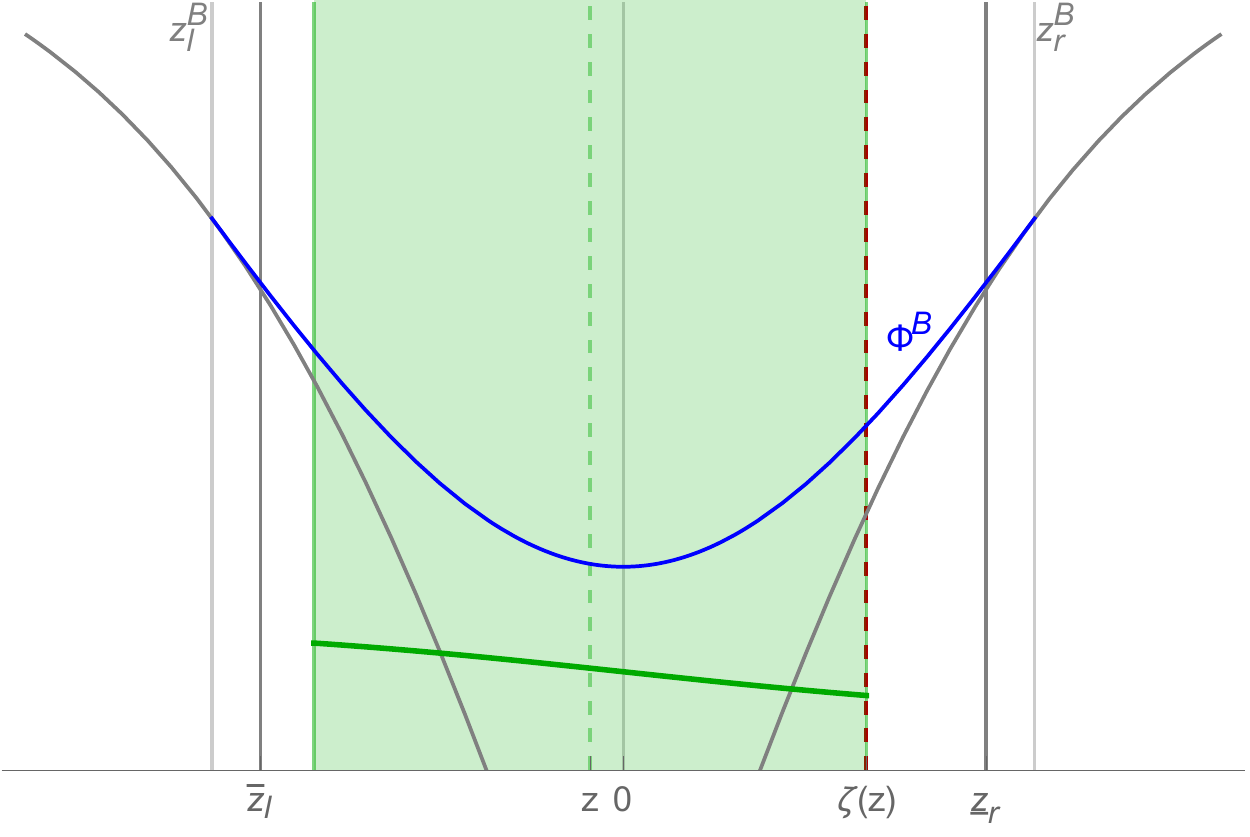}
\end{subfigure}
\begin{subfigure}{.49\textwidth}
    \centering
  \includegraphics[width=.9\textwidth]{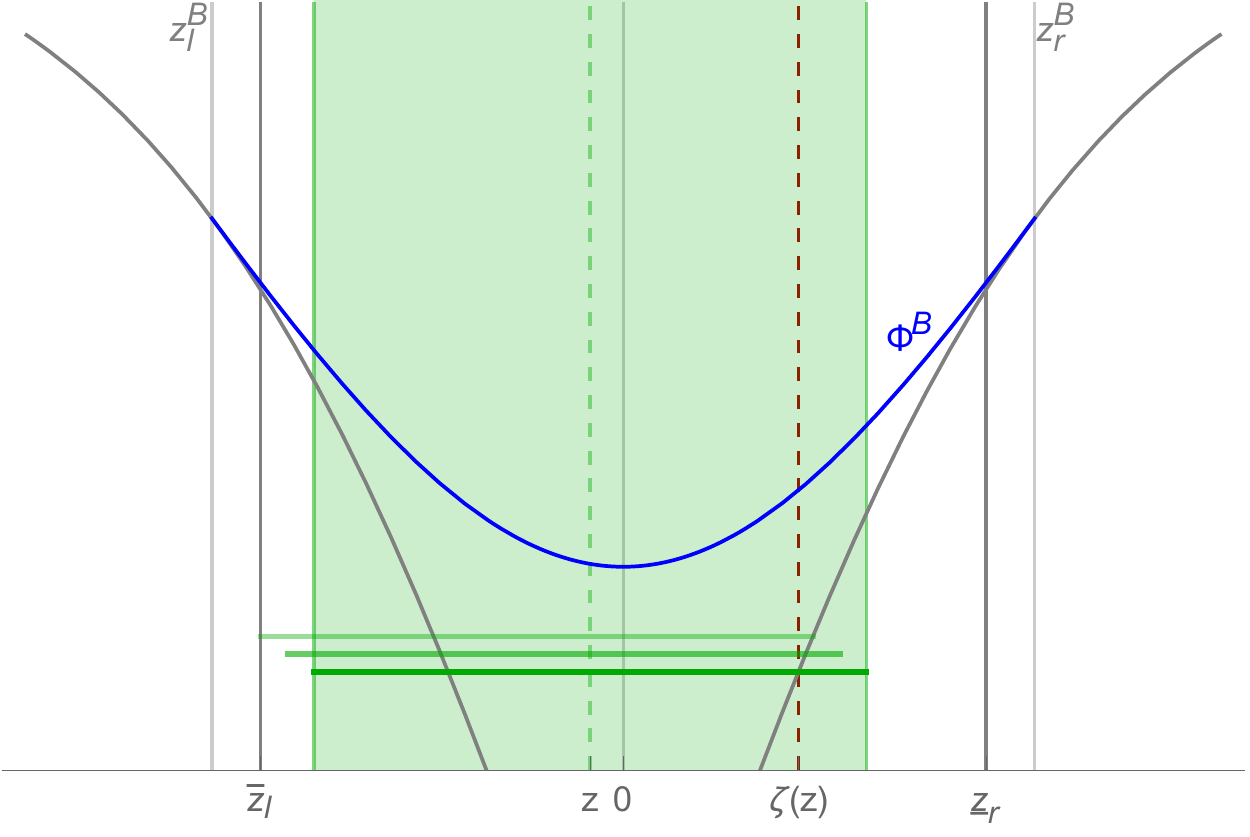}
  \end{subfigure}
\caption{The case of large $\Delta$: on the left, profitable deviation if the DM would experiment; on the right, intrapersonal equilibrium with mixed stopping (followed by $r$).}
\label{fig:Brownian-large}
\end{figure}

\noindent\emph{Large $\Delta$.} The above solution constitutes an equilibrium only when ambiguity is sufficiently small. Graphically, the value segment has to be sufficiently short so that its lower end does not fall below the stopping payoff.  
 Otherwise stopping becomes optimal under the worst-case belief, as we illustrate in \cref{fig:Brownian-large} (left panel). In such situations, the DM again randomizes between experimentation and stopping. Mixed stopping works here as follows: starting from $z=0$, if $z$ moves slightly to the left, the DM mixes between experimentation and stopping followed by action $r$; if instead $z$ moves slightly to the right, the DM mixes between experimentation and stopping followed by $\ell$.   This pattern is analogous to how  randomized stopping occurs in the baseline model. Intuitively, mixing with action $r$ when $z<0$ (resp. $\ell$ when $z>0$) hedges the DM against the possibility that the state is in fact $R$ (resp. $L$) and, hence, that experimentation will take a very long time.

The DM's value and equilibrium stopping rate in the mixed stopping region of the drift-diffusion model can be derived via the same approach we developed for the model with Poisson learning. The saddle-point HJB condition is now
\begin{align*}
	c& = \max_{\nu}\min_{y\in \mathcal{Y}(z)} \nu (U(y)-V(y,z))+ \frac{\psi^2}{2}(2p(y)-1)V_{y}(y,z)+\frac{\psi^2}{2}(2p(y)-1)V_{z}(y,z) \\
	& + \frac{\psi^2}{2}(V_{yy}(y,z)+V_{yz}(y,z)+V_{zz}(y,z)),
\end{align*}
where $p(y):=e^y/(1+e^y)$ is the probability of $R$ corresponding to the log-likelihood ratio $y$.
Letting $\zeta=\zeta(z)$ denote nature's choice in state $z$, we can derive $\zeta(z)$ and the DM's stopping rate $\nu(z)$ by following the same steps as in \cref{sec:main-result}. Let us consider the case $z<0$ where stopping is followed by $r$ (the case $z>0$ is analogous). Randomization between experimentation and $r$ requires that the coefficient of $\nu$ must vanish: 
\begin{equation}\label{eq:ind-DMz}
U_r(\zeta(z))=V(\zeta(z),z).
\end{equation}
To satisfy this condition, nature must choose an interior $\zeta(z)$, which requires that the value segment is flat. This means that $V_y, V_{yz}, V_{yy}$ all vanish and we can write $V(y,z)=\hat V(z)$ for all $y$. We then use \cref{eq:ind-DMz} to pin down nature's choice:
\begin{equation}\label{eq:zeta}
\zeta(z)=p^{-1}\left(\frac{\hat V(z)-u_r^L}{u_r^R-u_r^L}\right).
\end{equation}
For nature to choose an interior likelihood ratio, the coefficient of $y$ in the saddle-point condition must vanish, too:
\begin{equation}\label{eq:nu-z}
\nu(z)(u_r^R-u_r^L)+ \psi^2 V_{z}(\zeta(z),z)=0.
\end{equation}
 Using again \cref{eq:ind-DMz}, this requirement ties down the DM's stopping rate $\nu(z)$.
Finally, we need to determine the value function $\hat V(z)$. To this end, we substitute for $\zeta(z)$ and $\nu(z)$ to simplify the saddle-point condition and obtain the analog of \eqref{eq:ODE_Vhat}.
\begin{align*}
	c& = \psi^2\left(\frac{\hat V(z)-u_r^L}{u_r^R-u_r^L}-\frac{1}{2}\right)\hat V'(z)+\frac{\psi^2}{2}\hat V''(z).
\end{align*}
The main complication in the drift-diffusion model is that the boundary condition for this ODE is no longer pinned down by the minimum of the Bayesian value function, as it was before, but has to be found as a fixed point.

Apart from this technical difference, the solution method and qualitative features of the solution extend to the drift-diffusion model. In particular: 1) the DM has incentives to engage in excessive experimentation close to the stopping boundaries; 2) if ambiguity is large and the flow cost $c$ is small, the DM will exert self-control and preempt herself from experimenting excessively by randomly stopping at intermediate sets of beliefs.

\section{Concluding Remarks}

This paper studies the canonical Wald problem under the assumption that the DM faces uncertainty about the underlying probabilities of the different states. We consider a DM who seeks to maximize her payoff guarantee and updates beliefs prior by prior. Since  the DM's worst-case belief may change over time, the DM's preference (at the worst-case belief) may violate dynamic consistency. Rather than ruling out such violations, we seek to understand their behavioral implications on information acquisition. The main technical challenge is then to solve the DM's optimization problem in the presence of dynamic preference reversals. We develop a new method, based on a version of HJB conditions, and provide a micro-foundation for the approach. The method proves highly tractable and is applicable to dynamic optimization problems beyond information acquisition, such as real options theory, dynamic portfolio choice, or search models.

 We show that ambiguity gives rise to two behavioral traits---prolonged indecision and premature decisiveness. The inability by our DM to stop searching for $R$-evidence after many unsuccessful attempts is reminiscent of the \emph{sunk cost fallacy}, i.e., the tendency not to give up on projects or activities in which one has invested. In our case, the DM struggles to dismiss the possibility of the state being $R$ even after many failed bids to find evidence for it. The DM in that situation is guided by the right-most belief---namely, the belief that not only rationalizes the incurred experimentation costs so far but also calls for further continuation of learning. Ambiguity aversion could thus be viewed as an alternative mechanism behind the sunk cost fallacy, complementing other existing behavioral explanations. 

Our model can be extended to allow for a choice in the type of information the DM acquires. In \cref{sec:general-poisson} we  consider the general Poisson model introduced by \cite{Che2019}, where the DM has access to two news sources, one emitting $R$-evidence (as in the baseline model), the other emitting $L$-evidence. Conditional on acquiring information, the DM must then allocate her attention between  the two news sources. For instance, a theorist may try either to ``prove'' a theorem ($R$-evidence),  to find a ``counter-example'' disproving it ($L$-evidence), or to divide effort between the two endeavors. We show that  excessive learning and randomized stopping continue to arise in this setting with sufficiently large ambiguity and sufficiently small information costs.  In addition,  the DM may ``split'' her attention between the two news sources in order to hedge, even when a Bayesian DM would not do so under any prior. Whenever the DM employs the split-attention strategy, she continues learning until she fully resolves the uncertainty. This behavior thus provides another pattern of excessive learning, when it is seen from the perspective of a Bayesian observer with comparable beliefs. 

An open question is how our solution changes under different updating rules. Prior-by-prior updating entails that the DM updates on the state but not on the probability model. On the other extreme of the spectrum is  maximum-likelihood  updating \citep{gilboa1993updating}, according to which the DM only retains those priors which explain the data best. In our setting, this would not be very interesting, as the DM, in the absence of a breakthrough, would discard all priors but the left-most after the first instant of time. There are, however, less extreme versions that combine features of both updating rules (e.g. \cite{cheng2021relative}). It will be interesting to explore the implications of such rules on the incentives to acquire information in dynamic settings.

We focus on the case where the DM is fully sophisticated and thus correctly forecasts the behavior of her future selves. An open question is how the results extend to the case where the DM is partially naive. For instance, the DM may be paying attention only at certain points in time. This could be modeled by having a Poisson clock run in parallel, which then determines the times at which the DM contemplates the actions of her future selves and optimizes accordingly (otherwise she follows the naively optimal plan). A higher Poisson rate would thus capture a more sophisticated DM. Given such a model, one could ask how the DM's welfare (say with respect to initial preferences) would depend on the level of sophistication. We leave these interesting questions for future research.

\bibliographystyle{apalike}
\bibliography{ACM}

\newpage

\appendix

\section{Mathematical Appendix}

\subsection{Markov strategies and their values} \label{sec:strategies}

For any fixed prior $\mathcal{P}_0=[\pl_0,\ph_0]$, we first express the DM's strategy as specifying her action after each time $t$ of experimenting  without receiving any news.  (Recall that the latter are the only nontrivial subgames we focus on, since the DM chooses $r$ immediately following the arrival of $R$-evidence.)  

Specifically, a \emph{time-indexed strategy} is a function $\tilde \sigma:[0,\infty)\rightarrow[0,\infty)\times[0,1]^2$,
$\tilde \sigma_{t}=(\tilde \nu_{t},\tilde m_{t},\tilde \rho_{t})$,  where $\tilde\nu_{t}$ is the stopping rate and $\tilde m_{t}$ the instantaneous stopping probability used at time $t$ if no $R$-signal has been received up until time $t$. $\tilde \rho_{t}$
is the probability of action $r$ conditional on stopping at time
$t$ without receiving an $R$-signal.\footnote{\Cref{sec:stopping time} shows that the current formulation of a strategy is induced by a well-behaved stopping time distribution that the DM may choose.}  To ensure that a strategy admits a well-defined stopping time, we 
require that (a) any connected set of $t$'s in which   $\tilde m_{t}=1$ contains its infimum, and (b) the set $\{t\ge 0|\tilde m_{t}\in(0,1)\}$
is countable.  Time-indexed strategies satisfying these restrictions are called
\emph{admissible}, and $\widetilde\Sigma$ denotes the set of all admissible time-indexed strategies. Fix any  $\tilde\sigma\in \widetilde\Sigma$.  Then, starting at any time $s$, the probability that the strategy $\tilde \sigma$ prescribes the DM to stop by time $t>s$ is written as:
 $$F_s(t-s):=1-e^{-\int_{s}^{t}\tilde\nu_{s'}ds'}\prod_{s'\in [s,t]}(1-\tilde m_{s'}).$$
Using this function, we can write the expected payoff, or the {\it value}, of strategy $\tilde\sigma$ given belief $p$ at time $t$ as follows: 
\begin{align}
& \tilde{V}^{\tilde\sigma}(p,t)
\nonumber \\ =\, & p\left(\int_{t}^{\infty}(1-F_{t}(s))e^{-\lambda s}\left[\tilde\nu_{s}U^{R}(\tilde\rho_{s})+\lambda u_{r}^{R}-c\right]ds+\sum_{s\in   [t,\infty)}(1-F_{t}(s)) \tilde m_{s}\,e^{-\lambda s}U^{R}(\tilde\rho_{s})\right)\nonumber \\
 & +(1-p)\left(\int_{t}^{\infty}(1-F_{t}(s))e^{-\lambda s}\left[\tilde\nu_{s}U^{L}(\tilde\rho_{s})-c\right]\;ds+\sum_{s\in   [t,\infty)}(1-F_{t}(s))\tilde m_{s}\,U^{L}(\tilde\rho_{s})\right),\label{eq:def_Vtilde} 
\end{align}
where $U^{\omega}(\rho)=\rho u_{r}^{\omega}+(1-\rho)u_{\ell}^{\omega}$ 
denotes the utility from the mixed action $\rho$ if the state is
$\omega$. 

The law of motion for the belief in the absence of stopping is exogenous and deterministic, and
the belief always drifts downwards.  Hence, there is a one-to-one mapping
between time $t$ and the belief set $\mathcal{P}_t:=[\pl_t, \ph_t]$ for any initial prior belief set $\mathcal{P}_0:=[\pl_0, \ph_0]$ in the absence of stopping, {\it independently of the DM's strategy}.   That is, both $\pl_t$ and $\ph_t$ drift down according to the law of motion \cref{eq:law} given the initial values,  $\pl_0$ and $\ph_0$.   As in the text, we index the set $\mathcal{P}_t$ by the highest possible belief $\ph_t$, and call it the {\it state}. Noting that given $\ph$, $\pl$ is uniquely pinned down, we denote $\pl_t:=\pl(\ph_t)$. Any $\ph<\ph_0$ can then be uniquely mapped to the time $t\ge 0$ it takes for the state to drift down to $\ph$: 
\begin{align} \label{eq:time}
 t=\tau(\ph):=\frac{1}{\lambda}\ln\left(\frac{\ph}{1-\ph}\frac{1-\ph_0}{\ph_0}\right).
\end{align}
 Consequently, any time-indexed strategy $\tilde\sigma=(\tilde \sigma_t)$ induces a strategy 
$\sigma=(\sigma (\ph)):= (\tilde \sigma_{\tau(\ph)})$, and conversely, any strategy induces a time-indexed strategy via $\tilde \sigma_t :=\sigma(\tau^{-1}(t))$.   Importantly, note that the Markov structure of the strategy---the dependence of the strategy only on the state---involves no restriction on the strategy space.\footnote{Recall that this one-to-one mapping is a result of our environment.  Beyond our setting, it may be restrictive to focus on  Markov strategies.  That is, if   the mapping $t\mapsto\ph_{t}$ is not injective, then there can be strategies that depend
on calendar time that cannot be encoded as a Markov strategy.}
Clearly, the admissibility on $\tilde \sigma$ corresponds to the admissibility imposed on $\sigma$. The expected payoff from any $\sigma\in \Sigma$ can then be computed from \cref{eq:def_Vtilde} via a change of variable using \cref{eq:time}:   
\begin{align}\label{eq:value}
   V^{\sigma}(p,\ph):=\tilde V^{\tilde\sigma_{\tau(\cdot)}}(p,\tau(\ph)). 
\end{align}

\subsection{Microfoundation of the saddle-point HJB equations}  \label{sec:microfound-HJB}

We now justify our equilibrium concept or the use of saddle-point HJB equations.  The idea is to imagine that at each point in time, the DM can control, or commit to, her actions for a brief period of time, say $\e>0$. She thus acts dynamically consistently for a brief period of time, while taking as given the strategies of her future selves beyond that time interval. We then require the strategy to be a limit of the $\e$-dynamically consistent strategies as $\e\to 0$.

To make this formal, fix a candidate strategy $\sigma^*$.  Suppose that, at any state $\ph$ and any time $t$, the DM can commit to any strategy $\sigma$ (possibly different from $\sigma^*$) during a time interval $[t, t+\e)$, for some $\epsilon>0$, and the strategy switches to $\sigma^*$ thereafter.  
 
Let $(\sigma_{\e}(\cdot),\pi_{\e}(\cdot))$ be the saddle point strategy for such an $\e$ horizon problem (followed by $\s^*$), or an $\e$-commitment solution. We say that a pseudo equilibrium $(\s^*, \p^*)$ is {\it dynamically credible} if for each $\ph$ there is a sequence of $\e$-commitment solutions $(\sigma_{\e}(\ph),\pi_{\e}(\ph))$ that converges to $(\s^*,\p^*)$ as $\e\to 0$.  We further say that a pseudo equilibrium $(\s^*, \p^*)$ is {\it well-behaved} if, for each $\ph\in (0,1)$, $V^{\sigma^*}(\p^*(\ph),\ph), V^{\sigma^*}_p(\p^*(\ph),\ph)$ and $V^{\sigma^*}_p(\p^*(\ph),\ph_{-})$ are well-defined and  continuous at $(\p^*(\ph),\ph)$.
 We now provide the justification based on dynamic consistency.

Write $W^{\s}_{\e}(p,\ph)$ for the DM's value when the DM follows strategy $\s$ for $\e$-period, followed by the candidate strategy $\s^*$.  Then,
   \begin{align*}
 &  W^{\sigma}_{\e}(p,\ph)\\
= & \,  m(\ph) U_{\rho(\ph) }(p)+ (1-m(\ph) ) \left[  - p \int_{\ph^{\e}}^{\ph}e^{-\int_{0}^{\tau(\ph,\ph')}\left(\lambda+\nu(\ph^\tau)\right)d\tau}\left(-c+\lambda u_{r}^{R}+\nu(\ph')u_{\rho(\ph')}^R\right)\frac{1}{\eta(\ph')}d\ph' \right. \\
 & -(1-p) \int_{\ph^{\e}}^{\ph}e^{-\int_{0}^{\tau(\ph,\ph')}\nu(p^{\tau})d\tau}\left(-c+\nu(\ph')u_{\rho(\ph')}^{L}\right)\frac{1}{\eta(\ph')}d\ph'  \\  & \left.+\left(pe^{-\int_{0}^{\tau(\ph,\ph^{\e})}\left(\lambda+\nu(p^{\tau})\right)d\tau}+(1-p)e^{-\int_{0}^{\tau(\ph,\ph^{\e})}\nu(p^{\tau})d\tau}\right)V^{\sigma^*} (p^{\e},\ph^{\e})\right],
\end{align*}
where for any $p\in [0,1]$, $p^{\tau}$ denotes the $\tau$-ahead update of belief $p$, and
\begin{align*}  
 \tau(p,p'):=\frac{1}{\lambda}\ln\left(\frac{p}{1-p}\frac{1-p'}{p'}\right)
\end{align*}
is the time  it takes for the Bayesian update of $p$ to reach $p'$ in the absence of $R$-evidence.  (Note that the earlier definition of $\tau(\ph)$ in \Cref{sec:strategies} suppresses the first argument since it was fixed to $\ph_0$.) 
Let $(\sigma_{\e}(\cdot),\pi_{\e}(\cdot))$ denote an $\e$-commitment solution.  Then, it must satisfy the following saddle point condition: for each $\ph$
\begin{align}
    \sigma_{\e}(\ph) &\,  \in  \argmax_{\s} W_{\e}^{\s}(\pi_{\e}(\ph), \ph) \label{hgb1}\\ 
     \pi_{\e}(\ph) &\,  \in  \argmin_{p\in \mathcal{P}(\ph)} W_{\e}^{ \sigma_{\e}}(p, \ph). \label{hgb2} 
\end{align}
Let $W_{\e}(\p_{\e},\ph):=W^{\sigma_{\e}}_{\e}(\p_{\e},\ph)$ be the saddle point value of the $\e$-commitment solution at $\ph$.  
  
The $\e$-commitment saddle point value $W_{\e}(\p_{\e},\ph)$ is characterized as follows:  For any $\d\in (0,\e)$,
\begin{align*}
& W_{\e}(\p_{\e},\ph)\\
= & \, \max_{m,\nu, \rho}  m(\ph) U_{\rho(\ph)}(\p_{\e})+ (1-m(\ph)) \left[ - \p_{\e} \int_{\ph^{\d}}^{\ph}e^{-\int_{0}^{\tau(\ph,\ph')}\left(\lambda+\nu(p^{\tau})\right)d\tau}\left(-c+\lambda u_{r}^{R}+\nu(\ph')u_{\rho(\ph')}^{R}\right)\frac{1}{\eta(\ph')}d\ph' \right. \\
 & -(1-\p_{\e}) \int_{\ph^{\d}}^{\ph}e^{-\int_{0}^{\tau(\ph,\ph')}\nu(p^{\tau})d\tau}\left(-c+\nu(\ph')u_{\rho(\ph')}^{L}\right)\frac{1}{\eta(\ph')}d\ph'  \\  
 & \left.+\left(\p_{\e}e^{-\int_{0}^{\tau(\ph,\ph^{\d})}\left(\lambda+\nu(p^{\tau})\right)d\tau}+(1-\p_{\e})e^{-\int_{0}^{\tau(\ph,\ph^{\d})}\nu(p^{\tau})d\tau}\right)W_{\e-\d} (\pi_{\e}^{\d},\ph^{\d})\right].
\end{align*}
 
Following the dynamic programming principle, the RHS does not depend on any $\d\in (0, \e)$.  Hence, the derivative of the RHS with respect to $\d$ evaluated at $\d=0$ must equal zero.  So, we obtain the standard HJB equations:\footnote{Since 
 $$
 W_{\e-\d} (\pi_{\e}^{\d},\ph^{\d})= W_{\e}(\pi_{\e},\ph_-)+ \frac{\partial W_{\e}(\pi_{\e},\ph_-)}{\partial p} \eta(\pi_{\e}) \d+  \frac{\partial W_{\e}(\pi_{\e},\ph_-)}{\partial \ph} \eta(\ph) \d  + o(\d), 
 $$
 the derivative of $W_{\e-\d} (\pi_{\e}^{\d},\ph^{\d})$ with respect to $\d$ is:
$$ 
\lim_{\d\downarrow 0} \frac{W_{\e-\d} (\pi_{\e}^{\d},\ph^{\d})-W_{\e}(\pi_{\e},\ph_-)}{\d}=\frac{\partial W_{\e}(\pi_{\e},\ph_-)}{\partial p} \eta(\pi_{\e})+ \frac{\partial W_{\e}(\pi_{\e},\ph_-)}{\partial \ph} \eta(\ph).
$$}
   
\begin{align}
0 &=\max_{m,\nu,\rho}G(m,\nu,\rho,\p_{\e},\ph, W_{\e},dW_{\e}),\label{HJB0-e} 
\\  
\sigma_{\e}(\ph) & =(m_{\e}(\ph),\nu_{\e}(\ph),\rho_{\e}(\ph)) \in\arg\max_{m,\nu,\rho}G(m,\nu,\rho,\p_{\e},\ph,W_{\e},dW_{\e}).\label{HJBm-e} 
\end{align}

Likewise, by the saddle point condition, $W_{\e}(\p_{\e},\ph)$ is also equivalently characterized as: for any $\d\in (0,\e)$,
 \begin{align*}
& W_{\e}(\p_{\e},\ph)\\
= & \, \min_{p}   m_{\e}(\ph) U_{\rho_{\e}(\ph)}(p)+ (1-m_{\e}(\ph)) \left[  - p \int_{\ph^{\d}}^{\ph}e^{-\int_{0}^{\tau(\ph,\ph')}\left(\lambda+\nu_{\e}(p^{\tau})\right)d\tau}\left(-c+\lambda u_{r}^{R}+\nu_{\e}(\ph')u_{\rho(\ph')}^{R}\right)\frac{1}{\eta(\ph')}d\ph' \right. \\
 & -(1-p) \int_{\ph^{\d}}^{\ph}e^{-\int_{0}^{\tau(\ph,\ph')}\nu_{\e}(p^{\tau})d\tau}\left(-c+\nu_{\e}(\ph')u_{\rho(\ph')}^{L}\right)\frac{1}{\eta(\ph')}d\ph'  \\  & \left.+\left(pe^{-\int_{0}^{\tau(\ph,\ph^{\d})}\left(\lambda+\nu_{\e}(p^{\tau})\right)d\tau}+(1-p)e^{-\int_{0}^{\tau(\ph,\ph^{\d})}\nu_{\e}(p^{\tau})d\tau}\right)W^{\sigma}_{\e-\d} (p^{\d},\ph^{\d})\right].
\end{align*}

Following the dynamic programming principle, the RHS does not depend on any $\d\in (0, \e)$.  Hence, the derivative of the RHS with respect to $\d$ evaluated at $\d=0$ must equal zero.  So, we obtain the standard HJB equations:   
\begin{align}
\pi_{\e} \in\arg\min_{p\in \mathcal{P}(\ph)}G(m_{\e},\nu_{\e},\rho_{\e},p,\ph,W_{\e},dW_{\e}). \label{HJBpi-e} 
\end{align}

The proof is Appendix \cref{proof:refinement}. We begin with a useful observation:

\begin{lemma} \label{lem:refinement} Suppose a well-behaved pseudo equilibrium $(\s^*, \p^*)$ is dynamically credible. 
Then, for each  $\ph\in (0,1)$, for any sequence such that    $(\s_{\e}(\ph), \p_{\e}(\ph))\to (\s^*(\ph), \p^*(\ph))$ as $\e\to 0$,  $W_{\e}(\p_{\e}(\ph),\ph_{-}) \to V^{\sigma^*}(\p^*(\ph),\ph_{-})$, $\partial W_{\e}(\p_{\e}(\ph),\ph_-)/\partial p\to V^{\sigma^*}_p(\p^*(\ph),\ph_{-}) $, and $\partial W_{\e}(\p_{\e}(\ph),\ph_-)/\partial \ph\to V^{\sigma^*}_{\ph}(\p^*(\ph),\ph_{-}).$ 
   \end{lemma}

The result now follows.  

 \begin{theorem} \label{prop:HJB-justify}    
  If a well-behaved pseudo equilibrium  is dynamically credible,
then  it is an intrapersonal equilibrium.  
  Conversely, if there is a unique intrapersonal equilibrium, then it is dynamically credible.  
 \end{theorem}

\begin{proof}    Fix any $\ph$.   By the dynamic credibility of $(\s^*, \p^*)$, for each $\ph\in (0,1)$, there exists a sequence of $\e$ such that an associated $\e$-commitment solution $(\s_{\e}(\ph), \p_{\e}(\ph))$ converges to $(\s^*(\ph), \p^*(\ph))$ as $\e\to 0$. Given the continuity assumption, by \cref{lem:refinement}, $W_{\e}(\p_{\e}(\ph),\ph_{-}) \to V^{\sigma^*}(\p^*(\ph),\ph_{-})$, $\partial W_{\e}(\p_{\e}(\ph),\ph_-)/\partial p\to V^{\sigma^*}_p(\p^*(\ph),\ph_{-}) $, and $\partial W_{\e}(\p_{\e}(\ph),\ph_-)/\partial \ph\to V^{\sigma^*}_{\ph}(\p^*(\ph),\ph_{-}).$

Next, observe that $G$ is continuous.   Hence,  by Berge's maximum theorem,  the maximized value is continuous and the set of maximizers must be upper hemicontinuous, which in this case should mean a closed graph.  Hence, \cref{eq:HJB0} and \cref{eq:HJBm} must follow from \cref{HJB0-e} and \cref{HJBm-e} in the limit as $\e\to 0$.
 
 Applying the same argument, \cref{eq:HJBpi}  follows from \cref{HJBpi-e} in the limit as $\e\to 0$.  Again,  by Berge's maximum theorem, the minimized value is continuous and the set of minimizers must be upper hemicontinuous, which implies a closed graph.  Hence, the conditions must hold in the limit as $\e\to 0$.
 
 Conversely, suppose  $(\s^*, \p^*)$ is intrapersonal equilibrium.  Then, it satisfies  \cref{eq:HJB0}, \cref{eq:HJBm}, and  \cref{eq:HJBpi}, for each $\ph$.  Hence, for each $\ph$,  $V^{\sigma^*}(\p^*(\ph),\ph), V^{\sigma^*}_p(\p^*(\ph),\ph)$ and $V^{\sigma^*}_p(\p^*(\ph),\ph_{-})$ are well-defined and  continuous at $(\p^*(\ph),\ph)$.
Take a sequence (subsequence if necessary) of $\e>0$ such that $(\s_{\e}, \p_{\e},   W_{\e}(\p_{\e}(\ph),\ph),\partial W_{\e}(\p_{\e}(\ph),\ph)/\partial p, \partial W_{\e}(\p_{\e}(\ph),\ph)/\partial \ph)$ all converge to some limit, say $(\s_{0}, \p_{0},W_{0}(\p_{0}(\ph),\ph), \tilde W_{0, p}(\p_{0}(\ph),\ph), \tilde W_{0,\ph}(\p_{0}(\ph),\ph) )$.  The reason that such a sequence can be found is as follows.  First, 
  $(W_{\e}(\p_{\e}(\ph),\ph),\partial W_{\e}(\p_{\e}(\ph),\ph)/\partial p)$ is bounded within a compact set $\left[\min\{u^R_{\ell}, u^L_{r}\}, \max\{u^R_r,u^L_{\ell}\}\right]\times [u^R_{\ell}-u^L_{\ell}, u^R_r-u^L_{r}]$. This together with \cref{HJBm-e} means that   $\nu_{\e}(\ph)$ can be bounded within $[0, \overline \nu]$ for some large $\overline{\nu}$.  The boundedness of $\nu_{\e}(\ph)$ in turn means that $ \partial W_{\e}(\p_{\e}(\ph),\ph)/\partial \ph$ can be bounded within an compact set.  This proves that the converging sequence can be found for each $\ph$.  Next, since  $\partial^2 W_{\e}(\p_{\e}(\ph),\ph)/\partial p\partial \ph=0$ for all $\e$ (which follows from the fact that $W_{\e}(\cdot,\ph)$ is linear), we have $\tilde W_{0, p}(\p_{0}(\ph),\ph)= \partial W_{0}(\p_{0}(\ph),\ph)/\partial p$ and $\tilde W_{0, \ph}(\p_{0}(\ph),\ph)= \partial W_{0}(\p_{0}(\ph),\ph)/\partial \ph$.  It is immediate that the limiting strategy profile $(\s_{0}, \p_{0})$ is a pseudo equilibrium.  Also using the same argument as above, we conclude that the limit strategy profile $(\s_{0}, \p_{0})$, together with the value function $W_0$, satisfies \cref{eq:HJB0}, \cref{eq:HJBm},  and \cref{eq:HJBpi} for each $\ph$.  Therefore, $(\s_{0}, \p_{0})$ is also an intrapersonal equilibrium.  The uniqueness of the equilibrium then means that $(\s_{0}, \p_{0})=(\s^*,\p^*)$, so for each $\ph$, $(\s^*(\ph),\p_0^*(\ph))$ is a limit of some $\e$-commitment solution $(\s_{\e}(\ph),\p_{\e}(\ph))$ as $\e\to 0$.  Hence, $(\s^*,\p^*)$
  is dynamically credible.
  \end{proof}

\subsection{Proof of \cref{thm:main_result}} \label{proof:main}

   For our purpose, we first establish a preliminary result about the Bayesian value function.  Recall the solution $\phi$  to the ODE \cref{eq:Bayesian-ODE}.  Then, the following lemma holds. 

\begin{lemma} \label{lem:SCP}  
For any $p>  {p}^{B}_{\ell}$,  $\phi'(p; p, U_{\ell}(p)) -U_{\ell}'(p)> 0$ whenever $\phi(p; p, U_{\ell}(p))=U_{\ell}(p)$.
\end{lemma}

\begin{proof}
Define $\delta^{\omega}:=|\delta_r^{\omega}-\delta_{\ell}^{\omega}|,\omega=L,R$. It follows from \cref{eq:Bayesian-ODE} that
\begin{align*}
\lambda p(1-p)(\phi'(p; p, U_{\ell}(p)) -U_{\ell}'(p))&= \lambda p (u^R_r-\phi(p;p, U_{\ell}(p)))-c +\lambda (1-p) (u^L_{\ell}-U_{\ell}(p)) \\
&= \lambda \left( p  u^R_r + (1-p)  u^L_{\ell}-U_{\ell}(p)\right)-c \\
& = \lambda p \delta^R -c,
\end{align*}
where the second equality uses the fact that $ \phi(p;p, U_{\ell}(p)))
=U_{\ell}(p)$. Since the last line is strictly increasing in $p$ and equals $0$ when $p= {p}_{\ell}^{B}$, the above claim is proven.
\end{proof}
\medskip
 
For the proof, for each region of $\ph$'s, we will specify the strategy profile $(\sigma, \pi )$ and the associated value $V^{\sigma}(p,\ph)$ that are not fully specified in \cref{thm:main_result}. We will then verify the value function together with the strategy profile $(\sigma,\pi )$ satisfies the equilibrium conditions.   

For ease of presentation, we shall mainly focus on Case 1.  Further, we will assume that either $c<\underline c$ or $c\in (\cl, \ch]$ but $\Delta<\Delta_c$, where $\Delta_c$ will be specified later (in Region 3 in   \cref{sec:main-case}).  We call this case \emph{main case}.   We will first treat the main case in \cref{sec:main-case}.  The remaining cases will be treated in \cref{ssec:large-c} and \cref{ssec:case2}.

\subsubsection{Main Case} \label{sec:main-case}
 
\subsubsection*{Region 1: $\protect\ph\in[0,\protect\ph_{1}]$, where $\ph_1:=p_{\ell}^B$.}
 
\noindent\emph{Computation of equilibrium value.}
Recall that the strategy $\sigma$ calls for an immediate choice of $\ell$ for all state $\ph\le p_{\ell}^B= \ph_1$.  It thus immediately follows that the value associated with that strategy is:
$$
V^{\sigma}(p,\ph):=U_{\ell}(p), \mbox{ for all  } \ph\in [0,\ph_1].
$$

  \noindent\emph{Verification of equilibrium conditions.}  With $V^{\sigma}(p,\ph)=U_{\ell}(p)$, $m(\ph)=1$ and $\rho(\ph)=0$
  imply that  $\pi (\ph)=\ph$ is a minimizer in \cref{eq:HJBpi}.
  Since $U_{\ell}'(p)<0$ by assumption, $\pi  (\ph)=\ph$ is the
  unique minimizer in \cref{eq:worst-case-belief_proof1}. Substituting
  $V^{\sigma}(p,\ph)=U_{\ell}(p)$, $V_{p}^{\sigma}(p,\ph)=U'_{\ell}(p)$,
  $V_{\ph}^{\sigma}(p,\ph)=0$, and $\pi (\ph)=\ph$, \cref{eq:HJB0}
  simplifies to
  \[
  \max_{m,\nu,\rho}m\left[U_{\rho}(\ph)-U_{\ell}(\ph)\right]+(1-m)\left[-c+\nu(U_{\rho}(\ph)-U_{\ell}(\ph))+\ph\lambda(u_{r}^{R}-U_{\ell}(\ph))+U'_{\ell}(\ph)\eta (\ph)\right]=0.
  \]
  
  With $m=1$ and $\rho=0$, the LHS is zero. Moreover, for any $\rho,\nu$,
  simple algebra shows that the coefficient of $(1-m)$ is non-positive
  for all $\ph\le\ph_{1}$. Hence $\sigma=(1,0,0)$ is a maximizer and
  both \cref{eq:HJB0} and \cref{eq:HJBm}
  are satisfied for $V^{\sigma}(p,\ph)=U_{\ell}(p)$. Setting $m(\ph)=1$
  and $\rho(\ph)=0$, the objective in \cref{eq:HJBpi}
  is independent of $p$ and $\pi (\ph)=\ph$ is a minimizer.
  Hence we have shown that for the posited strategy profile $(\sigma,\pi )$ satisfy the equilibrium conditions, \cref{eq:worst-case-belief_proof1}--\cref{eq:HJBpi} for all $\ph\in[0,\ph_{1}]$.

\subsubsection*{Region 2: $\protect\ph\in ( \ph_{1}, \ph_2]$, where $\ph_2:=p_*$.}

\noindent\emph{Computation of equilibrium value.} Recall that the strategy $\sigma$ calls on the DM to experiment until  $\ph$ drifts to $\ph_1=p_{\ell}^B$.  Fix any state $\ph$ in this region. To compute the associated value   $V^{\sigma}(\cdot,\ph)$, it is useful to ask: \emph{for which belief $p\in \mathcal{P}(\ph)$ is the continuation strategy $\sigma_{\ph}$ Bayesian optimal?}  The answer is $p=\ph$, since the strategy is exactly what a Bayesian DM with belief $\ph$ will do---experimenting until $\ph$ reaches the Bayesian optimal stopping belief $p_{\ell}^B=\ph_1$. Consequently, we must have 
$$V^{\sigma}(\ph,\ph)=\Phi(\ph).$$
Since the strategy needs not be optimal for any DM with belief $p\ne \ph$ (including $p$ outside $\mathcal{P}(\ph)$), we must have 
$$V^{\sigma}(p,\ph)\le \Phi(p),  \forall p\in [0,1].$$
Finally, since $V^{\sigma}(p,\ph)$---the valuation of a fixed action path---must be linear in $p$, the preceding observations must imply:
\begin{equation} \label{eq:V-sigma_region2}
V^{\sigma}(p,\ph):= \Phi(\ph)+(p-\ph)\Phi^{\prime}(\ph), \mbox{ for all  } \ph\in  ( \ph_{1}, \ph_2].
\end{equation}

\noindent\emph{Verification of equilibrium conditions.}
Since $\Phi^{\prime}(\ph)\le 0$ for all $\ph\in(\ph_{1},\ph_2]$, $V_{p}^{\sigma}(p,\ph)\le 0$ and therefore
  $\pi  (\ph)=\ph$ is a minimizer in \cref{eq:worst-case-belief_proof1}.  With $\pi  (\ph)=\ph$ we have 
  \[
  V^{\sigma}(\pi (\ph),\ph)=\Phi (\ph),\quad V_{p}^{\sigma}(\pi  (\ph),\ph)=\Phi^{\prime}(\ph),\quad V_{\ph}^{\sigma}(\pi  (\ph),\ph)=0.
  \]
  Therefore \cref{eq:HJB0} simplifies to
  \[
  \max_{m,\nu,\rho}m\left[U_{\rho}(\ph)-\Phi (\ph)\right]+(1-m)\left[-c+\nu(U_{\rho}(\ph)-\Phi (\ph))+\ph\lambda(u_{r}^{R}-\Phi (\ph))+\Phi^{\prime}(\ph)\eta (\ph)\right]=0.
  \]
  Since $\Phi (\ph)\ge U(\ph)$ for any $\ph$, $\nu=0$ is optimal. Moreover,
  $m=0$ is optimal since we have from \cref{eq:Bayesian-ODE}
$$-c+\ph\lambda(u_{r}^{R}-\Phi (\ph))+\Phi^{\prime}(\ph)\eta (\ph)=0$$
  for the Bayesian value function.  Therefore \cref{eq:HJB0}
  and \cref{eq:HJBm} are satisfied for all $\ph\in(\ph_{1},\ph_2]$.
  Condition \cref{eq:HJBpi} holds since $m(\ph)=\nu(\ph)=0$
  implies that the objective in \cref{eq:HJBpi}
  is given by 
  \begin{align*}
  	& -c+p\lambda\left(u_{r}^{R}-\left(\Phi (\ph)+(p-\ph)\Phi^{\prime}(\ph)\right)\right)-\lambda p(1-p)\Phi^{\prime}(\ph)-\lambda\ph(1-\ph)(p-\ph)\Phi^{\prime\prime}(\ph)\\
  	= & -c+p\lambda\left(u_{r}^{R}-\left(\Phi (\ph)+(1-\ph)\Phi^{\prime}(\ph)\right)\right)-\lambda\ph(1-\ph)(p-\ph)\Phi^{\prime\prime}(\ph).
  \end{align*}
  Differentiating this with respect to $p$ yields 
  \begin{align*}
  	& \lambda\left(u_{r}^{R}-\left(\Phi (\ph)+(1-\ph)\Phi^{\prime}(\ph)\right)\right)-\lambda\ph(1-\ph)\Phi^{\prime\prime}(\ph).
  \end{align*}
  Differentiating the ODE for the Bayesian Value function (see \cref{eq:Bayesian-ODE})
  with respect to $\ph$ we obtain that this expression is equal
  to zero. Hence $\pi  (\ph)$ is a minimizer in \cref{eq:HJBpi}.
  
  To summarize, we have shown that for the posited profile $(\sigma, \pi )$, together with the value function
  $V^{\sigma}$, satisfy \cref{eq:worst-case-belief_proof1}--\cref{eq:HJBpi} for all $\ph\in(\ph_{1},\ph_2]$.

\subsubsection*{Region 3: $\protect\ph\in ( \ph_{2}, \ph_3]$.}

This region exists only when the ambiguity is sufficiently large so that $V^{\sigma}(\pl(\ph_2), \ph_2) < U_{\ell}(\pl(\ph_2))$.  If this inequality is reversed, then we set $\ph_3=\ph_2$, and Region 3 is empty. 
Assuming the inequality, we will specify the upper bound $\ph_3$, the critical cost $\cl$, and ambiguity level $\Delta_c$ referred to in the statement of the theorem.  Finally, we will specify $\nu(\ph)$ and $\pi(\ph)$ fully for $\ph$ in this region.\\ 
 
 \noindent\emph{Specification of $\nu$ and $\pi $ and equilibrium value.}
 The strategy $\sigma$ for this region involves randomization between experimentation and stopping for $\ell$, with the latter done at a Poisson rate $\nu(\ph)$. Meanwhile, nature chooses belief $\pi (\ph)\in \mathcal{P}(\ph)$.  Here, we specify $(\nu, \pi )$ precisely, together with the value $V^{\sigma}(p,\ph)$ associated with the strategy.  We first construct a value function $V(p,\ph)$ 
 that satisfies the equilibrium conditions, \cref{eq:worst-case-belief_proof1}-\cref{eq:HJBpi}, given the candidate strategy $\sigma$.  We will then establish that the constructed function $V(p,\ph)$ indeed coincides with the value of $\sigma$.
 
To begin, fix any $\ph>\ph_2$.  
First, the fact that the DM randomizes between experimentation and action $\ell$ means that the coefficient of $\nu$ in \cref{eq:HJBm} must vanish, which implies \eqref{eq:ind-DM}. For the required belief $\pi(\ph)$ to be nature's choice satisfying \cref{eq:worst-case-belief_proof1}, it is sufficient, and will be seen also necessary, to have  $V _p(p,\ph)=0$. Namely,  
 \begin{equation} \label{eq:R3-2}
V (p,\ph)=\hat V(\ph), \forall p.
 \end{equation}
Substituting this into \cref{eq:ind-DM}, we get  \eqref{eq:pi-r3}. The randomization by DM in turn implies that the derivative of the objective in   \cref{eq:HJBm} with respect to $p$ must vanish.  This fact, together with \cref{eq:R3-2} yields \eqref{eq:nu-r3}.
 
It now remains to specify the function $\hat{V} (\ph)$.  To this end, we use \cref{eq:pi-r3},\cref{eq:nu-r3} and \cref{eq:R3-2} to simplify \cref{eq:HJB0} and obtain the differential equation \eqref{eq:ODE_Vhat}. Together with the boundary condition that $\hat V(\ph_2)= \Phi(\ph_2)$, \eqref{eq:ODE_Vhat} admits a unique solution:\footnote{The derivation of the solution appears in  \cref{subsec:Solution-ODE}.}  
 \begin{equation}
 	\hat{V}(\ph):=\frac{C}{\left(\frac{1-\ph}{\ph}\right)^{\frac{u_{2}-u_{1}}{\delta_{\ell}}}+C}u_{1}
 	+\frac{\left(\frac{1-\ph}{\ph}\right)^{\frac{u_{2}-u_{1}}{\delta_{\ell}}}}{\left(\frac{1-\ph}{\ph}\right)^{\frac{u_{2}-u_{1}}{\delta_{\ell}}}+C}u_{2},\label{eq:Vhat_general_solution}
 \end{equation}
where
 \begin{align}
 	C & :=\frac{u_{2}-\Phi(\ph_2)}{\Phi(\ph_2)-u_{1}}\left(\frac{1-\ph_2}{\ph_2}\right)^{\frac{u_{2}-u_{1}}
 		{\delta_{\ell}}},\label{eq:Vhat_integration_constant}
 \end{align}
 and $u_1$ and $u_2$ are lower and higher roots of a quadratic equation; namely,
 \begin{equation}
 	u_{1,2}:=\frac{u_{r}^{R}+u_{\ell}^{L}}{2}\pm\sqrt{\left(\frac{u_{r}^{R}-u_{\ell}^{L}}{2}\right)^{2}+\frac{c}{\lambda}\delta_{\ell}}.\label{eq:u_values}
 \end{equation}

 It is instructive to visualize how nature's feasible ``set of values'' evolves on the equilibrium path as $\ph$ drifts down in this region.   Our construction so far indicates that at each state the feasible set forms a flat segment   $\{(p,\hat V(\ph)):p\in [\pl(\ph), \ph] \}$, and the belief at which the segment crosses $\ell$-payoff function $U_{\ell}(p)$ is precisely nature's choice $\pi (\ph)$.  As $\ph$ falls in this region, the segment shifts left. The starting state  of Region 3, or its right-most boundary, $\ph_3$, is the state such that its associated value segment just meets or ``touches'' the $\ell$-payoff function $U_{\ell}(p)$ at the former's  left-most end.  Formally, $\ph_3$ is defined by:
 \begin{equation}
\hat V(\ph_3)= U_{\ell}(\pl(\ph_3)).
\label{eq:pi3}
\end{equation}
 
Let $\hat u:=\min_{p\in[0,1]} U(p)$. The following observations are useful to establish:
  \begin{lemma}  
	\label{lem:properties_region3}
	\begin{description}
		\item [(i)] $\hat{V}(\cdot)$ is strictly decreasing, and $\hat{V}(\ph)\in(u_{1},u_{2})$ for all $\ph>\ph_2$.
		
		\item [(ii)]  If $V^{\sigma}(\pl(\ph_2), \ph_2) < U_{\ell}(\pl(\ph_2))$ (i.e., the condition for Region 3 to exist holds) and $u_1>u^R_{\ell}$, then there exists a unique solution $\ph_3\in (\ph_2, 1).$	
		
		\item [(iii)]  If $c\le\cl$, then $u_1\ge \hat{u}$, so $V^{\sigma}(\cdot, \ph_2) >\hat u$ for all $\ph\ge \ph_2$. If $c\in[\cl, \ch)$, there exists $\Delta_c>0$ such that 
		$V^{\sigma}(\cdot, \ph_3)=\hat V(\ph_3)\ge \hat u$ if and only if $\Delta<\Delta_c$.
	\end{description}
\end{lemma}
 \begin{proof}
 	The proof can be found in   \cref{subsec:Proof-of-Lemma1} below.
 \end{proof}
 \medskip
 
\cref{lem:properties_region3}-(i) implies that the value segment shifts up (as well as shifts left) as $\ph$ drifts down.\footnote{One can also see that $V^{\sigma}$ is continuous, including at the boundary.  This follows from the fact that $\hat V(\ph_2)=\Phi(\ph_2)$,  and that $\ph_2=p_*$ and
	$\Phi'(p_*)=0$.} Next, \cref{lem:properties_region3}-(ii) implies that $\ph_3$ is well defined and above $\ph_2$ whenever the region exists.  
Last, \cref{lem:properties_region3}-(iii) ensures that the value segment remains above $\hat u$ for all $\ph\in (\ph_2, \ph_3]$, if either $c\le \cl$ or if $c\in[\cl, \ch)$ but
$\Delta<\Delta_c$, for some $\Delta_c>0$.  Our verification below will use this fact, assuming the sufficient condition.

So far, our value function $V(p,\ph)$ is constructed from HJB equations along with \cref{eq:worst-case-belief_proof1}. We now claim the value function indeed represents the value of the candidate strategy $\sigma$:  
 \begin{lemma} \label{lem:r3-Vsigma}  
 For all $\ph\in (\ph_2, \ph_3)$, $V(p, \ph)=V^{\sigma}(p, \ph)$.
 \end{lemma}
\begin{proof}  
See   \cref{sec:r3-Vsigma}.
\end{proof}
\medskip

\noindent\emph{Verification of equilibrium conditions.}  To verify \cref{eq:HJB0}--\cref{eq:HJBm},
  note that the definition of $\pi  (\ph)$ in \cref{eq:pi-r3}
  implies that $U_{\ell}(\pi  (\ph))-V^{\sigma}(\pi  (\ph),\ph)=0$. We have already observed from \cref{lem:properties_region3}-(iii) that  $V^{\sigma}(\cdot,\ph)\ge \hat{u}$ for all $\ph\in(\ph_2,\ph_{3}]$.  This further implies that  $\pi  (\ph)\le\hat{p}$ for $\ph\in(\ph_2,\ph_{3}]$. Therefore $\rho(\ph)=0$ is a
  maximizer in \cref{eq:HJB0}--\cref{eq:HJBm}.
  By \cref{eq:ind-DM},
  $\nu(\ph)$ as defined in \cref{eq:nu-r3} is a maximizer as
  well. It remains to show that $m(\ph)=0$ is a maximizer and the RHS
  in \cref{eq:HJB0}--\cref{eq:HJBm}
  is equal to zero, both of which would follow if the coefficient of $(1-m)$ in the
  objective in \cref{eq:HJB0}--\cref{eq:HJBm}
  is equal to zero.  We can use \cref{eq:ind-DM} and \cref{eq:R3-2} to simplify the coefficient of $(1-m)$ to:
  \begin{equation}
  	-c+\pi  (\ph)\lambda(u_{r}^{R}-\hat{V}(\ph))-\lambda \ph (1-\ph)\hat V(\ph),\label{eq:coeff_1-m_region3}
  \end{equation}
  which vanishes precisely because $\hat V$ solves \cref{eq:ODE_Vhat}.
We thus conclude that
  $m(\ph)=0$ is a minimizer in \cref{eq:HJBm}.  We have thus verified \cref{eq:Vhat_general_solution}
  and \cref{eq:HJBm}.
  
  To verify \cref{eq:HJBpi} note that, with $m(\ph)=\rho(\ph)=0$
  and $\nu(\ph)$ defined in \cref{eq:nu-r3}, the derivative of
  the objective in \cref{eq:HJBpi} with respect
  to $p$ simplifies to:
  \begin{align*}
  	\nu(\ph)\left(u_{\ell}^{R}-u_{\ell}^{L}\right)+\lambda(u_{r}^{R}-V(\pi  (\ph),\ph))=\nu(\ph)\left(u_{\ell}^{R}-u_{\ell}^{L}\right)+\lambda\left(u_{r}^{R}-\hat{V}(\ph)\right)=0,
  \end{align*}
 where we have used $V_{p}(\pi  (\ph),\ph)=V_{p\ph}(\pi  (\ph),\ph)=0$.  Hence, we conclude that 
 $\pi(\ph)$ satisfies \cref{eq:HJBpi}.  It also satisfies \cref{eq:worst-case-belief_proof1} since $V^{\sigma}(p, \ph)$ is constant in $p$.
 
 \subsubsection*{Region 4: $\protect\ph\in ( \ph_{3}, \ph_4]$.}

 Here, $\ph_4$ will be defined as part of the analysis. \\

 \noindent\emph{Computation of equilibrium value.} Recall that the strategy $\sigma$ calls on the DM to experiment fully at each state $\ph>\ph_3$ until state $\ph_3$ is reached.   To compute the associated value $V^{\sigma}(p,\ph)$, it is useful to consider a (hypothetical) Bayesian DM who would find such a strategy optimal.  To this end, we represent the DM's problem  as a   stopping problem where the stopping payoff is the continuation value $v_{**}:=\hat V(\ph_3)$:\footnote{Recall that if Region 3 does not exist, then $\ph_3=\ph_2=p_*$, so 
 $\hat V(\ph_3)=\hat V(\ph_2)=\Phi(\ph_2)$, as defined before.  All subsequent results hold since $V_p(p,\ph_2)=\Phi'(\ph_2)=\Phi'(p_*)=0$.}
 
\begin{quote} \emph{Auxiliary Problem:}  Imagine a hypothetical Bayesian DM with any belief $p\ge \pl(\ph_3)$  who at each instant may experiment or stops.  She may experiment until either breakthrough occurs or her belief reaches $\pl(\ph_3)$. If she stops at any point, she collects the payoff
	$\hat V(\ph_3)$ (independent of $p$).   The optimal value of this stopping problem is:
\begin{equation} \label{eq:b-value}
	\Psi(p)=\max_{p'\in [\pl (\ph_3),p]}\phi  (p; p', v_{**} ).
\end{equation}
Let $p_{**}$ denote the optimal stopping belief.   We note that $\Psi(p_{**})=v_{**}$ and
$ \Psi'(p_{**})\ge 0$, with equality holding whenever $p_{**}>\pl (\ph_3)$ (a consequence of smooth pasting).  We can easily see that $p_{**}\in [\pl(\ph_3), \ph_3)$.\footnote{To see that $p_{**}<\ph_3$, it suffices to show that $p_{**}<p_*$ since $p_*\le\ph_3$.  To show     $p_{**}<p_*$, suppose otherwise.  Then,
	$$\lambda p_{**}(1-p_{**})\Psi'(p_{**}) = \lambda p_{**} (u^R_r-\Psi(p_{**}))- c= \lambda p_{**} (u^R_r-v_{**})- c
	> \lambda p_* (u^R_r-\Phi(p_*))- c  =\Psi'(p_*)\ge 0,$$
	where the first and second last equalities follow from the fact that both $\Psi$ and $\Phi$ solve \cref{eq:Bayesian-ODE}, the strict inequality from $v_{**}<\Phi(p)$ for all $p$, and the last inequality holds since $\ph_3\ge p_*$.} 
	\end{quote}

 Fix any state $\ph>\ph_3$.  We ask:  \emph{for what belief $p$ is the continuation strategy $\sigma_{\ph}$  optimal over Region 3}? Recall $\sigma_{\ph}$ prescribes:  ``experiment for the duration of $\tau(\ph, \ph_3)$ and, absent breakthrough by the end of the experimentation,  stop and collect   $\hat V(\ph_3)$,''  where $\tau(p,p')$ denotes the time it takes for a belief to drift from $p$ to $p'$.  Since, by definition, $p_{**}$ is the optimal stopping belief, the answer to the above question is precisely the belief, $q(\ph)$, such that
 \begin{equation}\label{eq:q}
     \tau(\ph, \ph_3)= \tau(q(\ph), p_{**}).
 \end{equation}
 In words, it is the belief that would be updated to $p_{**}$ after  the prescribed duration $\tau(\ph, \ph_3)$ of experimentation.\footnote{Obviously, $q(\ph_3)=p_{**}$.}  Since $p_{**}$ is the optimal stopping belief in the above Auxiliary Problem, a DM with belief $q(\ph)$ finds the prescribed strategy  optimal at state $\ph$ and thus will realize the value of $\Psi(q(\ph))$.  Consequently,
 $$V^{\sigma}(q(\ph), \ph)=\Psi(q(\ph)).$$
 What about a DM with belief $p\ne q(\ph)$?  Her value is weakly below $\Psi(q(p))$   and as noted before linear in $p$.  This pins down the value function for belief $p$ at state $\ph$:  
 \begin{equation}
 	V^{\sigma}(p,\ph):=\Psi(q(\ph))+(p-q(\ph))\Psi'(q(\ph)).  	\label{eq:V_region4}
 \end{equation}
 We define $\ph_4$ to be such that 
 \begin{equation}
 	V^{\sigma}(\pl(\ph_4),\ph_4)= U_{r}(\pl(\ph_4)).
 	\label{eq:pi4}
 \end{equation}
 
\noindent\emph{Verification of equilibrium conditions.} We first claim that $\pi (\ph)=\pl(\ph)$ is a worst-case belief satisfying \cref{eq:worst-case-belief_proof1}.  This follows from the fact, for any $p\in \mathcal{P}(\ph), \ph\ge \ph_3$,
 $$V_{p}^{\sigma}(p,\ph)=\Psi'(q(p))\ge 0,$$ where the equality follows from \cref{eq:V_region4} and the inequality from the convexity of $\Psi(\cdot)$ and $\Psi'(p_{**})\ge 0$. 
 
 We next verify \cref{eq:HJBm}.   First, we show that $\nu(\ph)=0$. To this end, we first observe:
\begin{lemma} \label{lem:uell-less} For all $\ph>\ph_3$, $V^{\sigma}(\pl(\ph),\ph)\ge U_{\ell}(\pl(\ph))$.
\end{lemma} 	
	
\begin{proof}  Note $V^{\sigma}(\pl(\ph_3),\ph_3)=\hat V(\ph_3)\ge U_{\ell}(\pl(\ph_3))$.\footnote{If $V^{\sigma}(\pl(\ph_2),\ph_2)< U_{\ell}(\pl(\ph_2))$, then the inequality holds with equality by definition. If $V^{\sigma}(\pl(\ph_2),\ph_2)\ge    U_{\ell}(\pl(\ph_2))$, then the inequality follows from the fact that $\ph_3=\ph_2$.}  Next, note that $\pl(\ph_3)>\underline p^B_{r}$ and $V^{\sigma}(\pl(\ph_3),\ph_3)=\phi(\pl(\ph_3); \pl(\ph_3), U_{\ell}(\pl(\ph_3)))$.  Since, by \cref{lem:SCP}, $\phi(p; p, U_{\ell}(p))$ can only cross  $U_{\ell}(p)$ from below, we have  
$$
\left.\frac{d V^{\sigma}(\pl(\ph),\ph)}{d\ph}\right |_{\ph=\ph_3}\ge U_{\ell}'(\pl(\ph_3)).
$$
Observe further $V^{\sigma}(\pl(\ph),\ph)= \phi(\pl(\ph); \pl(\ph_3), U_{\ell}(\pl(\ph_3)))$ is convex in $\pl(\ph)$ whereas $U_{\ell}(\pl(\ph))$ is linear in $\pl(\ph)$.  Combining the two facts leads to the desired conclusion.
 \end{proof}
 \medskip	
 
 Next, we observe that $U_{\ell}(\pl(\ph_3))=\hat V(\ph_3)>\hat u$. This means that 
  $$V^{\sigma}(\pl(\ph_3),\ph_3)= U_{\ell}(\pl(\ph_3))>U_{r}(\pl(\ph_3)).$$
  By definition of $\ph_4$, we have $V^{\sigma}(\pl(\ph),\ph)>U_{r}(\pl(\ph))$
 for all $\ph\in(\ph_{3},\ph_{4})$.  Combining this with \cref{lem:uell-less}, we conclude that $\nu(\ph)=0$.
 
 Next, we prove $m(\ph)=0$.   Substituting $\nu(\ph)=0$ and $\pi(\ph)=\pl(\ph)$,
 the objective in \cref{eq:HJBm} becomes  
 \[
 m\left[U_{\rho}(\pl)-V(\pl,\ph_{-})\right]+(1-m)\left[-c+ \lambda\pl (u_{r}^{R}-V(\pl,\ph_{-}))+V_{p}(\pl,\ph_{-})\eta (p)+V_{\ph}(\pl,\ph_{-})\eta (\ph)\right],
 \]
 where we suppress the argument of  $\pl(\ph)$ for notational ease.   Note that the coefficient of $m$ is negative (by the same argument
 as for $\nu=0$).  The coefficient of $(1-m)$ can be written
 \begin{align*}
& -c+\lambda \pl(u_{r}^{R}-V(\pl,\ph_{-}))+V_{p}(\pl,\ph_{-})\eta (\pl)+V_{\ph}(\pl,\ph_{-})\eta (\ph)\\
=& -c+ \lambda \pl\left(u_{r}^{R}-\Psi(q(\ph))-(\pl-q(\ph))\Psi^{\prime}(q(\ph))\right)\\
& \quad+\eta (\pl)\Psi^{\prime}(q(\ph))+\eta (\ph)\left[(\pl-q(\ph))\Psi^{\prime\prime}(q(\ph))q'(\ph)\right]\\
=& -c+ \lambda q(\ph)\left(u_{r}^{R}-\Psi(q(\ph))\right) + \eta (q(\ph))\Psi^{\prime}(q(\ph))\\
& \quad+\lambda(\pl-q(\ph))\left[u_{r}^{R}-\Psi(q(\ph)) -(1-q(\ph))\Psi^{\prime}(q(\ph))-q(\ph)(1-q(\ph))\Psi^{\prime\prime}(q(\ph))  \right]\\
=&0,
 \end{align*}
where we have used \cref{eq:V_region4} for the first equality and  $q'(\ph)=\eta (q(\ph))/\eta (\ph)$ for the second equality.\footnote{This holds since $\tau(q(\ph),q(\ph_{3}))=\tau(\ph,\ph_{3})$ implies that
	$\ln(q(\ph)/(1-q(\ph)))-\ln(p_*/(1- p_*))=\ln(\ph/(1-\ph))-\ln(\ph_{3}/(1-\ph_{3}))$.
	Hence the difference in log-likelihood ratios of $q(\ph)$ and $\ph$
	is constant 
	\[
	\ln\frac{q(\ph)}{1-q(\ph)}-\ln\frac{\ph}{1-\ph}=K,
	\]
	for some $K$.  
	Differentiating this with respect to $\ph$, we obtain $q'(\ph)=\eta (q(\ph))/\eta (\ph)$.} The last equality follows since $\Psi$ satisfies \cref{eq:Bayesian-ODE} and its derivative vanishes.
 We have shown that the coefficient of $m$ in the objective in \cref{eq:HJB0}
 and \cref{eq:HJBm} is negative and the coefficient
 of $(1-m)$ is zero. Therefore, $m=0$ is a maximizer and
 \cref{eq:HJB0} holds.
 
 It remains to verify \cref{eq:HJBpi}. Substituting
 $m(\ph)=\nu(\ph)=0$, the objective becomes the same as the coefficient of $1-m$ above, except that $\pl$ is replaced by $p$. Differentiating the expression with respect to $p$ yields: 
 \[
 \lambda\left(u_{r}^{R}-\Psi(q(\ph))-(1-q(\ph))\Psi^{\prime}(q(\ph))\right)-\lambda q(\ph)(1-q(\ph))\Psi^{\prime\prime}(q(\ph))=0,
 \]
 where the equality follows from the fact that $\Psi$ satisfies \cref{eq:Bayesian-ODE}, so its derivative, which coincides with the LHS of the above equation, must vanish.
 
We have thus shown that the objective in \cref{eq:HJBpi} is independent of $p$ and hence the requirement that $p=\pl(\ph)$ is a minimizer in
 \cref{eq:HJBpi} is satisfied.

\subsubsection*{Region 5: $\protect\ph\in[\ph_4,1]$.}

 \noindent\emph{Computation of equilibrium value.}
 Recall that the strategy $\sigma$ calls for an immediate choice of $r$ for all states $\ph\in[\ph_4,1]$.  It thus immediately follows that the value associated with that strategy is:
  $$
  V^{\sigma}(p,\ph):=U_{r}(p), \mbox{ for all  } \ph\in [ \ph_4,1].
  $$

 \noindent\emph{Verification of equilibrium conditions.}  Since $U_r(\cdot)$ is increasing, $\pi(\ph)=\pl(\ph)$ satisfies \cref{eq:worst-case-belief_proof1}; given $m(\ph)=\rho(\ph)=1$, the coefficient of $p$ in the objective of \cref{eq:HJBpi} vanishes, so $\pi(\ph)=\pl(\ph)$  is also a minimizer in \cref{eq:HJBpi}.

 Substituting $\pi  (\ph)=\pl(\ph)=\pl$ and $V^{\sigma}(p,\ph)=U_{r}(\ph)$
 in \cref{eq:HJB0}--\cref{eq:HJBm}
 we get the following expression for $\ph>\ph_{4}$
 \begin{align*}
 & m\left[U_{\rho}(\pl)-U_{r}(\pl)\right]+(1-m)\left[-c+\lambda\pl\left(u_{r}^{R}-U_{r}(\pl)\right)-\lambda\pl(1-\pl)U'_{r}(\pl)\right]\\
 = & m\left[U_{\rho}(\pl)-U_{r}(\pl)\right]+(1-m)\left[-c+\lambda\pl\left(U_{r}(1)-U_{r}(\pl)-(1-\pl)U'_{r}(\pl)\right)\right]\\
 = & m\left[U_{\rho}(\pl)-U_{r}(\pl)\right]+(1-m)(-c),
 \end{align*}
 where the last line follows since $U_{r}(\ph)$ is linear.  Recall that 
 $U_{\ell}(\pl(\ph_4))\le V^{\sigma}(\pl(\ph_4),\ph_4)=U_r(\pl(\ph_4))$, so 
 we note that $U_{\ell}(\pl)\le U_{r}(\pl)$ for $\pl\le \ph_4$.  
Therefore,  $m(\ph)=\rho(\ph)=1$ satisfies \cref{eq:HJBm}.  Substituting these, we also have \cref{eq:HJB0}.

\subsubsection{Case 1 with $c\ge \cl$  and $\Delta>\Delta_c$}   \label{ssec:large-c}
 
 In this case, Region 4, as well as its boundaries $\ph_3$ and $\ph_4$, need to be modified. 
 \cref{thm:main_result}  specifies
 \begin{equation}
 (m(\ph),\nu(\ph),\rho(\ph))=\mbox{$\left(1,0,\frac{\delta_{\ell}}{\delta_{r}+\delta_{\ell}}\right)$},\label{eq:sigma_region4-1}
 \end{equation}
 and $\pi  (\ph)=\hat{p},$ for $\ph\in[\ph_{3},\ph_{4})$, where $\ph_3$ is now set at $\ph_3'$, which satisfies $\hat V(\ph_3')=\hat u.$ 
 We note that this new $\ph_3'$ is smaller than the original $\ph_3$ and is still larger than $\ph_2$.\footnote{Recall $\hat V(\cdot)$ is strictly decreasing.   Since $\hat V(\ph_3)<\hat u$ when $c\in (\cl, \ch)$ and $\Delta>\Delta_c$, we have $\ph_3'<\ph_3$. We also have $\ph_3'>\ph_2$ since $\hat V(\ph_2)=\Phi(\ph_2)>\max\{U_{\ell}(\ph_2),U_r(\ph_2)\}\ge\hat u$.}  
 $\ph_4$ is now set at 
 $\ph_4'$, which uniquely satisfies $\pl(\ph_4')=\hat p.$ One can see that $\ph_4'>\ph_3'$. Given $\sigma(\ph)$ for this region, the value of the strategy is given by $V^{\sigma}(p,\ph)=\hat{u}$.
 
It is straightforward to verify that $V^{\sigma}(p,\ph)=\hat{u}$
together with $\pi  (\ph)=\hat{p}$ and \cref{eq:sigma_region4-1}
satisfy \cref{eq:worst-case-belief_proof1}--\cref{eq:HJBpi} for $\ph\in(\ph_{3},\ph_{4})$.
Indeed, inserting $\pi  (\ph)=\hat{p}$ and $V^{\sigma}(p,\ph)=\hat{u}$
as well as $V_{p}^{\sigma}(p,\ph)=0$ and $V_{\ph}^{\sigma}(p,\ph)=0$,
the objective in \cref{eq:HJB0} becomes $-(1-m)c$.
Hence $m=1$ is optimal and \cref{eq:HJB0} holds.
Since the objective is independent of $\nu$ and $\rho$, \cref{eq:HJBm}
also holds. At $\ph=\ph_{3}$, we have $V^{\sigma}(p,\ph_{-})=\hat{V}^{\sigma}(\ph_{3})=\hat{u}$
and $V_{p}^{\sigma}(p,\ph_{-})=0$, and $V_{\ph}^{\sigma}(p,\ph_{-})<0$,
but we can employ the same argument as in Region 3 to show that the
coefficients of $m$ and $(1-m)$, in the objective is zero and hence
$m=1$ is optimal. 

Finally, we verify \cref{eq:HJBpi}. Inserting
\cref{eq:sigma_region4-1}, and $V^{\sigma}(p,\ph_{-})=V^{\sigma}(p,\ph)=\hat{u}$
in the objective in \cref{eq:HJBpi}, we see that
the objective is equal to zero and hence $\pi  (\ph)=\hat{p}$
is a minimizer. Finally, $V_{p}^{\sigma}(p,\ph)=0$ implies that \cref{eq:worst-case-belief_proof1}
holds.

\newpage

\section{Online Appendix}

\subsection{Bayesian optimal behavior} \label{app:Bayesian}

\begin{proposition}\label{prop:Bayesian} The Bayesian DM takes action $\ell$ if $p_t\leq p^B_{\ell}$, takes action $r$ if $p_t\geq p^B_{r}$, and experiments if $p_t\in(p^B_{\ell},p^B_{r})$, where the threshold $p^B_{\ell}>0$ maximizes the value \cref{eq:Bayesian-Stopping Value} and $p^B_{r}\in [p^B_{\ell},1)$ equalizes the two terms in \cref{eq:Bayesian-Value}.
The experimentation region is non-empty, i.e. $p^B_{\ell}<p^B_{r}$, if $c<\overline{c}$,
where $\overline c$ is defined in \cref{eq:c-bar}.
\end{proposition}

\begin{proof}
In order to describe the Bayesian optimal stopping rule, consider the DM's value from experimentation under the assumption that she takes action $r$ if a breakthrough occurs: 
\[
\phi(p)= \lambda pu_r^Rdt+(1-\lambda p dt)\phi(p+dp)-cdt +o(dt).
\]
Taking the limit $dt\to0$, we obtain the following ODE
\begin{align}\label{eq:BayesianODE}
c=\lambda p(u_r^R-\phi(p))+\phi'(p)\eta(p).
\end{align}
Let $\phi(p;q,\Phi_0)$ denote the solution with boundary value $(p',\Phi_0)$, given by
\[
\phi(p\,;p',\Phi_0)=\frac{p-p'}{1-p'}u_{r}^{R}+\frac{1-p}{1-p'}\Phi_0-\left[\frac{p-p'}{1-p'}+(1-p)\log\left(\frac{p}{1-p}\frac{1-p'}{p'}\right)\right]\frac{c}{\lambda}.
\]
Conditional on taking action $\ell$ after stopping, the value of this stopping problem for a Bayesian DM is
\[
\Phi(p):=\max_{p'\in[0,p]}\phi(p;p',U_{\ell}(p')).
\]

We know that at $p=p_{\ell}^B$, we have $\phi'(p;p,U_{\ell}(p))=U'_{\ell}(p)=\delta_{\ell}$. Solving this condition for $p$ delivers
\begin{align*}
p^B_{\ell}&=\frac{1}{u_{r}^{R}-u_{\ell}^{R}}\frac{c}{\lambda}.
\end{align*}
The DM optimally follows the derived stopping rule as long as it delivers a higher payoff than taking action $r$. The right boundary $p^B_{r}$ is thus determined by the indifference condition $\Phi(p^B_{r})=U_r(p^B_{r})$. The Bayesian value function is
\[
\Phi^*(p)=\max\{\Phi(p),U_r(p)\}.
\]
Experimentation is optimal for some $p\in(0,1)$ if the left boundary $p^B_{\ell}$ is smaller than the belief at which the stopping payoffs $U_{\ell}(\cdot)$ and $U_r(\cdot)$ cross, given by
\[
\hat p=\frac{u_{\ell}^L-u_r^L}{\delta_{\ell}+\delta_r}.
\]
Substituting for $p^B_{\ell}$ and $\hat p$ and solving for $c$, the inequality $p_{\ell}^B<\hat p$ can be written as
\[
c<\overline{c}:=\lambda\frac{(u_{r}^{R}-u_{\ell}^{R})(u_{\ell}^{L}-u_{r}^{L})}{(u_{r}^{R}-u_{\ell}^{R})+(u_{\ell}^{L}-u_{r}^{L})}.
\]
\end{proof}

\subsection{Maxmin Commitment Solution}\label{app:commitment}

\begin{proposition}\label{prop:commitment} The maxmin value for the DM with a commitment ability is 
$$\min_{p\in \mathcal{P}_0}\Phi^*(p).$$
\begin{itemize}
\item If $p_*< p^B_r$ or $p^B_r\not\in \mathcal{P}_0$, the DM's maxmin strategy coincides with her Bayesian optimal strategy, as described in \cref{prop:Bayesian}, for prior belief $p_{min}\in \arg\min_{p\in \mathcal{P}_0}\Phi^*(p)$.
\item If $p_*= p^B_r\in \mathcal{P}_0$, then the DM randomizes as follows:
\begin{itemize}
\item  $c<\overline c$: the DM randomizes between action $r$ and acquiring information using the Bayesian optimal stopping rule for $p_r^B$ with probabilities $\xi$ and $1-\xi$, respectively, where $\xi$ is such that
\begin{eqnarray}\label{eq:xi}
\xi U_r' + (1-\xi) \Phi'(p^B_r)=0;
\end{eqnarray}
\item $c\geq\overline c$: the DM randomizes between actions $r$ and $\ell$ with probabilities $\hat\rho$ and $1-\hat\rho$, respectively, where $\hat\rho$ is specified by \eqref{eq:rho-hat}.
\end{itemize}
\end{itemize}
\end{proposition}

\begin{proof}
We consider a zero-sum game between the DM and adversarial nature seeking to minimize the DM’s expected payoff. In this game, the DM chooses a distribution over stopping times and a distribution over $\{\ell,r\}$ conditional on stopping, while nature chooses $p\in \mathcal{P}_0$. We want to show that the DM's strategy described in \cref{prop:commitment}, together with nature choosing $p_{min}\in\min_{p\in \mathcal{P}_0}\Phi^*(p)$, forms a saddle point.  As is well-known, if a saddle point exists, then the DM's strategy in the saddle point is maxmin-optimal.\footnote{ \label{fn:minmax} To see this, let $u(s_1, s_2)$ be player 1's payoff when she chooses $s_1\in S_1$ in a zero-sum game against player 2 who chooses $s_2$, where $S_1$ and $S_2$ are arbitrary (nonempty) sets.  Suppose there exists a saddle point $(s_1^*, s_2^*)$ such that 
  \begin{align*}
      u(s_1^*, s_2)\ge u(s_1^*, s_2^*)\ge u(s_1, s_2^*), \forall (s_1,s_2)\in S_1\times S_2.  \tag{*}\label{*}
  \end{align*} 
  Then,  
  \begin{align*}
      \inf_{s_2\in S_2} u(s_1^*, s_2)   \ge  u(s_1^*, s_2^*)  = \sup_{s_1\in S_1} u(s_1, s_2^*)  \ge  \inf_{s_2\in S_2} \sup_{s_1\in S_1} u(s_1, s_2) \ge  \sup_{s_1\in S_1} \inf_{s_2\in S_2}  u(s_1, s_2),
\end{align*}
  where the first inequality follows from the first inequality of \cref{*} and the last inequality follows from the min-max inequality.
 The above inequalities prove that $$s_1^*\in \arg\max_{s_1\in S_1}\inf_{s_2\in S_2}  u(s_1, s_2).$$
 Symmetrically, one can show that $s_2^*\in \arg\min_{s_2\in S_2}\sup_{s_1\in S_1}  u(s_1, s_2).$ Conversely, it can be proven that if $s_1^*$ is maxmin optimal and $s_2^*$ is minmax optimal, then $(s_1^*, s_2^*)$ is a saddle point.}  

Since the DM's strategy is Bayesian optimal against nature's choice $p_{min}$,\footnote{Note that the Bayesian DM with belief $p_r^B$ is indifferent between taking action $r$ and taking action $\ell$ if $c\ge \ch$ or acquiring information using the Bayesian optimal stopping rule for $p_r^B$ if $c<\ch$.} it suffices to show is that $p_{min}$ minimizes the DM's expected payoff from the strategy described in \cref{prop:commitment}.

To this end, we first examine the value $V_0(p)$ of the DM's strategy specified in the Proposition as a function of 
 nature's choice $p\in[0,1]$: 
 \[
V_0(p)=
\begin{cases}
\Phi^*(p_{min})+(p-p_{min}){\Phi^*}'(p_{min}) & \text{if }p_{min}\neq p_r^B\\
U_r(p) & \text{if } p_{min}=p_r^B>p_*\\
\xi U_r(p) + (1-\xi) (\Phi(p_*)+(p-p^*)\Phi^{'}(p_*)) & \text{if }p_{min}=p_r^B=p_{*}\text{ and }c<\overline c\\
\hat\rho U_r(p)+(1-\hat\rho) U_{\ell}(p) & \text{if }p_{min}=p_r^B=p_{*} \text{ and }c\ge\overline c.
\end{cases}
\]
 This function is obtained by means of several observations.  First, 
note that $V_0(p)$ is a convex combination of the expected payoff from the DM's strategy in state $R$ and in state $L$ with weight $p$. Hence, $V_0(p)$ is linear in $p$. 
Second,   the value of any strategy, including the one in question, is no higher than that of the Bayesian optimal strategy. Hence, we have $V_0(p)\le \Phi^*(p)$ for all $p\in[0,1]$.
Third, recall that the DM's strategy is Bayesian optimal for the minimizing belief $p_{min}$, so $V_0(p_{min})= \Phi^*(p_{min})$.  These three observations imply that $V_0(\cdot)$ is  tangent to $\Phi^*(\cdot)$ at $(p_{min},\Phi^*(p_{min}))$. Recalling that $\Phi^*$ is differentiable everywhere except possibly  at $p_r^B$ pins down the above value function.

We are now in a position to prove that $p_{\min} \in \arg\min_{p\in \mathcal{P}_0} V_0(p)$.   Note first that since $\Phi^*(\cdot)$ is a convex function, we have $p_{min}=\max \mathcal{P}_0$ if $\max \mathcal{P}_0<p_*$, $p_{min}=\min \mathcal{P}_0$ if $p_*<\min \mathcal{P}_0$, and $p_{min}=p_*$ otherwise.  Hence, it suffices to show that 
$V_0'(\cdot)\leq0$ if $p_{min}=\max \mathcal{P}_0<p_*$, $V_0'(\cdot)\geq 0$ if $p_{min}=\min \mathcal{P}_0>p_*$ and $V_0'(\cdot)=0$ if $p_{min}=p_*$.   

Suppose first $p_{min}\neq p_r^B$. In this case, $V_0'(\cdot)={\Phi^*}'(p_{min})$, which is negative if $p_{min}<p_*$, positive if $p_{min}>p_*$, and zero if $p_{min}=p_*$, as desired.  Suppose next $p_{min}=p_r^B>p_{*}$. Then, $V_0'(\cdot)=U_r'(\cdot)>0$, so nature's optimality condition is again satisfied.
Finally, if $p_{min}=p_r^B=p_*$, then, by definition of $\xi$ and $\hat\rho$, $V_0(p)$ is constant in $p$. Hence, nature is indifferent between all values of $p\in \mathcal{P}_0$.

\end{proof}

\subsection{Stopping-time formulation of strategies} \label{sec:stopping time}

The strategy the DM chooses can be formulated as a stopping time adapted to the filtration generated by the underlying state and the Poisson signal.  In the current setting, the DM faces a nontrivial decision only when no information is revealed; once breakthrough news is received, the DM stops immediately and chooses $r$.   Taking this as given, we can   describe the DM's strategy simply by the \emph{contingent stopping times}---a family of CDFs $\{F_t\}_{t\in [0,
  \infty)}$, where $F_t: [0, \infty]\to [0,1]$, for each $t\ge 0$,\footnote{We let the stopping time be $\infty$ if the DM never stops.}  and $F_t(\tau)$ denotes the probability of stopping by time $t+\tau$ starting at time $t$, conditional on the breakthrough news not being received by then.  The reason we consider a family of stopping times instead of a single stopping time is that the strategy must specify actions off (as well as on) the path.  For example, the original strategy may prescribe stopping by time $t$ with probability one, but if the DM deviates from the original plan and has not stopped by $t$, the strategy must prescribe her stropping plan from then on. 

We require a couple of  technical assumptions: 

\begin{enumerate}
    \item [(1)] (Admissibility) $F_t$ is nondecreasing and right-differentiable.  
    \item [(2)] (Consistency) For any $s<s'\le t$, 
    $$F_{s'}(t-s')= 
    \frac{F_s(t-s) -F_s(s'-s)}{(1-F_s(s'-s))},$$ 
    whenever $F_s(s'-s)<1$.
\end{enumerate}

Admissibility requires the distribution functions to be sufficiently well-behaved.  Note also that a monotonic function has countably many points of discontinuities, which will be invoked later. Consistency means that the stopping time distribution starting at $s'$ must form a consistent conditional stopping distribution starting at $s<s'$ according to Bayes formula, as long as the latter distribution does not prescribe stopping before $s$ with probability one.  Let  $\mathcal{F}$ be the set of all families of distributions satisfying these two requirements.  Let $\tilde\rho:[0,\infty)\to [0,1]$ be the choice strategy specifying the probability of choosing $r$ when stopping at time $t$.  Then, the strategy for the DM consists of a pair $(\{F_t\},\tilde\rho)$ such that $\{F_t\}\in \mathcal F$.

As an intermediate step, we next define the following notion of strategy. 
A \emph{time-indexed strategy} is a function $\tilde \sigma:[0,\infty)\rightarrow[0,\infty)\times[0,1]\times[0,1]$,
$\tilde \sigma_{t}=(\tilde \nu_{t},\tilde  m_{t},\tilde \rho_{t})$, where $\tilde \nu_{t}$ is the stopping
rate and $\tilde m_{t}$ is the instantaneous stopping probability used at time
$t$ if no $R$-signal has been received up until time $t$. $\rho_{t}$
is the probability of action $r$ conditional on stopping at time
$t$ without receiving an $R$-signal.  We require the following {\it admissibility} conditions on $\tilde \sigma$:
\begin{enumerate}
    \item [(a)] any connected set of $t$'s in which $m_t=1$  contains its infimum,
    \item [(b)] the set   $\{t\ge 0|m_{t}\in(0,1)\}$ is countable.
\end{enumerate}
 Condition (a) follows from the right continuity of $F_t$'s, which in turn follows from its right differentiability (assumed in (1)).  Condition (b) follows from the fact that each $F_t$ has countably many points of discontinuities. 
Let $\widetilde\Sigma$ denote the set of all admissible time-indexed strategies.  We now prove the equivalence between the two notions of strategies.

\begin{lemma}  Each $(\{F_t\},\tilde\rho), \{F_t\}\in \mathcal F$ induces an outcome-equivalent admissible strategy $\tilde\sigma\in \widetilde\Sigma$. Conversely, any admissible strategy $\tilde\sigma\in \widetilde\Sigma$ induces an outcome-equivalent $(\{F_t\},\rho), \{F_t\}\in \mathcal F$.
\end{lemma}

\begin{proof} For both statements,  $\tilde\rho$ is fixed to be the same, so we can focus on the specification of stopping times. 

Fix any $\{F_t\}\in \mathcal F$.  For each $t\ge 0$, we define:
$$\tilde \nu_t:=  F_t'(0),$$
where $F_t'(\cdot)$ is the right-derivative, which is well-defined by Admissibility (requirement (1) above), and 
$$\tilde m_t:=  F_t(0).$$ 
Then, $(\tilde\nu_t, \tilde m_t)_{t\ge 0}$  is outcome-equivalent to  $\{F_t\}_{t\in [0,
  \infty)}$.  To see this, fix any $s<t$. If $1-F_s(t-s)<1$, then by  Consistency (requirement (2) above), 
$$\frac{F_s'(t-s)}{1-F_s(t-s)}=F_t'(0)=\tilde \nu_t$$ 
and 
$$\frac{F_s(t-s)-F_s^{-}(t-s)}{1-F_s(t-s)}=F_t(0)=\tilde m_t,$$  where $F_s^-$ is the left limit, which is well defined.  
Hence, for any $s<t$,
\begin{align*}
& \Pr\{ \mbox{stopping by } t \mbox{ under } \tilde \sigma|\mbox{ starting at }s\}   \\
= & 1-e^{-\int_{s}^{t}\tilde\nu_{s'}ds'}\prod_{s'\in [s,t]}(1-\tilde m_{s'})\\
=&F_s(t-s).
\end{align*} 
Finally, the right-continuity and monotonicity of each $F_s(\cdot)$ imply that the admissibility of $\tilde \sigma$ is satisfied. In particular, $\tilde m_t>0$ for countably many $t$.  We thus conclude $(\tilde \nu, \tilde m, \tilde \rho)\in \widetilde \Sigma$.

Conversely, fix any   $\tilde\sigma\in  \widetilde \Sigma$.  Then, for any $s\le t$, define
 $$F_s(t-s):=1-e^{-\int_{s}^{t}\tilde\nu_{s'}ds'}\prod_{s'\in [s,t]}(1-\tilde m_{s'}).$$
 Again, the admissibility implies that $\{F_t\}$ satisfies (1).  Requirement (2) follows from the construction.  Hence, $\{F_t\}\in \mathcal F$.
\end{proof}

\subsection{Proof of \cref{lem:refinement}}\label{proof:refinement}

Since we are considering the left limit of the state $\ph$, a possible mass point at state $\ph$ does not appear in $W^{\sigma}_{\e}(p,\ph_-)$.  Specifically, we can write
  \begin{align*}
W_{\e}(\p_{\e},\ph_-)= & W_{\e}^{\s_{\e}}(\p_{\e},\ph_-)\\
=&- \p_{\e} \int_{\ph^{\e}}^{\ph}e^{-\int_{0}^{\tau(\ph,\ph')}\left(\lambda+\nu_{\e}(p^{\tau})\right)d\tau}\left(-c+\lambda u_{r}^{R}+\nu_{\e}(\ph')u_{\ell}^{R}\right)\frac{1}{\eta(\ph')}d\ph' \\
 & -(1-\p_{\e}) \int_{\ph^{\e}}^{\ph}e^{-\int_{0}^{\tau(\ph,\ph')}\nu_{\e}(p^{\tau})d\tau}\left(-c+\nu_{\e}(\ph')u_{\ell}^{L}\right)\frac{1}{\eta(\ph')}d\ph' \\
 & +\left(\p_{\e}e^{-\int_{0}^{\tau(\ph,\ph^{\e})}\left(\lambda+\nu_{\e}(p^{\tau})\right)d\tau}+(1-\p_{\e})e^{-\int_{0}^{\tau(\ph,\ph^{\e})}\nu_{\e}(p^{\tau})d\tau}\right)V^{\sigma^*}(\p^{\e}_{\e},\ph^{\e}).
\end{align*}
 
Note as $\e\to 0$, $\ph^{\e}\to \ph$ and $\tau(\ph,\ph^{\e})\to 0$.  Hence, as $\e\to 0$, $W_{\e}(\p_{\e},\ph_{-}) \to V^{\sigma^*}(\p^{*},\ph_{-})$.  
    
Next, consider  
 \begin{align*}
   \frac{\partial W_{\e}(\p_{\e},\ph_-)}{\partial p}= & -   \int_{\ph^{\e}}^{\ph}e^{-\int_{0}^{\tau(\ph,\ph')}\left(\lambda+\nu_{\e}(p^{\tau})\right)d\tau}\left(-c+\lambda u_{r}^{R}+\nu_{\e}(\ph')u_{\ell}^{R}\right)\frac{1}{\eta(\ph')}d\ph' \\
 &  + \int_{\ph^{\e}}^{\ph}e^{-\int_{0}^{\tau(\ph,\ph')}\nu_{\e}(p^{\tau})d\tau}\left(-c+\nu_{\e}(\ph')u_{\ell}^{L}\right)\frac{1}{\eta(\ph')}d\ph' \\
 &+\left(e^{-\int_{0}^{\tau(\ph,\ph^{\e})}\left(\lambda+\nu_{\e}(p^{\tau})\right)d\tau}-e^{-\int_{0}^{\tau(\ph,\ph^{\e})}\nu_{\e}(p^{\tau})d\tau}\right) V^{\sigma^*}(\p^{\e},\ph^{\e})\\
 & +\left(\p_{\e}e^{-\int_{0}^{\tau(\ph,\ph^{\e})}\left(\lambda+\nu_{\e}(p^{\tau})\right)d\tau}+(1-\p_{\e})e^{-\int_{0}^{\tau(\ph,\ph^{\e})}\nu_{\e}(p^{\tau})d\tau}\right) V^{\sigma^*}_p(\p^{\e},\ph^{\e})
\frac{\eta(\p_{\e}^{\e})}{\eta(\p^{\e})}\\
& \to V^{\sigma^*}_p(\p^*,\ph_-) \mbox{ as } \e\to 0,
\end{align*}
since $\p_{\e}\to \p^*, \ph^{\e}\to \ph$, and $\tau(\ph,\ph^{\e})\to 0$, as $\e\to 0$, provided that $V^{\sigma^*}_p(\cdot,\cdot)$ is continuous at $(\p^*,\ph_-)$.  

Finally, consider
 \begin{align*}
  &  \frac{\partial W_{\e}(\pi_{\e},\ph_-)}{\partial \ph}
  \\
  = &   -c+\nu_{\e}(U_{\rho}(\pi_{\e})-W^{\s_{\e}}_{\e}(\pi_{\e},\ph)) + \pi_{\e} \lambda(u_r^R-W^{\s_{\e}}_{\e}(\pi_{\e},\ph))+ \frac{\partial W^{\s_{\e}}_{\e}(\pi_{\e},\ph)}{\partial p} \eta(p)+ \frac{\partial W^{\s_{\e}}_{\e}(\pi_{\e},\ph)}{\partial \ph}\eta(\ph)  \\
 &  +  \left(\p_{\e}e^{-\int_{0}^{\tau(\ph,\ph^{\e})}\left(\lambda+\nu_{\e}(p^{\tau})\right)d\tau}+(1-\p_{\e})e^{-\int_{0}^{\tau(\ph,\ph^{\e})}\nu_{\e}(p^{\tau})d\tau}\right) V^{\sigma^*}_{\ph}(\p^{\e},\ph^{\e})
\frac{\eta(\ph^{\e})}{\eta(\ph)}\\
=& \left(\p_{\e}e^{-\int_{0}^{\tau(\ph,\ph^{\e})}\left(\lambda+\nu_{\e}(p^{\tau})\right)d\tau}+(1-\p_{\e})e^{-\int_{0}^{\tau(\ph,\ph^{\e})}\nu_{\e}(p^{\tau})d\tau}\right) V^{\sigma^*}_{\ph}(\p^{\e},\ph^{\e})
\frac{\eta(\ph^{\e})}{\eta(\ph)}\\
\to &  V^{\sigma^*}_{\ph}(\p^*,\ph_-) \mbox{ as } \e\to 0,
\end{align*}
where the second line holds because of the HJB for $\e$-commitment solution, i.e., by \cref{HJBm-e}.

\subsection{Proof of \cref{lem:properties_region3}\label{subsec:Proof-of-Lemma1}}

We first prove (i).  To show $\hat{V} (\ph)\in(u_{1},u_{2})$, it suffices to show
	that $C>0$ since in this case $\hat{V}^{\sigma}(\ph)$ is a convex
	combination of $u_{1}$ and $u_{2}$. We show that $\Phi (\ph_2)\in(u_{1},u_{2})$
	which implies that $C>0$. To see that $\Phi (\ph_2)>u_{1}$, note
	that by part (a) of this Lemma, $u_{1}$ is the value of a feasible
	strategy for the DM that differs from the Bayesian strategy. Therefore,
	the Bayesian value for any belief must strictly exceed $u_{1}$. To
	see that $\Phi (\ph_2)<u_{2}$, note that $u_{2}\ge\frac{1}{2}\left(u_{r}^{R}+u_{\ell}^{L}\right)+\frac{1}{2}\left|u_{r}^{R}-u_{\ell}^{L}\right|=u_{r}^{R}\vee u_{\ell}^{L}>\Phi (\ph_2)$. 
 Since $u_{1}<u_{2}$, the term $\left(\frac{1-\ph}{\ph}\right)^{\frac{u_{2}-u_{1}}{\delta_{\ell}}}$
	is decreasing in $\ph$ and therefore $\hat{V}^{\sigma\prime}(\ph)<0$. 
	
	We next prove (ii).  By the assumption, we have $V^{\sigma}(\pl(\ph_2), \ph_2) < U_{\ell}(\pl(\ph_2))$.  As $\ph\to 1$,  $\hat{V} (\ph)\to u_1$, whereas $U_{\ell}(\pl(\ph))\to u^R_{\ell}$, which by assumption is less than $u_1$.\footnote{This follows from the fact that $\lim_{\ph\rightarrow1}\left(\frac{1-\ph}{\ph}\right)^{\frac{u_{2}-u_{1}}{\delta_{\ell}}}=0$.  Note that $\pl(\ph)\to 1$ as $\ph\to 1$.}  Hence, by the intermediate value theorem, there exists $\ph_3\in (\ph_2, 1)$ such that $V^{\sigma}(\pl(\ph_3), \ph_3) =\hat V(\ph_3) =U_{\ell}(\pl(\ph_3))$.  The fact that such a $\ph_3$ is unique follows from the fact that $\hat V(\cdot)$ can only cross $U_{\ell}(\pl(\cdot))$ from below:
	 \begin{align*}
	\mbox{$\left.\frac{d\left(\hat{V}^{\sigma}(\ph)-U_{\ell}(\pl(\ph))\right)}{d\ph}\right|_{
			\tiny U_{\ell}(\pl(\ph))=\hat{V}(\ph)}$}  &  = \,
	 \mbox{$\frac{\left(u_{r}^{R}-\pl u_{\ell}^{R}-(1-\pl)u_{\ell}^{L}\right)\left(u_{\ell}^{L}-\pl u_{\ell}^{R}-(1-\pl)u_{\ell}^{L}\right)}{u_{\ell}^{L}-u_{\ell}^{R}}-\frac{c}{\lambda}-\left(u_{\ell}^{R}-u_{\ell}^{L}\right)\pl(1-\pl)$} >0\\
	\iff & \left(u_{r}^{R}-\pl u_{\ell}^{R}-(1-\pl)u_{\ell}^{L}\right)\pl-\frac{c}{\lambda}-\left(u_{\ell}^{R}-u_{\ell}^{L}\right)\pl(1-\pl)  >0\\
	  \iff & \left(u_{r}^{R}-u_{\ell}^{R}\right)\pl-\frac{c}{\lambda}   >0\\
 \iff & \pl  > p_{\ell}^B,
	\end{align*}
	where we have used the shorthand $\pl=\pl(\ph)$. Note that $U_{\ell}(\pl(\ph))=\hat{V}(\ph)$
	implies that $\pl>p_{\ell}^B$ since $U_{\ell}(p_{\ell}^B)=\Phi (p_{\ell}^B)>\Phi (\ph_2)=\hat{V}(\ph_2)>\hat{V}(\ph)= U_{\ell}(\pl)$ and since $U_{\ell}(\cdot)$ is decreasing.
	
	Finally, we prove (iii).  Simple algebra shows that $u_{1}>\hat{u}$ if and only if $c<\cl$:
		\begin{align*}
	u_{1}  >\hat{u} 
	\iff & \mbox{   $\frac{u_{r}^{R}+u_{\ell}^{L}}{2}-\sqrt{\left(\frac{u_{r}^{R}-u_{\ell}^{L}}{2}\right)^{2}+\frac{c}{\lambda}\delta_{\ell}}   >\frac{u_r^Ru_{\ell}^L-u_r^Lu_{\ell}^R}{\delta_{r}+\delta_{\ell}}$} \\
	\iff & \mbox{$\frac{\delta^R\delta_r+\delta^L\delta_{\ell}}{2(\delta_r+\delta_{\ell})}   >\sqrt{\left(\frac{u_{r}^{R}-u_{\ell}^{L}}{2}\right)^{2}+\frac{c}{\lambda}\delta_{\ell}}$}
	\\
	\iff & \mbox{$\left(\frac{\delta^R\delta_r+\delta^L\delta_{\ell}}{2(\delta_r+\delta_{\ell})}\right)^{2}   >\left(\frac{u_{r}^{R}-u_{\ell}^{L}}{2}\right)^{2}+\frac{c}{\lambda}\delta_{\ell}$ } \\
	\iff & \mbox{$c  <\frac{\lambda}{\delta_{\ell}}\left(\frac{\delta^R\delta_r+\delta^L\delta_{\ell}}{2(\delta_r+\delta_{\ell})}+\frac{u_{r}^{R}-u_{\ell}^{L}}{2}\right)\left(\frac{\delta^R\delta_r+\delta^L\delta_{\ell}}{2(\delta_r-\delta_{\ell})}+\frac{u_{r}^{R}-u_{\ell}^{L}}{2}\right)$}\\
	\iff & \mbox{$c  <\lambda\frac{\delta^R\delta^L\delta_r}{(\delta_r+\delta_{\ell})^2}=\frac{\delta_r}{\delta_r+\delta_{\ell}}\ch=\cl$}.
	\end{align*}
 For the last statement, fix any $c\in (\cl, \ch)$.  Then, $u_1<\hat u$, as shown above. Rewrite \cref{eq:pi3}
 $$\hat V (\ph_3(\Delta))= U_{\ell}(\pl(\ph_3(\Delta)))$$
so that  the dependence of $\ph_3$ is made explicit. Note first that neither $\hat V(\cdot)$ nor
$U_{\ell}(\cdot)$ depends on the size of ambiguity $\Delta$. The only source of dependence arises from the dependence of $\pl(\cdot)$ on $\Delta$.  Clearly $\pl(\ph)$ is strictly decreasing in $\Delta$ for each $\ph$. Since $\hat V(\cdot)$ can only cross $U_{\ell}(\pl(\cdot))$ from below (as was shown above), it follows that $\ph_3(\Delta)$ is strictly increasing in $\Delta$.  Further,  as $\Delta$ becomes sufficiently small, $\ph_3(\Delta)\to \ph_2$ and as $\Delta$ becomes sufficiently large, then  $\ph_3(\Delta)\to 1$.  Note that $\hat V(\ph_2)= \Phi(\ph_2)> \max\{U_r(\ph_2), U_{\ell}(\ph_2)\}\ge \hat u$ since $c<\ch$ and that $\lim_{\ph\to 1} \hat V(\ph)=u_1<\hat u$ since $c<\cl$.  We thus conclude that   there exists a unique $\Delta_c$ that satisfies 
$$ \hat V(\ph_3(\Delta_c))= \hat u.$$
Since $\hat V(\cdot)$ is strictly decreasing and $\ph_3(\cdot)$ is strictly increasing, it follows that $V^{\sigma}(\cdot, \ph_3)=\hat V(\ph_3)\ge \hat u$ if and only if $\Delta<\Delta_c$.

\subsection{Solution to \cref{eq:ODE_Vhat}\label{subsec:Solution-ODE}}

With the boundary condition $\hat{V}^{\sigma}(\ph_2)=\Phi (\ph_2)$
we obtain a candidate value function $\hat{V}^{\sigma}(\ph)$. 
\begin{align*}
\ph(1-\ph)\hat{V}^{\sigma\prime}(\ph)-\frac{\left(u_{r}^{R}-\hat{V}^{\sigma}(\ph)\right)\left(u_{\ell}^{L}-\hat{V}^{\sigma}(\ph)\right)}{\delta_{\ell}}+\frac{c}{\lambda} & =0\\
\ph(1-\ph)\hat{V}^{\sigma\prime}(\ph)-\frac{u_{r}^{R}u_{\ell}^{L}-\left(u_{r}^{R}+u_{\ell}^{L}\right)\hat{V}^{\sigma}(\ph)+\left(\hat{V}^{\sigma}(\ph)\right)^{2}}{\delta_{\ell}}+\frac{c}{\lambda} & =0\\
\ph(1-\ph)\hat{V}^{\sigma\prime}(\ph)+\frac{u_{r}^{R}+u_{\ell}^{L}}{\delta_{\ell}}\hat{V}^{\sigma}(\ph)-\frac{1}{\delta_{\ell}}\left(\hat{V}^{\sigma}(\ph)\right)^{2}+\frac{c}{\lambda}-\frac{u_{r}^{R}u_{\ell}^{L}}{\delta_{\ell}} & =0.
\end{align*}
Letting $\xih:=\ln\frac{\ph}{1-\ph}$ and $\tilde{V}(\xih)=\hat{V}^{\sigma}(\ph)$
we have $\ph(1-\ph)\hat{V}^{\sigma\prime}(\ph)=\tilde{V}'(\xih)$
so that the ODE can be written as 
\[
\tilde{V}'(\xih)+\frac{u_{r}^{R}+u_{\ell}^{L}}{\delta_{\ell}}\tilde{V}(\xih)-\frac{1}{\delta_{\ell}}\left(\tilde{V}(\xih)\right)^{2}+\frac{c}{\lambda}-\frac{u_{r}^{R}u_{\ell}^{L}}{\delta_{\ell}}=0
\]
\[
\tilde{V}'(\xih)=\frac{u_{r}^{R}u_{\ell}^{L}}{\delta_{\ell}}-\frac{c}{\lambda}-\frac{u_{r}^{R}+u_{\ell}^{L}}{\delta_{\ell}}\tilde{V}(\xih)+\frac{1}{\delta_{\ell}}\left(\tilde{V}(\xih)\right)^{2}.
\]
Let $t(\xi)=\tilde{V}(\xi)/\delta_{\ell}$. Then we have 
\begin{align*}
t'(\xi)=\frac{\tilde{V}'(\xi)}{\delta_{\ell}} & =\frac{u_{r}^{R}u_{\ell}^{L}}{\delta_{\ell}^{2}}-\frac{c}{\lambda\delta_{\ell}}-\frac{u_{r}^{R}+u_{\ell}^{L}}{\delta_{\ell}^{2}}\tilde{V}(\xih)+\left(\frac{\tilde{V}(\xih)}{\delta_{\ell}}\right)^{2}\\
 & =\frac{u_{r}^{R}u_{\ell}^{L}}{\delta_{\ell}^{2}}-\frac{c}{\lambda\delta_{\ell}}-\frac{u_{r}^{R}+u_{\ell}^{L}}{\delta_{\ell}}t(\xi)+\left(t(\xi)\right)^{2}\\
t'(\xi)-\left(t(\xi)\right)^{2} & =\frac{u_{r}^{R}u_{\ell}^{L}}{\delta_{\ell}^{2}}-\frac{c}{\lambda\delta_{\ell}}-\frac{u_{r}^{R}+u_{\ell}^{L}}{\delta_{\ell}}t(\xi)
\end{align*}.
Next suppose there is a function $\varphi(\xi)$ such that 
\[
t(\xi)=-\frac{\varphi'(\xi)}{\varphi(\xi)}.
\]
Then 
\[
t'(\xi)=-\frac{\varphi''(\xi)\varphi(\xi)-\left(\varphi'(\xi)\right)^{2}}{\left(\varphi(\xi)\right)^{2}}=-\frac{\varphi''(\xi)}{\varphi(\xi)}+\left(t(\xi)\right)^{2}.
\]
Hence 
\begin{align*}
-\frac{\varphi''(\xi)}{\varphi(\xi)} & =\frac{u_{r}^{R}u_{\ell}^{L}}{\delta_{\ell}^{2}}-\frac{c}{\lambda\delta_{\ell}}+\frac{u_{r}^{R}+u_{\ell}^{L}}{\delta_{\ell}}\frac{\varphi'(\xi)}{\varphi(\xi)}\\
\varphi''(\xi)+\frac{u_{r}^{R}+u_{\ell}^{L}}{\delta_{\ell}}\varphi'(\xi)+\left(\frac{u_{r}^{R}u_{\ell}^{L}}{\delta_{\ell}^{2}}-\frac{c}{\lambda\delta_{\ell}}\right)\varphi(\xi) & =0
\end{align*}.
The roots of $a^{2}+\frac{u_{r}^{R}+u_{\ell}^{L}}{\delta_{\ell}}a+\left(\frac{u_{r}^{R}u_{\ell}^{L}}{\delta_{\ell}^{2}}-\frac{c}{\lambda\delta_{\ell}}\right)$
are
\begin{align*}
a_{1,2} & =-\frac{u_{r}^{R}+u_{\ell}^{L}}{2\delta_{\ell}}\pm\sqrt{\left(\frac{u_{r}^{R}+u_{\ell}^{L}}{2\delta_{\ell}}\right)^{2}-\frac{u_{r}^{R}u_{\ell}^{L}}{\delta_{\ell}^{2}}+\frac{c}{\lambda\delta_{\ell}}}\\
 & =-\frac{u_{r}^{R}+u_{\ell}^{L}}{2\delta_{\ell}}\pm\sqrt{\left(\frac{u_{r}^{R}-u_{\ell}^{L}}{2\delta_{\ell}}\right)^{2}+\frac{c}{\lambda\delta_{\ell}}}.
\end{align*}
Note that this implies $a_{1}=-u_{2}/\delta_{\ell}$ and $a_{2}=-u_{1}/\delta_{\ell}$
where we use the subscript $1$ for the lower root in both $a_{i}$
and $u_{i}$. Given that the roots are real we have a general solution
\[
\varphi(\xi)=C_{1}e^{a_{1}\xi}+C_{2}e^{a_{2}\xi},
\]
where $C_{1},C_{2}$ are constants of integration. Setting $C=C_{2}/C_{1}$
we get the general solution for $t(\xi)$, $\tilde{V}(\xi)$, and
$\hat{V}^{\sigma}(\ph)$:
\[
t(\xi)=-\frac{C_{1}a_{1}e^{a_{1}\xi}+C_{2}a_{2}e^{a_{2}\xi}}{C_{1}e^{a_{1}\xi}+C_{2}e^{a_{2}\xi}}=-\frac{a_{1}e^{a_{1}\xi}+Ca_{2}e^{a_{2}\xi}}{e^{a_{1}\xi}+Ce^{a_{2}\xi}}=-\frac{a_{1}e^{\left(a_{1}-a_{2}\right)\xi}+Ca_{2}}{e^{\left(a_{1}-a_{2}\right)\xi}+C}
\]
\[
\tilde{V}(\xi)=-\delta_{\ell}\frac{a_{1}e^{\left(a_{1}-a_{2}\right)\xi}+Ca_{2}}{e^{\left(a_{1}-a_{2}\right)\xi}+C}
\]
\[
\hat{V}^{\sigma}(\ph)=-\delta_{\ell}\frac{a_{1}\left(\frac{\ph}{1-\ph}\right)^{a_{1}-a_{2}}+Ca_{2}}{\left(\frac{\ph}{1-\ph}\right)^{a_{1}-a_{2}}+C},
\]
where 
\[
a_{1}-a_{2}=-\frac{1}{\delta_{\ell}}\sqrt{\left(u_{r}^{R}-u_{\ell}^{L}\right)^{2}+4\frac{c}{\lambda}\delta_{\ell}}<0.
\]
We note that 
\[
\lim_{\ph\rightarrow1}\hat{V}^{\sigma}(\ph)=-\delta_{\ell}a_{2}=u_{1}.
\]
Boundary condition: 
\begin{align*}
\hat{V}^{\sigma}(\ph_2) & =\Phi (\ph_2)\\
\iff-\delta_{\ell}\frac{a_{1}\left(\frac{\ph}{1-\ph}\right)^{a_{1}-a_{2}}+Ca_{2}}{\left(\frac{\ph}{1-\ph}\right)^{a_{1}-a_{2}}+C} & =\Phi (\ph_2)\\
\iff-\delta_{\ell}\left(a_{1}\left(\frac{\ph}{1-\ph}\right)^{a_{1}-a_{2}}+Ca_{2}\right) & =\Phi (\ph_2)\left(\left(\frac{\ph}{1-\ph}\right)^{a_{1}-a_{2}}+C\right)\\
\iff\left(\Phi (\ph_2)+\delta_{\ell}a_{2}\right)C & =-\left(\Phi (\ph_2)+\delta_{\ell}a_{1}\right)\left(\frac{\ph}{1-\ph}\right)^{a_{1}-a_{2}}\\
C & =-\frac{\Phi (\ph_2)+\delta_{\ell}a_{1}}{\Phi (\ph_2)+\delta_{\ell}a_{2}}\left(\frac{\ph_2}{1-\ph_2}\right)^{a_{1}-a_{2}}\\
 & =\frac{u_{2}-\Phi (\ph_2)}{\Phi (\ph_2)-u_{1}}\left(\frac{\ph_2}{1-\ph_2}\right)^{a_{1}-a_{2}}.
\end{align*}

\subsection{Proof of \cref{lem:r3-Vsigma}} \label{sec:r3-Vsigma}

The value of strategy $\sigma$ is given by:
\begin{align*}
V^{\sigma}(p,\ph)= & p \int_{\ph_2}^{\ph}e^{-\int_{0}^{\tau(\ph,\ph')}\left(\lambda+\nu(p_{\tau})\right)d\tau}\left(-c+\lambda u_{r}^{R}+\nu(\ph')u_{\ell}^{R}\right)\frac{1}{\lambda\ph'(1-\ph')}d\ph' \\
 & +(1-p) \int_{\ph_2}^{\ph}e^{-\int_{0}^{\tau(\ph,\ph')}\nu(p_{\tau})d\tau}\left(-c+\nu(\ph')u_{\ell}^{L}\right)\frac{1}{\lambda\ph'(1-\ph')}d\ph' \\
 & +\left(pe^{-\int_{0}^{\tau(\ph,\ph_2)}\left(\lambda+\nu(p_{\tau})\right)d\tau}+(1-p)e^{-\int_{0}^{\tau(\ph,\ph_2)}\nu(p_{\tau})d\tau}\right)V^{\sigma}(p_{\tau(\ph,\ph_2)}(p),\ph_2).
\end{align*}
Using the change of variables $\tau=\tau(\ph,\ph'')$ and hence $\ph_{\tau}=\ph''$
and $d\tau=-\frac{1}{\lambda\ph''(1-\ph'')}d\ph''$ we can rewrite
this as follows:\footnote{Remember that $\ph_{t}=\frac{\ph_{0}}{\ph_{0}+(1-\ph_{0})e^{\lambda t}}$.
Setting $\ph_{0}=\ph$ and $\ph_{t}=\ph'$ we get 
\begin{align*}
\tau(\ph,\ph') & =\frac{1}{\lambda}\left(\log\frac{\ph}{(1-\ph)}-\log\frac{\ph'}{1-\ph'}\right).
\end{align*}
Hence,
\[
\frac{d\tau(\ph,\ph')}{d\ph}=\frac{1}{\lambda\ph(1-\ph)}\qquad\text{and}\qquad\frac{d\tau(\ph,\ph')}{d\ph'}=-\frac{1}{\lambda\ph(1-\ph)}.
\]
 }

\begin{align*}
V^{\sigma}(p,\ph)= & p \int_{\ph_2}^{\ph}e^{-\int_{\ph'}^{\ph}\left(\lambda+\nu(\ph'')\right)\frac{d\ph''}{\lambda\ph''(1-\ph'')}}\left(-c+\lambda u_{r}^{R}+\nu(\ph')u_{\ell}^{R}\right)\frac{1}{\lambda\ph'(1-\ph')}d\ph'  \\
 & +(1-p) \int_{\ph_2}^{\ph}e^{-\int_{\ph'}^{\ph}\nu(\ph'')\frac{d\ph''}{\lambda\ph''(1-\ph'')}}\left(-c+\nu(\ph')u_{\ell}^{L}\right)\frac{1}{\lambda\ph'(1-\ph')}d\ph' \\
 & +\left(pe^{-\int_{\ph_2}^{\ph}\left(\lambda+\nu(\ph'')\right)\frac{d\ph''}{\lambda\ph''(1-\ph'')}}+(1-p)e^{-\int_{\ph_2}^{\ph}\nu(\ph'')\frac{d\ph''}{\lambda\ph''(1-\ph'')}}\right)V^{\sigma}(p_{\tau(\ph,\ph_2)}(p),\ph_2).
\end{align*}

 We wish to show that $V(p, \ph)$ defined in Region 3 coincides with $V^{\sigma}(p, \ph)$ above.  To this end, note first that $V(\cdot, \ph_2)=\hat V(\ph_2)=V^{\sigma} (\cdot, \ph_2)$, by the definition of $\hat V(\ph)$ at $\ph=\ph_2$.   Recall that we defined $V(p,\ph)$ in Region 3 so that    $V_{p}(p,\ph)=0$
and that
\begin{equation}
c=\lambda p\left(u_{r}^{R}-V(p,\ph)\right)+\nu(\ph)\left(U_{\ell}(\ph)-V(p,\ph)\right)+\eta (p)V_{p}(p,\ph)+\eta (\ph)V_{\ph}(p,\ph)\label{eq:HJB_with_strategy}
\end{equation}
for all $p\in[0,1]$ and $\ph\in(\ph_2,\ph_{3}
]$.

To show that  $V(p,\ph)=V^{\sigma}(p,\ph)$, substitute $p=0$
and $p=1$ in \cref{eq:HJB_with_strategy} to obtain two ODEs:
\begin{align*}
V_{\ph}(0,\ph) & =\frac{c-\nu(\ph)u_{\ell}^{L}}{\eta (\ph)}+\frac{\nu(\ph)}{\eta (\ph)}V(0,\ph);\\
V_{\ph}(1,\ph) & =\frac{c-\lambda u_{r}^{R}-\nu(\ph)u_{\ell}^{R}}{\eta (\ph)}+\frac{\lambda+\nu(\ph)}{\eta (\ph)}V(1,\ph).
\end{align*}
Integrating these with the boundary condition given by $V(\cdot,\ph_2)=V^{\sigma}(\cdot,\ph_2)$
we get 
\begin{align*}
V(0,\ph) & =e^{\int_{\ph_2}^{\ph}\frac{\nu(x)}{\eta (x)}dx}V^{\sigma}(0,\ph_2)+e^{\int_{\ph_2}^{\ph}\frac{\nu(x)}{\eta (x)}dx}\int_{\ph_2}^{\ph}e^{-\int_{\ph_2}^{\ph'}\frac{\nu(x)}{\eta (x)}dx}\frac{c-\nu(\ph')u_{\ell}^{R}}{\eta (\ph')}d\ph'\\
 & =e^{\int_{\ph_2}^{\ph}\frac{\nu(x)}{\eta (x)}dx}V^{\sigma}(0,\ph_2)+\int_{\ph_2}^{\ph}e^{\int_{\ph'}^{\ph}\frac{\nu(x)}{\eta (x)}dx}\frac{c-\nu(\ph')u_{\ell}^{R}}{\eta (\ph')}d\ph'\\
 & =V^{\sigma}(0,\ph),
\end{align*}
and
\begin{align*}
V(1,\ph) & =e^{\int_{\ph_2}^{\ph}\frac{\lambda+\nu(x)}{\eta (x)}dx}V^{\sigma}(1,\ph_2)+e^{\int_{\ph_2}^{\ph}\frac{\lambda+\nu(x)}{\eta (x)}dx}\int_{\ph_2}^{\ph}e^{-\int_{\ph_2}^{\ph'}\frac{\lambda+\nu(x)}{\eta (x)}dx}\frac{c-\lambda u_{r}^{R}-\nu(\ph')u_{\ell}^{R}}{\eta (\ph')}d\ph'\\
 & =e^{\int_{\ph_2}^{\ph}\frac{\lambda+\nu(x)}{\eta (x)}dx}V^{\sigma}(1,\ph_2)+\int_{\ph_2}^{\ph}e^{\int_{\ph'}^{\ph}\frac{\lambda+\nu(x)}{\eta (x)}dx}\frac{c-\lambda u_{r}^{R}-\nu(\ph')u_{\ell}^{R}}{\eta (\ph')}d\ph'\\
 & =V^{\sigma}(1,\ph).
\end{align*}

Since $V(p,\ph)=p V(1,\ph)+(1-p)V(0,\ph)$ and $V^{\sigma}(p,\ph)=p V^{\sigma}(1,\ph)+(1-p)V^{\sigma}(0,\ph)$, we have proven that 
  $V(p,\ph)=V^{\sigma}(p,\ph)$.

\subsection{Case 2:  $\Phi'(p_*)<0$}   \label{ssec:case2}

\begin{theorem}\label{thm:case2}
Suppose $c<\overline c$ and $\Phi'(p_*)<0$. For each $c\in(0,\overline c)$ there is a threshold $\Delta_c\in(0,+\infty]$ with $\Delta_c=+\infty$ if and only if $c\leq\underline c$ such that the intrapersonal equilibrium is described by:
\medskip
\begin{itemize}
\item[(a)] If $\Delta\leq\Delta_c$, there exist cutoffs $0<\ph_1<\ph_2\le\ph_3<\ph_4<1$ with $\ph_1=p_{\ell}^B$ and $\ph_2=p_*$ such that
\[
(m(\ph),\nu(\ph),\rho(\ph))=
\begin{cases}
(1,0,0)\\
(0,0,\cdot)\\
(m(\ph_2),0,1)\\
(0,\nu^*(\ph),0)\\
(0,0,\cdot)\\
(1,0,1)
\end{cases}\;\pi(\ph)=\begin{cases}
\ph & \text{if }\ph\in[0,\ph_{1}]\\
\ph & \text{if }\ph\in(\ph_{1},\ph_{2})\\
\ph & \text{if }\ph=\ph_2\\
\pi^*(\ph)\quad & \text{if }\ph\in(\ph_{2},\ph_{3})\\
\pl(\ph) & \text{if }\ph\in[\ph_{3},\ph_{4})\\
\pl(\ph) & \text{if }\ph\in[\ph_{4},1]
\end{cases}
\]
with $m(\ph_2)\in(0,1)$, $\nu^*(\ph)>0$ and $\pi^*(\ph)\in(\pl(\ph),\ph)$, for $\ph\in(\ph_2,\ph_3)$, and $\ph_{2}<\ph_{3}$ if and only if $\Phi(p^{*})<U_{\ell}(\pl(p^{*}))$.
\medskip
\item[(b)] If $\Delta>\Delta_c$, the equilibrium has the same structure as in (a) except that for $\ph\in[\ph_3,\ph_4)$, where
\[
(m(\ph),\nu(\ph),\rho(\ph))=\left(1,0,\hat\rho\right)\quad\text{and}\quad \pi(\ph)=\hat p.
\]
\end{itemize}
\end{theorem}

In this case, the characterization remains the same except at one state $\ph=\ph_2$.  In the main case, recall that $\ph_2$ was part of Region 2.  When $\Phi'(p_*)<0$, Region 2 strategy and verification apply only for $\ph\in (\ph_1, \ph_2)$.  (The strategies and verification for all other regions remain valid.)  The case of $\ph=\ph_2$ will change as follows.

 When $\Phi'(p_*)<0$,  we have $\ph_2=p_*=p_r^B$, where recall $p_r^B$ is the stopping boundary for the Bayesian DM. Therefore, $\Phi (\ph_2)=U_{r}(\ph_2)$.  The strategy $\sigma(\ph_2)$ specifies $\nu(\ph_2)=0$ and $\rho(\ph_2)=1$.  We now specify $m(\ph_2)$ below.  Given $\nu(\ph_2)=0$ and $\rho(\ph_2)=1$, we have
\begin{align*}
V^{\sigma}(p,\ph_2) & =U_{r}(\ph_2)+(p-\ph_2)\left[m(\ph_2)U_{r}'(\ph_2)+(1-m(\ph_2))V_{p}^{\sigma}(p,\ph_{-}^{2})\right]\\
 & =U_{r}(\ph_2)+(p-\ph_2)\left[m(\ph_2)U_{r}'(\ph_2)+(1-m(\ph_2))\Phi^{\prime}(\ph_2)\right].
\end{align*}
To understand the first line, note that conditional on stopping, the
DM gets a payoff of $U_{r}(p)=U_{r}(\ph_2)+(p-\ph_2)U_{r}'(\ph_2)$,
and conditional on not stopping, she gets $V^{\sigma}(p,\ph_{2-})=V^{\sigma}(\ph_{2-},\ph_{2-})+(p-\ph_2)V_{p}^{\sigma}(p,\ph_{2-})=U_{r}(\ph_2)+(p-\ph_2)\Phi'(\ph_2)$.\footnote{Here we have to distinguish carefully between $V^{\sigma}(p,\ph_{2-})$,
$V_{p}^{\sigma}(p,\ph_{2-})$ and $V^{\sigma}(p,\ph_2)$, $V_{p}^{\sigma}(p,\ph_2)$.} 

We set $m(\ph_2)$ so that the term in the square brackets vanishes:
\[
m(\ph_2)=\frac{-\Phi^{\prime}(\ph_2)}{U'_{r}(\ph_2)-\Phi^{\prime}(\ph_2)}\in(0,1),
\]
where $m(\ph_2)\in(0,1)$ holds since   $\Phi^{\prime}(\ph_2)<0$. With this definition of $m(\ph_2)$,
we have $V_{p}^{\sigma}(p,\ph_2)=0$ so that $\pi(\ph_2)=\ph_2$
is a minimizer in \cref{eq:worst-case-belief_proof1}.

To verify \cref{eq:HJB0} at $\ph_2,$ we substitute
$\pi  (\ph_2)=\ph_2$, $V(\pi  (\ph_2),\ph_{2-})=\Phi (\ph_2)=U_{r}(\ph_2)$,
$V_{p}^{\sigma}(\pi  (\ph_2),\ph_{2-})=\Phi^{\prime}(\ph_2)$,
and $V_{\ph}^{\sigma}(\pi  (\ph_2),\ph_{2-})=0$. Since $\ph_2>\hat{p}$ if $\Phi (\ph_2)=U_{r}(\ph_2)$,  $\rho=1$ is optimal, and hence the objective is independent
of $\nu$, we obtain the following simplified version of \cref{eq:HJB0}:

\[
\max_{m}(1-m)\left[-c+p\lambda(u_{r}^{R}-\Phi (\ph_2))+\Phi^{\prime}(\ph_2)\eta (\ph_2)\right]=0.
\]
The terms inside the square brackets are equal to zero (from \cref{eq:BayesianODE}), so \cref{eq:HJB0} holds, and  $\sigma(\ph_2)=(m(\ph_2),0,1)$
is a maximizer of  \cref{eq:HJBm}. Finally,
note that the objective in \cref{eq:HJBpi} is
\begin{align*}
 & m(\ph_2)\left(U_{r}(p)-V^{\sigma}(p,\ph_{2-})\right)\\
 & +(1-m(\ph_2))\left[-c+p\lambda\left(u_{r}^{R}-V^{\sigma}(p,\ph_{2-})\right)+V_{p}^{\sigma}(p,\ph_{2-})\eta (p)+V_{\ph}^{\sigma}(p,\ph_{2-})\eta (\ph_2)\right]\\
= & m(\ph_2)\left(U_{r}(p)-\left(\Phi (\ph_2)+(p-\ph_2)\Phi^{\prime}(\ph_2)\right)\right)-(1-m(\ph_2))c\\
 & +(1-m(\ph_2))\left[\begin{array}{l}
p\lambda\left(u_{r}^{R}-\left(\Phi (\ph_2)+(p-\ph_2)\Phi^{\prime}(\ph_2)\right)\right)-\lambda p(1-p)\Phi^{\prime}(\ph_2)\\
-\lambda\ph_2(1-\ph_2)(p-\ph_2)\Phi^{\prime\prime}(\ph_2)
\end{array}\right]\\
= & m(\ph_2)\left(U_{r}(p)-\left(\Phi (\ph_2)+(p-\ph_2)\Phi^{\prime}(\ph_2)\right)\right)-(1-m(\ph_2))c\\
 & +(1-m(\ph_2))\left[p\lambda\left(u_{r}^{R}-\Phi (\ph_2)\right)+p\lambda(1-\ph_2)\Phi^{\prime}(\ph_2)-\lambda\ph_2(1-\ph_2)(p-\ph_2)\Phi^{\prime\prime}(\ph_2)\right],
\end{align*}
where we have used \cref{eq:V-sigma_region2} to obtain the second
line. Differentiating this with respect to $p$ yields 
\begin{align*}
 & m(\ph_2)\left(U'_{r}(p)-\Phi^{\prime}(\ph_2)\right)\\
 & +(1-m(\ph_2))\left[\lambda\left(u_{r}^{R}-\left(\Phi (\ph_2)+(1-\ph_2)\Phi^{\prime}(\ph_2)\right)\right)-\lambda\ph_2(1-\ph_2)\Phi^{\prime\prime}(\ph_2)\right]\\
= & -\Phi^{\prime}(\ph_2)+(1-m(\ph_2))\left[\lambda\left(u_{r}^{R}-\Phi (\ph_2)\right)+\lambda(1-\ph_2)\Phi^{\prime}(\ph_2)-\lambda\ph_2(1-\ph_2)\Phi^{\prime\prime}(\ph_2)\right],
\end{align*}
where we have used the definition of $m(\ph_2)$ to obtain the second
line. Since the derivative of \cref{eq:Bayesian-ODE} vanishes, the terms inside the brackets vanish. Given
$\Phi^{\prime}(\ph_2)<0$, $\pi(\ph)=\ph$ is thus the unique
maximizer in \cref{eq:HJBpi}.

To summarize, we have shown that for the posited $\sigma$, \cref{eq:worst-case-belief_proof1}--\cref{eq:HJBpi} hold for $\ph=\ph_2$.

\subsection{Uniqueness} \label{sec:uniqueness}

Here we prove the intrapersonal equilibrium, $\sigma=(\nu,\mu,\rho)$, together with nature's choice $\pi$, described in \cref{thm:main_result} and \cref{thm:case2}, is unique for the case $c<\cl$ and $c\ge\ch$. Towards a contradiction, suppose there is a different equilibrium $\tilde\sigma=(\tilde\nu,\tilde m,\tilde\rho)$, with $\tilde\pi$ describing nature's choice.   Let
\[
\tilde p:=\inf\{\ph\in(0,1):\tilde\sigma(\ph)\neq\sigma(\ph)\}.
\]
Assuming $c<\cl$, we will start with Case 1 and move through the different parameter regions, excluding each of them.

\begin{enumerate}
    \item $\tilde p\in[0,\ph_1)$: Recall $m(\ph)=1$ and $\rho(\ph)=0$ for all $\ph\in[0,\ph_1)$. Assuming $\tilde p<\ph_1$, the admissibility of $\tilde \sigma$ means that there must exist an open interval $(\tilde p,\tilde p+\varepsilon)\subset[0,\ph_1]$ such that $\tilde m(\ph)=0$ for all $\ph\in(\tilde p,\tilde p+\varepsilon)$.\footnote{Our admissibility restrictions rule out the possibilities of $m(\ph)<1$ on an isolated point as well as $m(\ph)\in(0,1)$ for all $\ph$ belonging to an interval of states.\label{f:admiss}} Consider $\ph\in(\tilde p,\tilde p+\varepsilon)$. Since $V^{\tilde\sigma}(p,\ph)<\Phi^*(p)$ for all $p\in[\pl(\ph),\ph]$ and since $\Phi^*(p)\le U_{\ell}(p)$ for all $p\le \ph_1=p^B_r$, we have $V^{\tilde\sigma}(p,\ph)<U_{\ell}(p),\forall p\in [\pl(\ph),\ph]$ and hence $V^{\tilde\sigma}(\tilde\pi(\ph),\ph)<U_{\ell}(\tilde\pi(\ph))$, which violates condition \eqref{eq:UvsValue} of \cref{def:pseudo}.
    \item $\tilde p\in[\ph_1,\ph_2)$: Recall $m(\ph)=\nu(\ph)=0$ for all $\ph\in(\ph_1,\ph_2)$.
    \begin{enumerate}
        \item Suppose there is some $\varepsilon>0$ such that $\tilde m(\ph)=0$ for all $\ph\in[\tilde p,\tilde p+\varepsilon)$. We then have $V^{\tilde\sigma}(p,\tilde p)=V^{\sigma}(p,\tilde p)$ for all $p$, with the value $V^{\tilde\sigma}(p,\ph)$ being continuous in $\ph$ at $\ph=\tilde p$. Since $V^{\sigma}_p(p,\ph)<0$, it follows $V^{\tilde\sigma}_{p}(p,\ph)<0$ and hence $\tilde\pi(\ph)=\ph$ for $\ph\geq\tilde p$ sufficiently close to $\tilde p$. Next, since $V^{\sigma}(\ph,\ph)>U_{\rho}(\ph)$ for all $\ph\in(\ph_1,\ph_2)$ and all $\rho\in[0,1]$, the same property holds under strategy $\tilde\sigma$ for $\ph\geq\tilde p$ sufficiently close to $\tilde p$, that is, $V^{\tilde\sigma}(\ph,\ph)>U_{\rho}(\ph)$ for all $\rho\in[0,1]$. Condition \eqref{eq:opt-stop} of \cref{def:pseudo} then implies $\tilde\nu(\ph)=\nu(\ph)=0$ for all $\ph$ belonging to a right neighborhood of $\tilde p$, which is a contradiction.
        \item Suppose next that for every $\varepsilon>0$ there is some $\ph\in[\tilde p,\tilde p+\varepsilon)$ such that $\tilde m(\ph)>0$. Since $\tilde p<\ph_2=p_*<\hat p$, we have $\ph<\hat p$ for all $\ph\geq\tilde p$ sufficiently close to $\tilde p$. Hence, by condition \eqref{eq:opt-action}, if $\tilde m(\ph)>0$ for such values of $\ph$, then $\tilde\rho(\ph)=0$. We start by considering the DM's strategy at state $\tilde p$. With probability $\tilde m(\tilde p)$ the DM stops and collects payoff $U_{\ell}(p)$ for some $p\in \mathcal{P}(\tilde p)$; and with the complementary probability she obtains the continuation payoff of strategy $\sigma$. Her value at state $\tilde p$ is thus
        \begin{eqnarray}\label{eq:mtilde}
        V^{\tilde\sigma}(p,\tilde p)=\tilde m(\tilde p)U_{\tilde\rho(\tilde p)}(p)+(1-\tilde m(\tilde p))V^{\sigma}(p,\tilde p),
        \end{eqnarray}
        with $U_{\tilde\rho(\tilde p)}(p)=U_{\ell}(p)$. Since both $U_{\ell}(p)$ and $V^{\sigma}(p,\tilde p)$ are strictly decreasing in $p$, we have $\tilde\pi(\tilde p)=\tilde p$ by condition \eqref{eq:worst-case-belief_proof1}. Since $U_{\ell}(\tilde p)<V^{\sigma}(\tilde p,\tilde p)=\Phi(\tilde p)$, $V^{\tilde\sigma}(\tilde\pi(\tilde p),\tilde p)$ is strictly decreasing in $\tilde m(\tilde p)$ and hence $\tilde m(\tilde p)=0$. Since we assumed that for every $\varepsilon>0$ there exists some $\ph\in(\tilde p,\tilde p+\varepsilon)$ such that $\tilde m(\ph)>0$ and since we just showed that $\tilde m(\tilde p)=0$, the remaining possibility is to have an interval to the right of $\tilde p$ on which $\tilde m(\ph)=1$ (and $\tilde\rho(\ph)=0$). For all states $\ph$ belonging to this interval, we thus have $V^{\tilde\sigma}_{p}(p,\ph)=U_{\ell}'(p)<0$ and hence $\tilde\pi(\ph)=\ph$. The HJB functional then simplifies to
        \begin{align*}
        G(m,\nu,0,\tilde\pi(\ph),\ph,V^{\tilde\sigma},dV^{\tilde\sigma})=(1-m)\left(-c+\ph\lambda\delta^R\right).
        \end{align*}
        Since $\ph>p_{\ell}^B=\frac{c}{\lambda\delta^R}$, the functional is strictly decreasing $m$. Condition \eqref{eq:HJBm} thus requires $\tilde m(\ph)=0$ for all $\ph\geq\tilde p$ sufficiently close to $\tilde p$, which yields a contradiction.
    \end{enumerate}
    
    \item $\tilde p\in[\ph_2,\ph_3)$: Recall $m(\ph)=0$, $\nu(\ph)>0$ and $\rho(\ph)=0$ for all states $\ph$ in Region 3.
    \begin{enumerate}
        \item Suppose again there is some $\varepsilon>0$ such that $\tilde m(\ph)=0$ for all $\ph\in[\tilde p,\tilde p+\varepsilon)$. For every $\varepsilon'\in(0,\varepsilon)$, we can then find some $\ph\in[\tilde p,\tilde p+\varepsilon')$ such that $\tilde\nu(\ph)\neq\nu(\ph)$.\footnote{The possibility where there is an $\varepsilon'\in(0,\varepsilon)$ such that $\tilde\nu(\ph)=\nu(\ph)$ for all $\ph\in[\tilde p,\tilde p+\varepsilon')$, but then $\tilde\rho(\ph)\neq\rho(\ph)=0$ for $\ph$ close or equal to $\tilde p$ would clearly violate the HJB conditions.} As before, we have $V^{\tilde\sigma}(p,\tilde p)=V^{\sigma}(p,\tilde p)$, with $V^{\tilde\sigma}(p,\ph)$ being continuous in $\ph$ at $\ph=\tilde p$. If $\tilde\nu(\ph)>0$ on a right neighborhood of $\tilde p$, the same HJB conditions pinning down $\nu(\ph)$ must hold. Uniqueness of the solution of these conditions then implies $\tilde\nu(\ph)=\nu(\ph)$ for all $\ph$ belonging to this neighborhood. We are left with the following possibility: for every $\varepsilon'\in(0,\varepsilon)$, we can find some $\ph\in[\tilde p,\tilde p+\varepsilon')$ such that $\tilde\nu(\ph)=0$. As $\varepsilon'\to0$, the probability that $R$-evidence arrives within time $\tau(\ph,\tilde p)$ vanishes, so both $V^{\tilde\sigma}_p(p,\ph)$ and $V^{\tilde\sigma}_{p\ph}(p,\ph)$ vanish (recall that at $\ph=\tilde p$, the value segment $\{V^{\sigma}(p,\tilde p)\}_{p\in \mathcal{P}(\tilde p)}$ is flat.) Hence, for $\varepsilon'$ sufficiently small, the right-hand side of the HJB functional is strictly increasing in $p$ for all $\ph$ such that $\tilde\nu(\ph)=0$. Nature thus chooses $\tilde\pi(\ph)=\pl(\ph)$. Since however  $U_{\ell}(\pl(\ph))>V^{\tilde\sigma}(\pl(\ph),\ph)$ for $\ph$ sufficiently close to $\tilde p$, this violates condition \eqref{eq:opt-stop}.
        \item Consider next the case where for every $\varepsilon>0$ there is some $\ph\in[\tilde p,\tilde p+\varepsilon)$ such that $\tilde m(\ph)>0$. As before, we start by showing $\tilde m(\tilde p)=0$. At state $\tilde p$ the DM's value is described by \eqref{eq:mtilde}.
        \begin{itemize}
            \item Suppose $\tilde\pi(\tilde p)>\pi(\tilde p)$. Then, since $V^{\sigma}(\tilde\pi(\tilde p),\tilde p)>U(\tilde\pi(\tilde p))$, $V^{\tilde\sigma}(p,\tilde p)$ strictly decreases in $\tilde m(\tilde p)$, which implies $\tilde m(\tilde p)=0$.
            \item Suppose $\tilde\pi(\tilde p)\leq\pi(\tilde p)$. Since $\pi(\tilde p)<\hat p$, condition \cref{eq:opt-action} implies $\tilde\rho(\tilde p)=0$.\footnote{Unless $\tilde m (\tilde p)=\tilde\nu(\tilde p)=0$, in which case $\tilde\rho(\tilde p)$ can take an arbitrary value.} For nature to optimally choose $\tilde\pi(\tilde p)<\tilde p$, the DM's value $V^{\tilde\sigma}(p,\tilde p)$ must be weakly increasing in $p$, that is, $V^{\tilde\sigma}_p(p,\ph)\geq0$. However, $V^{\sigma}_p(p,\tilde p)=0$ and $U_{\ell}'(p)<0$, so $V^{\tilde\sigma}_p(p,\ph)\geq0$ requires $\tilde m(\tilde p)=0$, as can be seen from \eqref{eq:mtilde}.
        \end{itemize} 
        Having shown $\tilde m(\tilde p)=0$, this leaves the possibility that there is an interval $(\tilde p,\tilde p+\varepsilon],\varepsilon>0$, such that   $\tilde m(\ph)=1$ for all $\ph\in (\tilde p,\tilde p+\varepsilon]$. If $\ph\leq\hat p$ for all $\ph$ in that interval, then \eqref{eq:opt-action} implies $\tilde\rho(\ph)=0$, in which case we have $V^{\tilde\sigma}(p,\ph)=U_{\ell}(p)$ and hence $\tilde\pi(\ph)=\ph$. We can then follow the same argument as in point 2(b) to show that $\tilde m(\ph)=1$ violates the HJB condition \eqref{eq:HJBm}. Suppose instead $\tilde p\geq\hat p$. Given $\pl(\tilde p)<\pl(\ph_3)<\hat p$, the saddle point conditions \eqref{eq:worst-case-belief_proof1} and \eqref{eq:opt-action} of \cref{def:pseudo} imply $\tilde\rho(\ph)=\hat\rho$ and $\tilde\pi(\ph)=\hat p$ for all $\ph>\tilde p$ sufficiently close to $\tilde p$ such that $\tilde m(\ph)=1$, with $V^{\tilde\sigma}(p,\ph)=\hat u$ for all $p\in \mathcal{P}(\ph)$ as the DM's value. Substituting for nature's choice and $V^{\tilde\sigma}$,\footnote{Note that, given $V^{\tilde\sigma}(p,\ph)=\hat u$, we have $V^{\tilde\sigma}_p(p,\ph)=V^{\tilde\sigma}_{\ph}(p,\ph)=0$.} the HJB functional simplifies to
        \begin{align*}
        G(m,\nu,0,\tilde\pi(\ph),\ph,V^{\tilde\sigma},dV^{\tilde\sigma})=(1-m)\left(-c+\hat p\lambda(u_r^R-\hat u)\right).
        \end{align*}
        The coefficient of $(1-m)$ is strictly positive if
        \begin{eqnarray*}
        \hat p>\frac{c}{\lambda(u_r^R-\hat u)}&\iff&\frac{\delta^L}{\delta^R+\delta^L}>\frac{c}{\lambda\left(u_r^R-\frac{u_r^Ru_{\ell}^L-u_r^Lu_{\ell}^R}{\delta^R+\delta^L}\right)}\\
        &\iff& c<\lambda\frac{\delta^L}{\delta^R+\delta^L}\frac{u_r^R(\delta^R+\delta^L)-u_r^Ru_{\ell}^L+u_r^Lu_{\ell}^R}{\delta^R+\delta^L}\\
        &\iff& c<\lambda\frac{\delta^L}{\delta^R+\delta^L}\frac{\delta^R\delta_r}{\delta_r+\delta_{\ell}}=\cl
        \end{eqnarray*}
        Since the latter inequality is satisfied by assumption, we ruled out the possibility $\tilde m(\ph)=1$ on an interval $(\tilde p,\tilde p+\varepsilon)$. Taken together, this shows that $\tilde p$ cannot belong to $[\ph_2,\ph_3)$.
    \end{enumerate}
    \item $\tilde p\in[\ph_3,\ph_4)$: Recall $m(\ph)=\nu(\ph)=0$ for all $\ph\in[\ph_3,\ph_4)$. The argument is analogous to that in Point 2. In the current case, $U_{\tilde\rho(\tilde p)}(p)<V^{\sigma}(p)$ for all $p\in \mathcal{P}(\tilde p)$ if $\tilde\rho(\tilde p)\leq\hat\rho$. If instead $\tilde\rho(\tilde p)>\hat\rho$, then $\tilde\pi(\tilde p)=\pl(\tilde p)$ and the argument follows from $U_{\tilde\rho(\tilde p)}(\pl(\tilde p))<V^{\sigma}(\pl(\tilde p),\tilde p)$.
    \item $\tilde p\in[\ph_4,1]$: Recall $m(\ph)=1$ and $\rho(\ph)=1$ for all $\ph\in[\ph_4,1]$. For every $\ph\geq\ph_4$, we have $\pl(\ph)>\hat p$ and hence $U_{\rho}(p)< U_r(p)$ for all $\rho<1$ and all $p\in \mathcal{P}(\ph)$. Condition \eqref{eq:HJBm} thus requires $\tilde\rho(\ph)=1$ for all $\ph\geq\tilde p$. At $\ph=\ph_4$, $U_r(\pl(\ph))=V^{\sigma}(\pl(\ph),\ph)$, so any $\tilde m(\ph_4)\in[0,1]$ solves the DM's optimization problem. Leaving aside this point of indifference, we want to show that for any $\ph>\ph_4$, $\tilde m(\ph)=m(\ph)=1$ must hold. Assuming this property fails, we must have an open interval $(\tilde p,\tilde p+\varepsilon),\varepsilon>0$ such that $\tilde m(\ph)=0$ for all $\ph\in(\tilde p,\tilde p+\varepsilon)$.\footnote{See  \cref{f:admiss}.} If $\tilde p>\ph_4$, then the DM always ends up choosing $r$, whether a breakthrough occurs or not. This strategy is thus dominated by an immediate stopping with action $r$. Suppose instead $\tilde p=\ph_4$ and $\tilde m(\ph)=0$ for all $\ph\in(\ph_4,\ph_4+\varepsilon)$ and some $\varepsilon>0$. If $\tilde\nu(\ph)=0$ on a right neighborhood of $\tilde p$, the DM's value on that neighborhood is described by
    \[
    V^{\tilde\sigma}(p,\ph)=\Psi(q(\ph))+(p-q(\ph))\Psi'(q(\ph)),
    \]
    with $\Psi$ and $q$ as defined in \eqref{eq:b-value} and \eqref{eq:q}. But, by definition of $\ph_4$ in \eqref{eq:pi4}, $V^{\tilde\sigma}(p,\ph)<U_r(\ph)$ holds for all $\ph>\ph_4=\tilde p$ and $p\in \mathcal{P}(\ph)$, so the strategy violates condition \eqref{eq:opt-stop}. For interior stopping rates $\tilde\nu(\ph)>0$ on a right neighborhood of $\tilde p$, \eqref{eq:HJBm} requires $V^{\tilde\sigma}(\pl(\ph),\ph)=U_r(\pl(\ph))$ for all $\ph$ belonging to that neighborhood, which in turn requires $\tilde m(\ph)=1$ on that neighborhood, a contradiction.
\end{enumerate}

 \noindent\emph{Case 2.} Case 2 is distinguished from Case 1 only by the DM's strategy at state $p_*$ (the boundary between Regions 2 and 3), where the DM mixes instantaneously between experimentation and action $r$. The arguments above clearly apply to Regions 1 and 2. They also apply to Regions 3-5 if we can show $\tilde\sigma(p_*)=\sigma(p_*)$, so that the DM's value at the left boundary of Region 3 has the same properties as in Case 1. To this end, we distinguish two possibilities based on whether the value segment is upward sloping or downward sloping at $p_*$.
 \begin{enumerate}
     \item Suppose $\tilde m(p_*)$ and $\tilde\rho(p_*)$ are such that $V^{\tilde\sigma}_p(p,p_*)<0$. Then $\tilde\pi(p_*)=p_*$. Since $\hat p<p_*$ in Case 2, we have $U_{\rho}(p_*)<U_r(p_*)$ for all $\rho<1$ and hence $\tilde\rho(p_*)=1$. Consider a right neighborhood of $p_*$ and suppose $\tilde m(\ph)=0$ for all $\ph$ belonging to the neighborhood. Under this assumption the value $V^{\tilde\sigma}(p,\ph)$ is right-continuous in $\ph$ at $\ph=p_*$.\footnote{  As we approach $p_*$ from the right, the probability of stopping before $p_*$ vanishes.} For $\ph>p_*$ sufficiently close to $p_*$ we thus have $\tilde\pi(\ph)=\ph$. Since stopping and taking action $r$ is strictly Bayesian optimal for all $\ph>p_*$, we then have $U_r(\tilde\pi(\ph))>V^{\tilde\sigma}(\tilde\pi(\ph),\ph)$ for all $\ph>p_*$ sufficiently close to $p_*$, thus violating \eqref{eq:opt-stop} of \cref{def:pseudo}. Suppose next $\tilde m(\ph)=1$ for all $\ph$ belonging to an interval $(p_*,p_*+\varepsilon),\varepsilon>0$. We distinguish two cases. If $\underline p(p_*)\geq\hat p$, then \eqref{eq:opt-action} implies $\tilde\rho(\ph)=1$ for all $\ph\in(p_*,p_*+\varepsilon)$ and thus $V^{\tilde\sigma}_p(p,p_*)>0$, a contradiction. If $\underline p(p_*)<\hat p$, then there exists an $\varepsilon>0$ such that $\tilde\rho(\ph)=\hat\rho$ and $\tilde\pi(\ph)=\hat p$ for all $\ph\in(p_*,p_*+\varepsilon)$. This possibility is ruled out by the argument in point 3(b) above.
    \item Suppose $\tilde m(p_*)$ and $\tilde\rho(p_*)$ are such that $V^{\tilde\sigma}_p(p,p_*)>0$. Nature's choice thus satisfies $\tilde\pi(p_*)=\pl(p_*)$. If $\tilde\pi(p_*)=\pl(p_*)\leq p_{\ell}^B$, then, given nature's choice, it is Bayesian optimal to stop and take action $\ell$, so we have $\tilde m(p_*)=1$ and $\tilde\rho(p_*)=0$. This implies $V^{\tilde\sigma}(p,p_*)=U_{\ell}(p)$ and thus $V^{\tilde\sigma}_p(p,p_*)<0$, a contradiction. Suppose instead $\tilde\pi(p_*)=\pl(p_*)> p_{\ell}^B$. The DM's value at $p_*$ under strategy $\tilde\sigma$ is now described by 
    \begin{eqnarray*}
        V^{\tilde\sigma}(\tilde\pi(p_*),p_*)=\tilde m(p_*)U_{\tilde\rho(p_*)}(\tilde\pi(p_*))+(1-\tilde m(p_*))\Phi(\tilde\pi(p_*)).
    \end{eqnarray*}
    Since $U_{\tilde\rho(p_*)}(\tilde\pi(p_*))<\Phi(\tilde\pi(p_*))$ for all $\tilde\rho(p_*)\in[0,1]$, this value is strictly decreasing in $\tilde m(p_*)$. We thus have $\tilde m(p_*)=0$, which in turn implies $V^{\tilde\sigma}(p,p_*)=\Phi(p)$ and thus $V^{\tilde\sigma}_p(p,p_*)<0$, again a contradiction.
\end{enumerate}
Taken together, we have shown that $\tilde\sigma$ must be such that 
\begin{eqnarray}\label{eq:deriv0}
V^{\tilde\sigma}_p(p,p_*)=\tilde m(p_*)U_{\tilde\rho(p_*)}'(p)+(1-\tilde m(p_*))\Phi'(p)=0.
\end{eqnarray}
Since $\Phi'(p)<0$ for all $p\in \mathcal{P}(p_*)$, this equality requires $\tilde\rho(p_*)\in[\hat\rho,1]$ and $\tilde m(p_*)\in[m(p_*),1]$.\footnote{Recall that $m(p_*)$ is determined by
\[
m(p_*)U_{r}'(p_*)+(1-m(p_*))\Phi'(p_*)=0.
\]
} For the DM to optimally choose $\tilde m(p_*)\geq m(p_*)$, it must further hold $U_{\tilde\rho(p_*)}(\tilde\pi(p_*))=\Phi(\tilde\pi(p_*))$. Since for all $p\in \mathcal{P}(p_*)$ and all $\tilde\rho(p_*)\geq\hat\rho$, we have $U_{\tilde\rho(p_*)}(p)\leq\max\{\hat u,U_r(p)\}\leq\Phi(p)$, with the second equality being strict unless $p=p_*$ and the first inequality  being strict at $p=p_*$ unless $\tilde\rho(p_*)=1$. Hence, $U_{\tilde\rho(p_*)}(\tilde\pi(p_*))=\Phi(\tilde\pi(p_*))$ holds if and only if $\tilde\rho(p_*)=\rho(p_*)=1$ and $\tilde\pi(p_*)=\pi(p_*)=p_*$. The stopping probability $\tilde m(p_*)$ is then pinned down by \eqref{eq:deriv0} and equal to $m(p_*)$. We thus have $\tilde\sigma(p_*)=\sigma(p_*)$. Given this property, the arguments used to prove uniqueness in Case 1, in particular those regarding Region 3, apply to the current case.\\

\noindent\emph{High experimentation costs.} When $c\geq\ch$, uniqueness of the intrapersonal equilibrium follows from the argument in the main text following \cref{thm:main_result}, namely the facts that the commitment solution involves no experimentation and that implementing this solution requires no commitment. Combining these properties with the fact that the commitment solution is unique (see \cref{app:commitment}) implies that under any alternatively strategy there must be a state at which the DM can profitably deviate to the commitment solution.

\subsection{Proof of \cref{prop:Knightian}}  \label{sec:proof-Knightian}

Consider first $c<\cl$. The candidate strategy profile has stationary actions $\tilde \s=(0,\tilde\nu, 0)$ and  $\p= \tilde{p}:=\frac{u_{\ell}^{L}-\tilde{u}}{u_{\ell}^{L}-u_{\ell}^{R}}$,  where $\tilde u$ is defined in the statement. The associated value of the stationary strategy is $V^{\tilde\s}(p)=\tilde u$ for all $p$.\footnote{That this is the correct value of $\tilde \s$ can be seen by confirming that $\tilde u= \frac{\tilde\nu U_{\ell}(\tilde p)+ \tilde p\lambda u_r^R}{\tilde\nu  + \tilde p\lambda}.$} 
We observe that  $\tilde u> U_r(\hat p)$ and $\tilde p< \hat p$ if and only if $c<\cl$.
Hence, for $c<\cl$, we have   $U_{\ell}(\tilde{p})=\tilde{u}\ge U_{r}(\tilde p)$.  It then follows  
$$
(0,\tilde \nu, 0) \in\argmax_{(m,\nu, \rho)} m (U_{\rho}(\tilde{p})-\tilde{u}) +(1-m) \left[-c+ \nu(U_{\rho}(\tilde{p})-\tilde{u})+\lambda\tilde{p}(u_{r}^{R}-\tilde{u})\right],
$$
which proves that  both \cref{eq:HJB0} and \cref{eq:HJBm} are satisfied.  Next, given $\tilde \nu$, the derivative of 
\[
\tilde\nu(U_{\rho}(p)-\tilde{u})+\lambda p (u_{r}^{R}-\tilde{u})
\]
with respect to $p$ vanishes, so  
\[
\tilde p\in \argmin_{p} \tilde m (U_{\tilde\rho}({p})-\tilde{u}) +(1-\tilde m) \left[-c+ \tilde\nu(U_{\tilde\rho}( {p})-\tilde{u})+\lambda {p}(u_{r}^{R}-\tilde{u})\right],
\]
where $(\tilde m, \tilde \rho)=(0,0).$  We have thus verified   \cref{eq:HJBpi}.  
 
Consider next $c\ge \cl$. As noted above, we have $\tilde u\le \hat u$ in this case. The proof is then identical to that of \cref{ssec:large-c}, which applies here since $\Delta=\infty$ if $[\pl_0, \ph_0]=[0,1]$.   

To see that the equilibrium is the limit of the equilibrium strategies as $(\pl_0, \ph_0)\to (0,1)$, note that for $c<\cl$ and $(\pl_0, \ph_0)$ sufficiently close to $(0,1)$, $\ph_0$ falls into Region 3. Indeed, as $(\pl_0,\ph_0)\to(0,1)$, $\ph_3$ rises to 1 faster than $\ph_0$ does. To see this, suppose $(\pl_0,\ph_0)$ is sufficiently close to $(0,1)$  so that $\ph_0>p_*=\ph_2$ and $\pl_0<p_{\ell}^B$.  Recall  $\ph_3$ is given by $V^{\sigma}(\pl(\ph_3),\ph)=U_{\ell}(\pl(\ph_3))$, that is, as the value of $\ph$ for which the left end of the value segment touches $U_{\ell}$. Since $V^{\sigma}(\pl(\ph_3),\ph_3)$ lies below the Bayesian value function, we clearly have $\pl(\ph_3)\geq p_{\ell}^B$.  Since $\pl_0=\pl(\ph_0)<p_{\ell}^B$, $\ph_0$ is strictly smaller than $\ph_3$ (and of course strictly greater than $p_*$). Given that $\ph_0$ falls into Region 3, it is then routine to check that $(\nu(\ph_0), \p(\ph_0))\to (\tilde\nu, \tilde p)$ as $\ph_0\to 1$.  Similarly, when $c\ge \cl$, the same conclusion follows since the condition for \cref{ssec:large-c} is satisfied for $\ph_0$ sufficiently close to 1.

\subsection{Proof of \cref{prop:random}}

In the symmetric case, the Bayesian value function can be written as
\[
\Phi(p)=\delta-\frac{c}{\lambda}-(1-p)\ln\left(\frac{p}{1-p}\frac{\delta-c/\lambda}{c/\lambda}\right)\frac{c}{\lambda}
\]
with derivative
\[
\Phi'(p)=\left(\ln\left(\frac{p}{1-p}\frac{\delta-c/\lambda}{c/\lambda}-\frac{1}{p}\right)\right)\frac{c}{\lambda}.
\]

It is easy to see that $\lim_{p\uparrow1}\Phi'(\ph)=+\infty$. For all states $\ph\in(p_{\ell}^B,p_*]$ experimentation until the Bayesian update of $\ph$ reaches the Bayesian stopping boundary $p_{\ell}^B$ (or $R$-evidence arrives) is clearly optimal: by the assumption $c<\ch$, the worst-case payoff associated to this strategy, $\Phi(\ph)$, is strictly greater than the worst-case payoff for action $\ell$, given by $U_{\ell}(\pl(\ph))$ and strictly greater than the worst-case payoff for action $r$, given by $U_r(\ph)$. 

Fixing this part of the strategy and moving backwards in time from $\ph=p_*$, the value segment becomes upward sloping and the experimentation strategy is evaluated at the left-most belief. The DM's value at this belief lies on the tangency line of $\Phi$ touching at $\ph$, given by
\[
l(p;\ph)=\Phi(\ph)+\Phi'(\ph)(p-\ph).
\]
Convexity of $\Phi$ implies that the value $l(0,\ph)$ is strictly decreasing in $\ph$ with $\lim_{\ph\uparrow1}l(0;\ph)=-\infty<U_{\ell}(p)$ for all $p$. Given $l(0,p_*)=\Phi(p_*)>U_{\ell}(p_*)$, the difference $U_{\ell}(\ph)-l(0;\ph)$ as a function of $\ph$ must then have an intersection with zero. Notice next that for any given value of $\ph$, we can find a $\Delta$ sufficiently large such that $\pl(\ph)$ is arbitrarily close to zero. Together with the previous property, this implies that for $\Delta$ sufficiently large, there is a state $\ph>p_*$ such that $l(\pl(\ph);\ph)=U_{\ell}(\ph)$. Let $\ph_s$ denote the smallest of such states. For all $\ph\in(\p_{\ell}^B,\ph_s)$, the DM will then experiment, whereas at $\ph=\ph_s$ she stops to take action $\ell$. \footnote{For $\pl(\ph_s)$ sufficiently close to zero, we must have $U_{\ell}(\ph_s)\geq U_r(\pl(\ph_s))$, so action $\ell$ is indeed optimal.}

What remains to be shown is that there is an $\varepsilon>0$ sufficiently small such that the DM prefers experimentation over stopping at all states $\ph\in(\ph_s,\ph_s+\varepsilon)$. The payoff from experimentation at $\ph>\ph_s$ evaluated at belief $p$ is\footnote{Recall that $\tau(\ph,\ph_s)$ denotes the time is takes for belief $\ph>\ph_s$ to be updated to $\ph_s$ in the absence of a breakthrough.}
\[
p\left(1-e^{-\lambda\tau(\ph,\ph_s)}\right)\delta+(1-p)\delta-c\tau(\ph,\ph_s).
\]
As $\ph\downarrow\ph_s$, this payoff converges to $(1-p)\delta$. Hence, for $\ph$ sufficiently close to $\ph_s$, the worst-case belief is $\ph$. According to belief $\ph$ experimentation for time $\tau(\ph,\ph_s)$ followed by action $\ell$ is preferred to taking action $\ell$ immediately if and only if 
\begin{align}\label{eq:exp-eps}
\ph\left(1-e^{-\lambda\tau(\ph,\ph_s)}\right)\delta-c\tau(\ph,\ph_s)>0,
\end{align}
where
\[
\tau(\ph,\ph_s)=\frac{1}{\lambda}\ln\left(\frac{\ph}{1-\ph}\frac{1-\ph_s}{\ph_s}\right).
\]
It can be easily verified that the left-hand side of \eqref{eq:exp-eps} approaches zero from above as $\ph\downarrow\ph_s$ if and only if $\ph_s>p_{\ell}^B$, which is satisfied since $\ph_s>p_*>p_{\ell}^B$.

\subsection{Proof of \Cref{prop:prolong}}

It is useful to characterize the learning time through the following recursion equation.\footnote{This equation is derived as follows.  Take a short length $dt>0$ of time, then 
\begin{align*}
    T(\t;\Delta) & = (\lambda \theta+ \nu_{\Delta}(\theta))dt \cdot dt +(1-(\lambda \theta+ \nu_{\Delta}(\theta))dt )(dt + T(\t_{dt})) +o(dt)\\
    &= (\lambda \theta+ \nu_{\Delta}(\theta))dt \cdot dt +(1-(\lambda \theta+ \nu_{\Delta}(\theta))dt )(dt + T(\t;\Delta)+ T'(\t;\Delta) \dot \theta dt) +o(dt). 
\end{align*}
 Collecting terms, and letting $dt\to 0$ while using $\dot\t=-\lambda \theta(1-\theta)$, we obtain the equation.}  
$$\lambda \theta (1-\theta) T'_{\Delta}(\theta)= 1 - T_{\Delta}(\theta)(\theta\lambda + \nu_{\Delta}(\theta)),$$
where $\nu_{\Delta}(\theta)$ is the stopping rate for ambiguity level $\Delta$ when the state $\ph$ is such that $\ln\left(\frac{\ph}{1-\ph}\right)=\ln\left(\frac{\theta}{1-\theta}\right)+\frac{\Delta}{2}$. Using the corresponding equation for $\Delta'>\Delta$, one can write:
$$\lambda \theta (1-\theta)[T'_{\Delta}(\theta)- T'_{\Delta'}(\theta)]  = T_{\Delta'}(\theta)(\theta\lambda + \nu_{\Delta'}(\theta)) - T_{\Delta}(\theta)(\theta\lambda + \nu_{\Delta}(\theta)).$$
For any $\theta$ such that $\ln\left(\frac{\theta}{1-\theta}\right)+\frac{\Delta'}{2}\le p_*$, we have $\nu_{\Delta'}(\theta)=\nu_{\Delta}(\theta)=0$, and $T_{\Delta'}(\theta)\ge  T_{\Delta}(\theta)$, with the inequality being strict whenever $T_{\Delta'}(\theta)>0$.  This proves that there exists a $\hat \theta$  such that $T_{\Delta'}(\theta)\geq T_{\Delta}(\theta)$ if $\theta<\hat \theta$. The above equation implies that for such  $\theta$, we have $T'_{\Delta'}(\theta)\leq T'_{\Delta}(\theta)$. 

To see the only if part, observe first that $\nu_{\Delta}(\theta)\le \nu_{\Delta'}(\theta)$ for all $\theta$, which follows from the fact that $\ln\left(\frac{\theta}{1-\theta}\right)+\frac{\Delta'}{2}>\ln\left(\frac{\theta}{1-\theta}\right)+\frac{\Delta}{2}$ and from the equation determining $\nu$ in Region 3.  This means that whenever $T_{\Delta'}(\theta)=T_{\Delta}(\theta)$, $T'_{\Delta}(\theta)\ge T'_{\Delta'}(\theta)$; namely, $T_{\Delta'}$ crosses $T_{\Delta}$ at most once and from above.

\subsection{Proof of \cref{prop:CS}}

We indicate by $\ph_n$ and $\qh_n$ for $n=1,...,4$ the boundaries of the different regions when the initial sets of priors are, respectively, $\mathcal{P}$ and $\mathcal{Q}$. Let $\bar t_{\mathcal{P}}$ denote the supremum of $\text{Supp}(F_{\mathcal{P}})$, i.e., the latest time the DM stops.

Since $F_{\mathcal{P}}(t)=1$ for all $t>\bar t_{\mathcal{P}}$, the property $F_{\mathcal{Q}}(t)\leq F_{\mathcal{P}}(t)$ for all $t>\bar t_{\mathcal{P}}$ is trivially satisfied. What remains to be shown is thus $F_{\mathcal{Q}}(t)\geq F_{\mathcal{P}}(t)$ for all $t<\bar t_{\mathcal{P}}$. Clearly, if $\mathcal{P}$ is such that the DM stops deterministically, this property is satisfied, as $F_{\mathcal{P}}(t)=0$ for all $t<\bar t_{\mathcal{P}}$. We thus focus on the case where the DM randomizes under $\mathcal{P}=[\pl_0,\ph_0]$, which requires $\ph_0\in(p_*,\ph_4)$ as well as $\Phi(p_*)<U_{\ell}(\hat p(\tau(\ph_0,p_*),\pl_0))$.

We show first $\tau(\ph_0,\ph_3)\geq\tau(\qh_0,\qh_3)$. Towards a contradiction, suppose $\tau(\ph_0,\ph_3)<\tau(\qh_0,\qh_3)$, so the intrapersonal; equilibrium under $\mathcal{P}$ prescribes randomization at time $\tau(\qh_0,\qh_3)$. By \cref{lem:properties_region3}, $\hat V(\ph)$ is strictly decreasing in $\ph$ in Region 3. Recalling that $p^{\delta}$ denotes the $\delta$ update of $p$,\footnote{That is, $p^{\delta}=\frac{pe^{-\lambda\delta}}{pe^{-\lambda\delta}+1-p}$} we have
\begin{eqnarray*}
U_{\ell}(\pl_0^{\tau(\qh_0,\qh_3)})&\geq& \hat V(\ph_0^{\tau(\qh_0,\qh_3)})>\hat V(\ph_0^{\tau(\ph_0,\ph_3)})>\hat V(\qh_0^{\tau(\qh_0,\qh_3)})=U_{\ell}(\ql_0^{\tau(\qh_0,\qh_3)}).
\end{eqnarray*}
The first inequality follows from the optimality of randomization at time $\tau(\qh_0,\qh_3)$, the second inequality is due to $\tau(\ph_0,\ph_3)<\tau(\qh_0,\qh_3)$, and the third inequality follows from $\hat V(\ph_3)>\hat V(\qh_3)$ (since $\ph_3<\qh_3$). Since, however, $U_{\ell}(\cdot)$ is strictly decreasing  and $\pl_0\geq\qh_0$, $U_{\ell}(\pl_0^{\tau(\qh_0,\qh_3)})\leq U_{\ell}(\ql_0^{\tau(\qh_0,\qh_3)})$ must hold, which yields a contractions. Hence, $\tau(\ph_0,\ph_3)\geq\tau(\qh_0,\qh_3)$.

Next, we prove that for each $t<\bar t_{\mathcal{P}}$, the DM stops at a smaller rate when her initial set of priors is $\mathcal{P}$ than when it is $\mathcal{Q}$. For each $t\geq0$, denote by $\hat\nu(t;\mathcal{P}):=\nu(\ph^{t};\mathcal{P})$ the DM's stopping rate as a function of time. We want to show for all $t\leq \bar t_{\mathcal{P}}$,
\begin{equation}\label{eq:nu-t}
\hat\nu(t;\mathcal{P})\leq\hat\nu(t;\mathcal{Q}).
\end{equation}
When $t<\tau(\ph_0,\ph_3)$ or $t\in(\tau(\ph_0,\ph_*),\bar t_{\mathcal{P}}]$, $\hat\nu(t;\mathcal{P})=0$, so \eqref{eq:nu-t} trivially holds. Considering $t\in[\tau(\ph_0,\ph_3),\tau(\ph_0,p_*)]$, we have
\begin{align}\label{eq:nu-t-d}
\hat\nu'(t;\mathcal{Q})=\nu'(\hat p(t,\mathcal{Q}))\frac{d\hat p(t,\qh_0)}{dt}<0,
\end{align}
since for all $\ph$ belonging to Region 3, $\nu'(\ph)=\lambda\hat V'(\ph)/\delta^{\ell}<0$ (see \eqref{eq:nu-r3}). \eqref{eq:nu-t-d} directly implies
\[
\hat\nu(t;\mathcal{P})=\nu(\hat p(t,\ph_0))=\nu(\hat p(t+\tau(\qh_0,\ph_0),\qh_0))=\hat\nu(t+\tau(\qh_0,\ph_0);\mathcal{Q})\leq \hat\nu(t;\mathcal{Q})
\]
and thus \cref{eq:nu-t}. Since \cref{eq:nu-t} holds for all $t<\bar t_{\mathcal{P}}$, it follows
\begin{eqnarray*}
F_{\mathcal{P}}(t)&=&1-e^{-\int_0^t\hat\nu(u;\mathcal{P})du}\leq1-e^{-\int_0^t\hat\nu(u;\mathcal{Q})du}=F_{\mathcal{Q}}(t)
\end{eqnarray*}
for all $t<\bar t_{\mathcal{P}}$, as desired.

\section{General Poisson Learning}  \label{sec:general-poisson}

In this extension, we consider a general Poisson model introduced by \cite{Che2019}: at each instant, the DM may seek either $R$-evidence (as before) or $L$-evidence.  One interpretation is that there are two news sources emitting the two types of evidence. The DM may then allocate a share $\alpha\in [0,1]$ of her attention to the $R$-evidence news source and a share $1-\alpha$  to  the $L$-evidence news source, and receive the evidence proportionately at rates $\alpha \lambda$ in state $R$ and at rates $(1-\alpha) \lambda$ in state $L$. For instance, a theorist may try either to ``prove'' a theorem ($R$-evidence),  to find a ``counter-example'' disproving it ($L$-evidence), or to divide effort between the two endeavors. We assume throughout that the stopping payoffs are symmetric:  $u^R_r=u^L_{\ell}=\delta>0$ and $u^R_{\ell}=u^L_{r}=0$.

\subsection{Bayesian optimal strategy.}

\begin{figure}[htb]
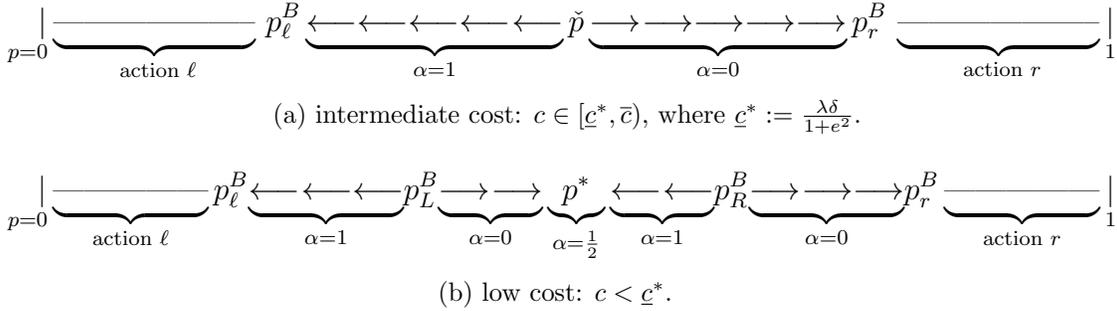


\noindent \begin{centering}
\[
\underset{p=}{\phantom{|}}\underset{0}{|}\underbrace{\text{\ensuremath{\vphantom{p^{*}}}--------------------}}_{\text{ action }\ell}\,p^B_{\ell}\underbrace{\longleftarrow\longleftarrow\longleftarrow\longleftarrow\vphantom{p^{*}}\negthickspace\negmedspace\longleftarrow}_{\alpha=1}\check{p}\underbrace{\longrightarrow\vphantom{p^{*}}\negthickspace\negmedspace\longrightarrow\longrightarrow\longrightarrow\longrightarrow}_{\alpha=0}p^B_{r}\,\underbrace{\text{\ensuremath{\vphantom{p^{*}}}--------------------}}_{\text{ action }r}\underset{1}{|}
\]
\par\end{centering}
\noindent \begin{centering}
{\footnotesize{}{}{}{}{}{}{}(a) intermediate cost:  $c\in [\cl^*, \ch)$, where $\cl^*:= \frac{\lambda \delta}{1+e^2}$.} 
\par\end{centering}

\noindent \begin{centering}
\[
\underset{p=}{\phantom{|}}\underset{0}{|}\underbrace{\text{\ensuremath{\vphantom{p^{*}}}---------------}}_{\text{action }\ell}p^B_{\ell}{\underbrace{\longleftarrow\vphantom{p^{*}}\negthickspace\negmedspace\longleftarrow\longleftarrow}_{\alpha=1}}p^B_{L}{\underbrace{\longrightarrow\vphantom{p^{*}}\negthickspace\negmedspace\longrightarrow}_{\alpha=0}\underbrace{p^{*}}_{\alpha=\frac{1}{2}}\underbrace{\longleftarrow\vphantom{p^{*}}\negthickspace\negmedspace\longleftarrow}_{\alpha=1}}p^B_{R}{\underbrace{\longrightarrow\vphantom{p^{*}}\negthickspace\negmedspace\longrightarrow\longrightarrow}_{\alpha=0}}p^B_{r}\underbrace{\text{\ensuremath{\vphantom{p^{*}}}---------------}}_{\text{action }r}\underset{1}{|}
\]
{\footnotesize{}{}{}{}{}{}{}{}(b) low cost:   $c<\cl^*$.} 
\end{centering}
\centering{}\protect\caption{\label{fig:structure_of_optimal_solution} Bayesian optimal strategy.}
\end{figure}

Consider the Bayesian optimal strategy,  assuming $c<\ch=\frac{\lambda \delta}{2}$ so that experimentation is optimal for some beliefs.\footnote{Otherwise, the DM simply chooses action $a=r,\ell$ that maximizes $U(p)$.} 
The optimal strategy, characterized in \cite{Che2019}, is succinctly described in \Cref{fig:structure_of_optimal_solution} for two different ranges of learning costs. Suppose first the cost is {\it intermediate}; i.e., $c\in [\cl^*, \ch)$, where $\cl^*:= \frac{\lambda \delta}{1+e^2}$.  Then, the DM seeks contradictory evidence; namely she chooses $\alpha=0$ if $p\in (1/2, p_r^B)$ and $\alpha=1$ if $p\in (p_{\ell}^B, 1/2)$.  See panel (a) of  \Cref{fig:structure_of_optimal_solution}.

For a {\it low cost} $c<\cl^*$, the contradictory-evidence seeking is still optimal near the stopping boundaries, as before, but a new learning strategy emerges in the middle region.  In that region, the DM seeks {\it confirmatory evidence}, $L$-evidence for $p\in (p_L^B, 1/2)$ and $R$-evidence for $p\in (1/2, p_R^B)$; see panel (b) of \Cref{fig:structure_of_optimal_solution}.   When such evidence does not arrive, the DM's belief drifts inwardly toward $p^*=1/2$.  Once $p^*=1/2$ is reached, the DM splits her attention between the two news sources with $\alpha=1/2$; her belief then never moves, and learning  continues indefinitely until either evidence is obtained. Intuitively, at that belief, the DM finds both types of experimentation equally tempting and acts like   a ``Buridan's donkey,'' unable to drift away from the belief that causes the dilemma. The value of this {\it split-attention} learning can be computed:
$$
u^*= \frac{1}{2}u^R_r+ \frac{1}{2}u^L_{\ell}- \frac{2c}{\lambda}= \delta- \frac{2c}{\lambda}.
$$

With a little abuse of notation, let $\Phi^*(p)$ denote the value of the Bayesian optimal strategy for belief $p$. Given the symmetry of payoffs, $\Phi^*$ is symmetric around, and attains its minimum at, $1/2$.  One implication of the above characterization is that $\Phi^*(1/2)=u^*$ for $c\le \cl^*$ and $\Phi^*(1/2)>u^*$ for $c>\cl^*$, which explains why splitting attention is part of Bayesian optimal strategy if and only if $c<\cl^*$.
   
\subsection{Ambiguity aversion.} 

Suppose now the DM is ambiguity averse, endowed with an interval of priors.  How would the presence of the additional news source affect her behavior? Will see that the solution exhibits several patterns of behavior established from the baseline model as well as some additional features that are linked to the DM's imformation choice in the generalized model.\\

\noindent\emph{Low cost or small ambiguity.} Consider  first  the case in which either the cost is  low with $c< \cl^*$ or the ambiguity is small.   As before, a ``small'' ambiguity means that $\Phi^*(\frac{1}{2})\ge U_{\ell}(\pl(\frac{1}{2}))$, where we recall that $\pl(\ph)$ denotes the lowest possible belief when the state is $\ph$. Below, we suppress the dependence of $\pl$ on $\ph$ when no confusion arises.  The equilibrium for this case is characterized as follows.

 \begin{theorem} \label{thm:2news-L}  Suppose the cost is low or the ambiguity is small. The intrapersonal equilibrium of the ambiguity-averse DM is as follows.
 \begin{itemize}
 \item If $\ph\le 1/2$, the DM employs the Bayesian optimal strategy of type $\ph$, with the worst-case belief $\pi(\ph)=\ph$.
 
 \item If $\pl\ge 1/2$,  the DM employs the Bayesian optimal strategy of type $\pl$, with the worst-case belief $\pi(\ph)=\pl$. 
 
 \item If $\pl< 1/2< \ph$, then the DM employs the split-attention learning strategy or a hedged action, whichever gives a higher payoff, with the worst-case belief $\pi(\ph)=1/2$.
 \end{itemize}
 
 \end{theorem}

\begin{proof}
The proof is completely analogous to the baseline case for either $\ph<p_*=1/2$ or $\pl>p_*$.  Specifically, consider the former case. With $\pi(\ph)=\ph$, the DM's HJB coincides with that of the Bayesian DM for belief $\ph$, so the result follows from \cite{Che2019}.  For nature's optimality, since the Bayesian value function is decreasing, clearly $\pi(\ph)=\ph$ is optimal.  For nature's HJB, its first-order condition with respect to $p$ reduces to the second derivative of the Bayesian HJB, which of course is zero, so the condition holds. 
  
Next, consider the case in which $\pl<1/2<\ph$. Nature chooses $\pi(\ph)=1/2$. If $u^*<\max\{U_{\ell}(1/2), U_{r}(1/2)\}$, then randomizing between $r$ and $\ell$ with equal probabilities satisfies the HJB conditions. Given this, the value function $V(p, \ph)=\hat u=\frac{1}{2}\delta$ for all $p$, so nature does not wish to deviate from $\pi(\ph)=1/2$. 
If $u^*\ge \max\{U_{\ell}(1/2), U_{r}(1/2)\}$, the HJB conditions hold with $\nu=0$ and $\alpha=1/2$, so split-attention learning is optimal.  Given this choice, since $V(p, \ph)=u^*$ for all $p$, nature does not wish to deviate from $\pi(\ph)=1/2$. 
\end{proof}

\begin{figure}
  \centering
\begin{subfigure}{.45\textwidth}
  \centering
  \includegraphics[width=.9\textwidth]{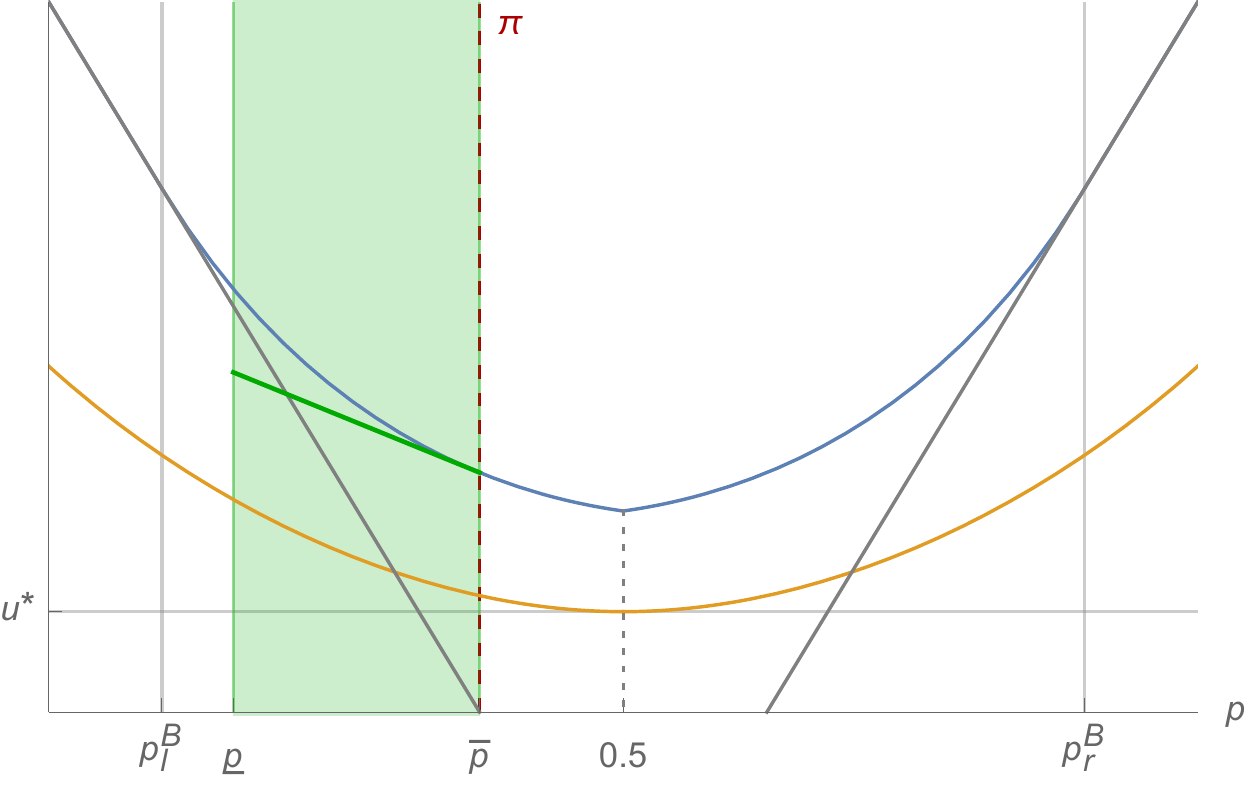}
\end{subfigure}
\begin{subfigure}{.45\textwidth}
    \centering
  \includegraphics[width=.9\textwidth]{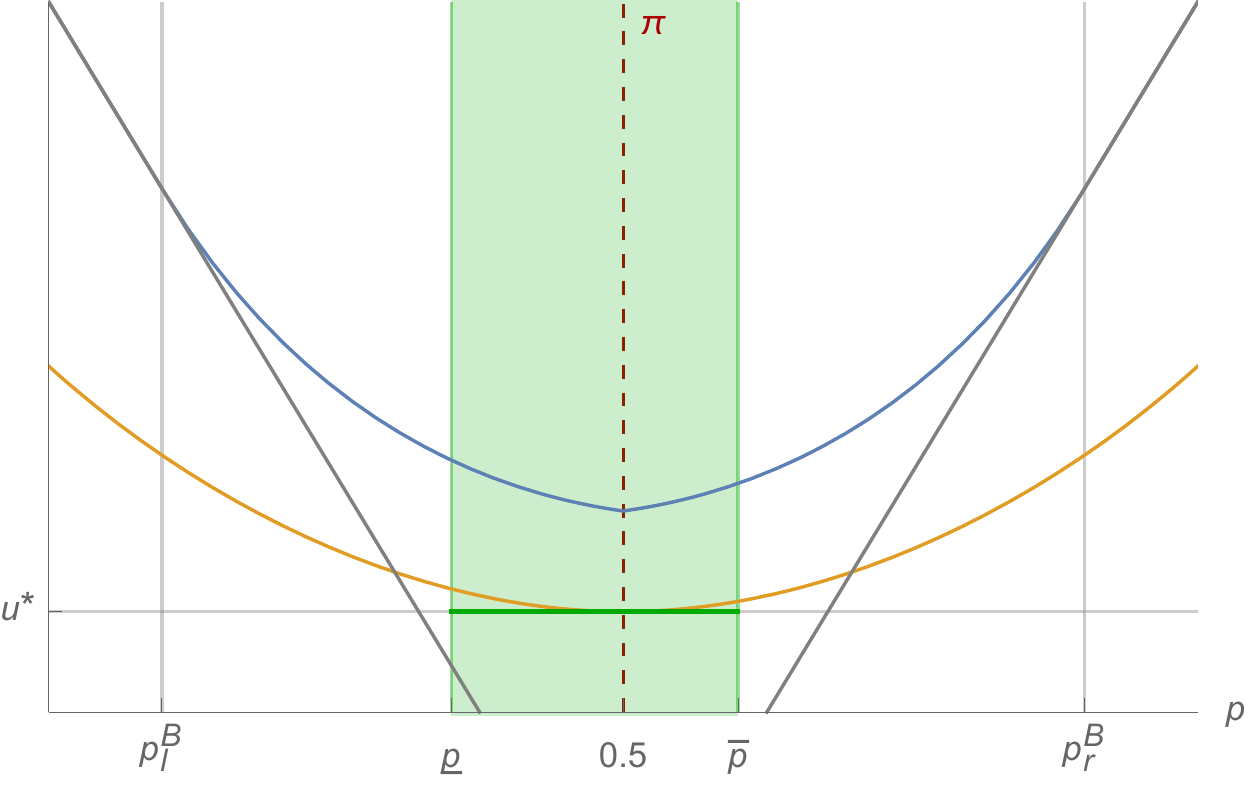}
\end{subfigure}
\caption{The case of intermediate cost and small ambiguity.}
\label{fig:2news1}
\end{figure}

To explain, suppose first the highest belief $\ph$ is less than $1/2$. Then, as before, the worst-case belief is the right-most belief $\ph$, and the DM adopts the Bayesian optimal strategy for that belief. See the left panel of \Cref{fig:2news1}.  In case $c\in [\cl^*, \ch)$, the worst-case belief $\ph$ drifts leftward, and the experimentation lasts until the belief reaches the left stopping boundary. In other words, \Cref{thm:main_result} applies precisely in this case with the same implication---including prolonged learning---as before.  In case $c<\cl^*$, the belief $\ph$, actually drifts rightward when $\ph>p^B_L$. Once it reaches $p^*=1/2$, split-attention learning begins, and it lasts until the state is learned.

The case where the lowest belief $\pl$ is greater than $1/2$ is the mirror image of the preceding case. The worst-case belief is now the left-most belief $\pl$, and the DM adopts the Bayesian optimal belief.  Equilibrium is precisely the same as above, except that the type of evidence the DM seeks is the opposite to the above case.

Finally, suppose $\pl< 1/2< \ph$.  In this case, the Bayesian optimal strategy for $p\ne 1/2$ cannot occur. To see this, suppose the worst-case belief is  $\pi>1/2$. Then, the value of the Bayesian optimal strategy for belief $\pi$---i.e., the value segment tangent to $\Phi^*(\cdot)$ at $p=\pi$---is increasing in the belief, so the worst-case belief is $\pl$, which, however, is strictly less than $1/2$.  The worst-case belief cannot be $\pl$ either, since, if so, the value of the Bayesian optimal strategy for $\pl$ is decreasing in the belief, suggesting that the worst-case belief is $\ph$, which now exceeds $1/2$. 

This dilemma is resolved by a form of hedging.  If $u^*>\hat u$, then split-attention learning serves that purpose, with the worst-case belief $\pi=1/2$.  As can be seen in the right panel of \Cref{fig:2news1}, the value segment touches the Baysian value function associated with confirmatory evidence seeking (orange curve) and is flat, justifying the interior choice of the worst-case belief.  Note  that the belief is absorbing since no updating from $p=1/2$ can occur; in short, once the ambiguity set of beliefs $\mathcal{P}$ contains $1/2$, the set never moves, and the DM is destined forever to split her attention between two news sources.

If the cost is low, then the split-attention learning is Bayesian optimal at belief $p=1/2$, so the strategy is relatively cheap; this is why it prevails as the dominant form of hedging no matter how large the ambiguity is.  If the cost is intermediate with  $c\in [\cl^*, \ch)$, then the split-attention payoff $u^*$ is strictly below the Bayesian value function, so hedging in this way is relatively costly. In fact, the Bayesian DM would never adopt split-attention learning in this case (see \Cref{fig:structure_of_optimal_solution}-(a)).  This explains the discontinuity in the right panel of \Cref{fig:2news1}.\\

\noindent\emph{Intermediate cost and large ambiguity.}  Suppose next that the cost is {\it intermediate}, namely $c\in [\cl^*, \ch)$, and the ambiguity is {\it large} in the sense that $U_{\ell}(\pl(\frac{1}{2}))> \Phi^*(\frac{1}{2})$. Recall the value  $\hat V(\ph)$  defined in \Cref{thm:main_result} for the randomized stopping region (Region 3).  This value solves   \cref{eq:ODE_Vhat} with the boundary condition $(\ph_2, \Phi^*(\ph_2))=(\frac{1}{2}, \Phi^*(\frac{1}{2}))$ and is explicitly  expressed in \cref{eq:Vhat_general_solution} (with $\ph_2=\frac{1}{2}$). Let
$p_+ := \hat V^{-1}(u^*\vee \hat u)$ and $p_-:= 1-p_+$. We note that $0<p_-<1/2< p_+<1$. The intra-personal equilibrium is then characterized as follows.

\begin{theorem} \label{thm:2news-Int-stop}  Suppose the cost is intermediate, and the ambiguity is large.   Then, the following is an equilibrium.
 \begin{itemize}
 \item For any $\ph\le 1/2$, the DM employs the Bayesian optimal strategy for belief $\pi(\ph)=\ph$. 
 
 \item For any $\pl\ge 1/2$, the DM employs the Bayesian optimal strategy for belief $\pi(\ph)=\ph$. 
 
 \item Suppose $[\pl, \ph]\supset [p_-,p_+]$.  Then, the DM hedges either by employing the split-attention learning or by mixing between $r$ and $\ell$, with the worst-case belief $\pi(\ph)=1/2$.
 
 \item Suppose $\pl<1/2$  and $\ph\in (1/2, p_+]$.  Then, the DM randomizes between $\ell$ and $R$-evidence seeking.  Namely, she stops according to a Poisson rate or else she seeks $R$-evidence, as characterized in \Cref{thm:main_result} and \Cref{proof:main}.
 
 \item Suppose $\pl\in [p_-, 1/2)$  and $\ph>1/2$.  Then, the DM randomizes between $r$ and $L$-evidence seeking.  Namely, she stops according to a Poisson rate or else she seeks $L$-evidence, as characterized in \Cref{thm:main_result} and \Cref{proof:main}.\footnote{Given the symmetric payoffs, the characterization is a mirror image of the preceding case.  More precisely, with the state now indexed by $\pl$, the stopping rate is $\nu^*(1-\pl)$ and the worst-case belief is $\pi^*(1-\pl)$, where $\nu^*$ and $\pi^*$ are defined respectively in \cref{eq:nu-r3} and \cref{eq:pi-r3}, respectively. }
 \end{itemize}
 \end{theorem}

\begin{proof} The proof for all cases, except for the last two are the same as above. Of the latter two, since the case of $\pl\in [p_-, 1/2)$ is an exact mirror image of the case $\ph\in (1/2, p_+]$, we simply focus on that case.  This case in turn corresponds precisely to Region 3 of \Cref{thm:main_result}.  The construction of $(\nu(\ph), \pi(\ph), \hat V(\ph))$ follows exactly without any modification. The verification of HJB is also the same except that $\alpha(\ph)=1$ needs to be verified now; namely, we additionally need to show that the DM wishes to experiment by confirming state $R$ (rather than state $L$).  To this end, we take the derivative of $G$ with respect to $\alpha$: 
$$\frac{\partial G}{\partial \alpha}\bigg|_{\alpha=1} =\pi (u_r^R-\hat V(\ph))-(1- \pi)(u_{\ell}^L-\hat V(\ph))-2\hat V'(\ph)\ph(1-\ph).$$
 It suffices to show that this derivative is nonnegative when  evaluated at $(\nu(\ph), \pi(\ph), \hat V(\ph))$.
Indeed, when 
substituting $(\nu(\ph), \pi(\ph), \hat V(\ph))$  from  \cref{eq:nu-r3}, \cref{eq:pi-r3}, and \cref{eq:ODE_Vhat}, this expression is nonnegative if and only if 
$$\hat V(\ph)\ge u^*,$$
which holds since $\ph\le p_+$.
\end{proof}

If $\ph\le 1/2$ or $\pl\ge 1/2$, then just as before, the DM employs the Bayesian optimal strategy for the inner-most belief, which is to seek contradictory evidence until the most skeptical belief reaches a stopping boundary.  

\begin{figure}
  \centering
\begin{subfigure}{.49\textwidth}
  \centering
  \includegraphics[width=.9\textwidth]{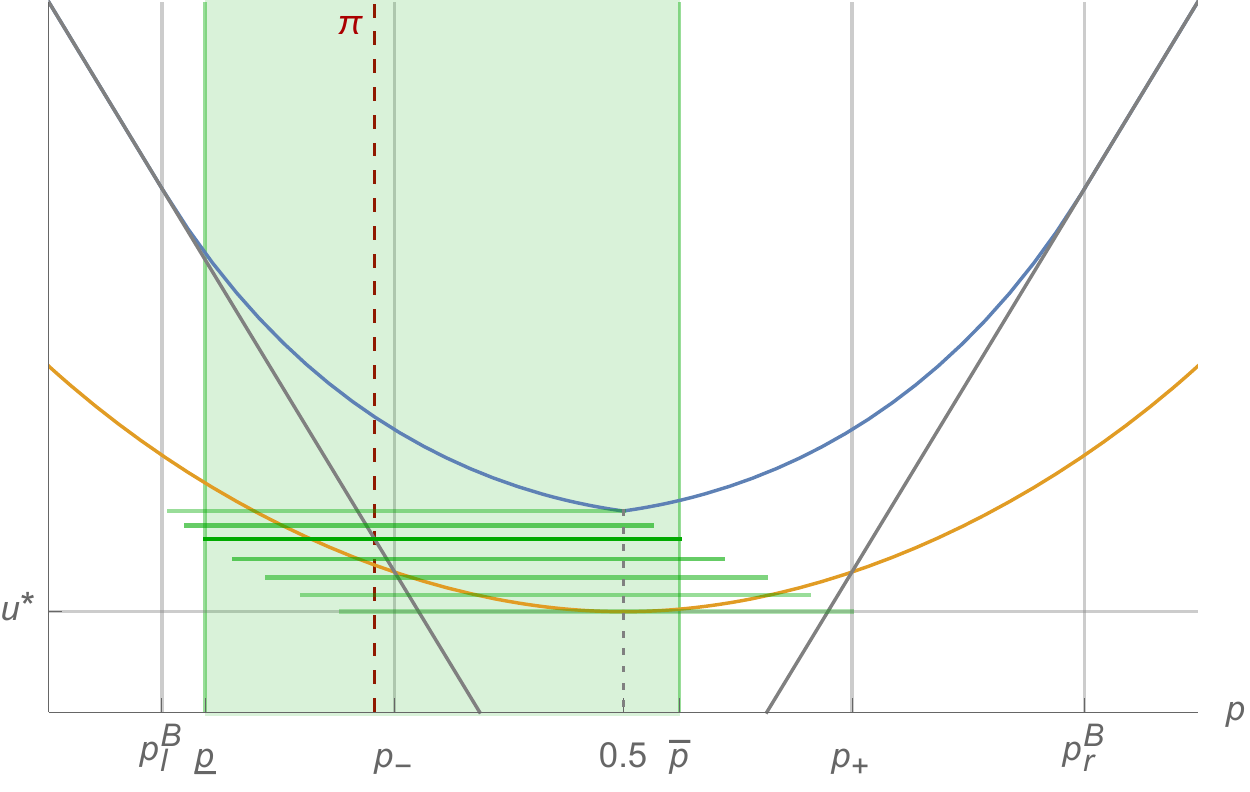}
  \end{subfigure}
\begin{subfigure}{.49\textwidth}
    \centering
  \includegraphics[width=.9\textwidth]{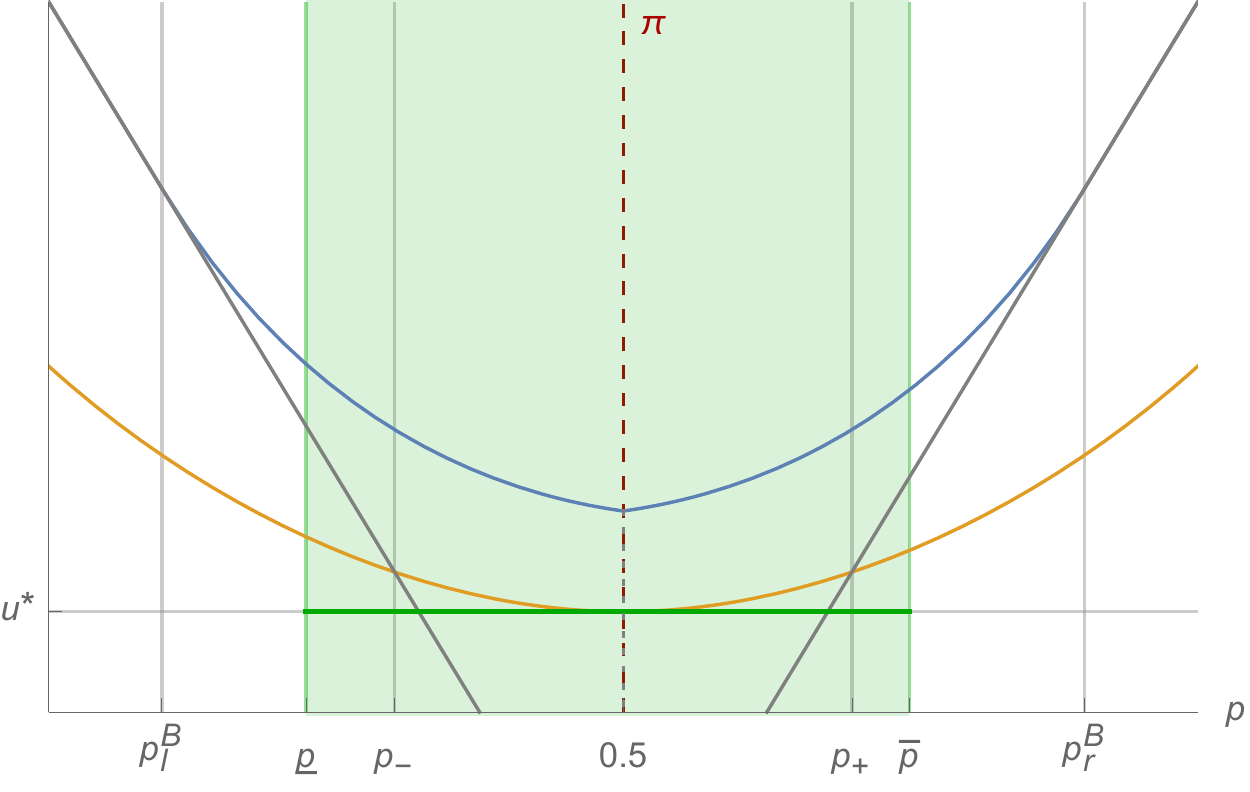}
\end{subfigure}
\caption{The case of intermediate $c$ and large $\Delta$.}
\label{fig:2news2}
\end{figure}

If $\pl< 1/2<\ph$, the DM employs a hedging strategy. If $[\pl, \ph]\supset [p_-,p_+]$ the DM either adopts split attention learning (if $u^*>\hat u$) or randomizes between $\ell$ and $r$ (if $u^*\leq\hat u$). As mentioned above, since the cost is intermediate, split-attention learning involves a strict welfare loss, so the Bayesian DM would never adopt that strategy. See the right panel of \Cref{fig:2news2}. If instead $\ph\in (1/2, p_+]$, the DM randomizes between $R$-evidence seeking and immediate action $\ell$.  Namely, she stops according to a Poisson rate or else she seeks $R$-evidence, as characterized in \Cref{thm:main_result}.   See the left panel of \Cref{fig:2news2}.  Symmetrically if  $\pl\in [p_-, 1/2)$, then the DM adopts randomized stopping, now, between $L$-evidence seeking and action $r$.

In sum, the equilibrium exhibits the behavioral patterns observed in the baseline model, such as {\it excessive learning} and {\it premature stopping}, but it also features a new strategy, namely split attention.

\end{document}